\documentclass[10pt]{article}
\usepackage{amssymb, amsmath}
\begin{document}

\makeatletter
\@addtoreset{equation}{section}
\def\theequation{\thesection.\arabic{equation}}
\def\@maketitle{\newpage
 \null
 {\normalsize \tt \begin{flushright} 
  \begin{tabular}[t]{l} \@date  
  \end{tabular}
 \end{flushright}}
 \begin{center} 
 \vskip 2em
 {\LARGE \@title \par} \vskip 1.5em {\large \lineskip .5em \begin{tabular}[t]{c}\@author 
 \end{tabular}\par} 
 \end{center}
 \par
 \vskip 1.5em} 
\makeatother
\topmargin=-1cm
\oddsidemargin=1.5cm
\evensidemargin=-.0cm
\textwidth=15.5cm
\textheight=22cm
\setlength{\baselineskip}{16pt}
\title{Metric-Like Formalism for Matter Fields Coupled to  3D Higher Spin Gravity}
\author{
Ippei~{\sc Fujisawa} and
Ryuichi~{\sc Nakayama}
       \\[1cm]
{\small
    Division of Physics, Graduate School of Science,} \\
{\small
           Hokkaido University, Sapporo 060-0810, Japan}
}
\date{
EPHOU-13-003  \\
April  2013 @
}
%
%
\maketitle

\begin{abstract} 
Action integral for a matter system composed of 0- and 2-forms, $C$ and $B_{\mu\nu}$,  
topologically coupled to 3D spin-3 gravity  is considered first in the frame-like formalism. The field $C$ satisfies an eq of motion, $\partial_{\mu} \, C+A_{\mu} \, C-C \, \bar{A}_{\mu}=0$, where  $A_{\mu}$ and $\bar{A}_{\mu}$ are the Chern-Simons gauge fields. With a suitable gauge fixing of a new local symmetry and  diffeomorphism, only one component of $B_{\mu\nu}$, say $B_{\phi r}$, remains non-vanishing and satisfies $\partial_{\mu} \, B_{\phi r}+\bar{A}_{\mu} \, B_{\phi r}-B_{\phi r} \, A_{\mu}=0$. These eqs are the same  as those for 3d (free) Vasiliev scalars, $C$ and $\tilde{C}$. The spin connection is eliminated by solving the eq of motion 
for the total action, and it is shown that in the resulting metric-like formalism, $(BC)^2$ interaction terms are induced 
because of the torsion. The world-volume components of the matter field, $C^0$, $C^{\mu}$ and 
$C^{(\mu\nu)}$, are introduced by contracting the local-frame index of $C$ with those of the inverse vielbeins, 
$E_a^{\mu}$ and $E_a^{(\mu\nu)}$, which were defined by the present authors in ArXiv:1209.0894 [hep-th]. 

3D higher spin gravity theory contains various metric-like fields. These metric-like fields, as well as 
the new connections and the generalized curvature tensors,  introduced in the above mentioned paper, 
are explicitly expressed in terms of the metric $g_{\mu\nu}$ and the spin-3 field $\phi_{\mu\nu\lambda}$ by 
means of the $\phi$-expansion. The action integral for the pure spin-3 gravity in the metric-like formalism up to ${\cal O}(\phi^2)$, 
obtained before in the literature, is re-derived. Then the matter action is re-expressed in terms of $g_{\mu\nu}$,  
$\phi_{\mu\nu\rho}$ and the covariant derivatives for spin-3 geometry.  Spin-3 gauge transformation is extended to the matter fields. 

It is also found that the action of the matter-coupled theory in the metric-like formalism has larger symmetry than that of the pure spin-3 gravity. The matter-coupled theory in the metric-like formalism is invariant under the ordinary diffeomorphisms, because the vielbein and the spin connection are covariant vectors of diffeomorphisms. They are also gauge fields for the local translation. 
In the pure spin-3 gravity this symmetry provides the ordinary diffeomorphism and the spin-3 transformation in the metric-like formalism. When the matter is coupled, the local translation yields a new symmetry in the metric-like formalism, which does not contain diffeomorphism.

\end{abstract}
\newpage
\setlength{\baselineskip}{18pt}

\newcommand{\bm}[1]{\mbox{\boldmath $#1$}}
\section{Introduction}
\hspace{5mm}
In these several decades, study of higher-spin gauge theories has made steady progress. In the 
frame-like approach, Vasiliev proposed non-linear equations of motion for infinite tower of 
higher-spin gauge fields.\cite{FV1}\cite{FV2}\cite{Vasiliev}\cite{Blencowe} Although its 
description based on an action principle is still under investigation, it was conjectured that 
the higher-spin gravity in 3 dimensions is dual to the 2D W minimal CFT 
models,\cite{GG}\cite{GaHa}\cite{GGHR}\cite{Ahn}\cite{GGreview} and this duality 
(correspondence) has been studied in the version of the model with 
linearized scalar fields.\cite{CY}\cite{KP2}\cite{AKP2}\cite{HKP}\cite{CAhn} 

It was also noticed that in 3 dimensions great simplifications occur.\cite{Campoleoni}\cite{C2} 
The higher-spin fields can be truncated to only those with spin $s \leq N$ and the theory with 
negative cosmological constant in the frame-like approach can be defined in terms of the 
$SL(N,R) \times SL(N,R)$ Chern-Simons action. Various black hole solutions were found and their 
properties were 
studied.\cite{GK}\cite{AGKP}\cite{BCT}\cite{KP}\cite{KP2}\cite{GHJ}\cite{review}\cite{CLW}\cite{CLW2} 
In this frame-like approach the gravity and the higher-spin fields are described in terms of the 
vielbein $e^a_{\mu} \, (a=1,..,N^2-1)$ and the spin connection $\omega^a_{\mu}$ and the action integral 
is first order in the derivatives of these fields. 

In gravity theories, there also exists a metric-like approach. In this approach the fields are metric 
tensor $g_{\mu\nu}$ and higher-spin gauge fields, and an action for massless higher-spin fields was 
proposed by Fronsdal.\cite{Fronsdal} Correlation functions on the boundary CFT are studied by using 
holographic renormalization.\cite{LS}Cubic interaction vertices were also constructed.\cite{FT}  
In this approach the action is second order in the derivatives. 
It is more suitable for understanding of geometric properties. 

In 3D spin-3 gravity such an action was constructed in \cite{Campometric}  by perturbation in powers 
of the spin-3 field $\phi_{\mu\nu\rho}$ up to ${\cal O}(\phi^2)$ . In \cite{FN} the present authors 
eliminated the spin connection  from the $SL(3,R) \times SL(3,R)$ Chern-Simons theory by solving 
the torsion-free condition and then substituting the solution into the action integral. In this way 
we obtained an action integral which is quadratic in the derivatives of fields without using the 
perturbative methods. For this purpose we introduced a subsidiary vielbein $e^a_{(\mu\nu)}$, which 
is expressed in terms of the ordinary vielbein $e^a_{\mu}$, in order to define inverse vielbeins, 
$E_a^{\mu}$, $E_a^{(\mu\nu)}$. This allowed us to solve the torsion-free conditions.  We defined 
generalizations of the Christoffel symbols and curvature tensors. However, the action still contains 
metric-like fields expressed in terms of the vielbeins and the structure constants of the Lie algebra. 
In this sense, although the theory is in the second-order formalism, it is not a complete metric-like 
formalism.  On the other hand, the existence of the second-order formalism is explicitly demonstrated. 
Advantage of our formalism is that this can be established to all orders in $\phi$. 
To express the action only in terms of the metric-like fields, it is necessary to express the vielbein 
in terms of the metric and the spin-3 field, and substitute the result into the action integral. This 
can be performed only by perturbation in powers of the spin-3 field. This is one of the purposes of 
this paper. We will express $\Gamma^M_{\mu N}$, ${R^M}_{N\mu\nu}$ and other quantities in terms of 
$g_{\mu\nu}$ and $\phi_{\mu\nu\lambda}$ explicitly, and then express the action integral for spin-3 
gravity in terms of the metric-like fields, and justify our formalism. The action for the spin-3 
field turned out to agree with the result of \cite{Campometric}. 

Now, although 3D higher-spin gauge theory can be formulated in terms of the Chern-Simons theory, 
this is just a \lq pure spin-3 gravity' theory. It is desirable to include matter fields.  Actually, 
there must be \lq scalar fields' in Vasiliev theory. It is, however, a very difficult task to 
construct action integrals for matter fields interacting with higher-spin gravity fields  in an invariant manner 
in the metric-like approach. To our knowledge, this has never been done in the literature.  In the second
 half of this paper, we will extend the 3D spin-3 gravity theory by topologically coupling  matter 
fields. Matter fields are  a 0-form $C$ and a 2-form $B=\frac{1}{2} \, B_{\mu\nu} \, dx^{\mu} \wedge dx^{\nu}$. 
Hamiltonian analysis of this model will be performed and it is shown that for fixed flat Chern-Simons gauge fields, 
$A, \, \bar{A}=\omega \pm \frac{1}{\ell} \, e$, this model provides Lagrangian formulation of the scalars in (free) 3d Vasiliev theory.

These matter fields are 3 $\times $ 3 matrices and have local frame indices: $C^a$, $C^0$, ${B_a}_{\mu\nu}$,
 ${B_0}_{\mu\nu}$. ($a=1,2, ..,8$) This is a frame-like approach. By contracting these indices with 
those of the vielbeins $e^a_{\mu}$, $e^a_{(\mu\nu)}$, or their inverse $E^{\mu}_a$, $E^{(\mu\nu)}_a$, we 
obtain matter fields with world-volume indices in the metric-like approach: $C^0$, 
$C^{\mu}=E^{\mu}_a \, c^a$, $C^{(\mu\nu)}=E^{(\mu\nu)}_a \, C^a$, {\em etc}. The $C$ field is not just a 
scalar field, but turned into a set of scalar, vector and tensor fields. Similarly for $B$. 

In order to construct the spin-3 gravity theory interacting with the matter fields in the metric-like 
formalism, we need to eliminate the spin connection $\omega^a_{\mu}$ by using some of the equations 
of motion. Due to the matter-coupling the equation of motion with respect to the spin connection is different from the one for the pure spin-3 gravity: a torsion tensor appears. 
The result $\tilde{\omega}^a_{\mu}(e,B,C)$ differs from the one 
 $\omega_{\mu}^a(e)$ obtained by solving the equation of motion in the pure spin-3 gravity.  Accordingly, the Christoffel-like connections and the spin connection must be replaced by new ones. We will show that the action integral for the pure spin-3 gravity sector and that for the matter sector get extra interaction terms, which are quadratic in the torsion tensor and of quartic order of matter fields, $(BC)^2$. The action for the gravity sector is given by
\begin{eqnarray}
S_{\text{CS}} &=& \frac{k}{48\pi \ell} \, \int d^3x \, \left(-\epsilon^{\mu\nu\lambda} \, {F_{\mu M}}^N \, {R^M}_{N \nu \lambda}+\frac{24}{\ell^2} \, \tilde{e}-2 \, \epsilon^{\mu\nu\lambda} \, {F_{\mu M}}^N \, \Delta \, \Gamma^M_{\nu K} \, \Delta \, \Gamma^K_{\lambda N}\right) 
\end{eqnarray}
$\tilde{e}$ is a generalized cosmological term (\ref{cosmological}). ${F_{\mu M}}^N$ is a metric-like tensor defined in (\ref{Fabc}) and appendix E. 
The first and the second terms can be rewritten as a sum of Einstein-Hilbert action and  Fronsdal's spin-3 gauge action. 
The matter action in the metric-like formulation is given by
\begin{eqnarray}
S_{\text{matter}} &=& \int d^3x \, \sqrt{-g} \, \left[
{B_N}^{\lambda} \,  \left( \nabla_{\lambda}\, C^N + \frac{1}{\ell} \, {K^N}_{M\lambda}  \, C^M
\right)+\frac{2}{\ell} \, {B_{\lambda}}^{\lambda} \, C^0  \right.      \nonumber \\  
&& \left. +
{B_0}^{\lambda} \, \left(\partial_{\lambda} \, C^0+\frac{4}{3\ell} \,  C_{\lambda}  \right)   + {B_N}^{\lambda} \, C^M \, \Delta \, \Gamma^N _{\lambda M} \right] \label{MatterMetriclike}
\end{eqnarray}
Here $\nabla_{\lambda}$ is a covariant derivative associated with the connections $\Gamma^M_{\mu N}$. ${B_N}^{\lambda}$ and ${B_0}^{\lambda}$ are defined in (\ref{BNl})-(\ref{B0l}), and another metric-like tensor ${K^N}_{M\lambda}$ is given in (\ref{tensorK}) and appendix C.   $\Delta \, \Gamma^N _{\lambda M}$, (\ref{DeltaGamma}), is a shift of the connection $\Gamma^N _{\lambda M}$ due to the torsion. Those terms which include $\Delta \Gamma$ yield  fourth order  interactions of the form $(BC)^2$. 
Under spin-3 transformation, fields $C^0$, $C^{\mu}$ and $C^{(\mu\nu)}$ transform into each other, and the transformation rule will be obtained explicitly. 

The novel feature of the spin-3 gravity with matter coupling is that local translation of the metric $g_{\mu\nu}$ and the spin-3 gauge field $\phi_{\mu\nu\lambda}$ contains terms which depend on the matter fields $C$, $B$ through the torsion tensor. This symmetry has an origin in the gauge symmetry of the frame-like formalism. However, our spin-3 gravity with matter coupling is still invariant under diffeomorphism, because the vielbein $e^a_{\mu}$ is a covariant vector under diffeomorphism. 
Hence, the difference of the two transformations is also a symmetry transformation, and the symmetry of the spin-3 gravity theory in the metric-like formalism is enhanced by the matter coupling.  
We also find that the transformation rules of the matters, $C$ and $B$, become non-linear in the matter fields.

This paper is organized as follows. In sec.2 our second order formalism will be reviewed briefly. In sec.3 
various metric-like fields will be expressed in terms of the metric $g_{\mu\nu}$ and the spin-3 field $\phi_{\mu\nu\lambda}$ by using perturbation in $\phi$. The four kinds of the generalized Christoffel connection $\Gamma^M_{\mu N}$, the generalized curvature tensor ${R^M}_{N\mu\nu}$ and the action integral for the pure spin-3 gravity, are then expressed in terms of these fields. The transformation properties of $g_{\mu\nu}$ and $\phi_{\mu\nu\lambda}$ are then studied.  In sec.4 an action integral for matter fields coupled to the spin-3 gravity is explicitly written down.  Hamiltonian analysis of this model will be performed.The action integral is then converted into a metric-like form by contracting the indices of the matter 
fields with those of the vielbeins.  Transformation rules of the matter fields under spin-3 transformation will be worked out explicitly. Sec.5 is a summary. In appendix A some formulae for $sl(3,R)$ algebra are presented. This appendix is the same as appendix A of \cite{FN}, but 
included for convenience. 
In appendices B-H, some equations for tensors and transformations are presented. We performed various computations by using {\it xAct} packages for Mathematica\cite{xAct}. 

\section{Brief Review of the Formalism}
\hspace{5mm}
In \cite{FN} we defined a subsidiary field $e^a_{(\mu\nu)}$ in terms of the vielbein field $e^a_{\mu}$. 
\begin{equation}
e^a_{(\mu\nu)} = \frac{1}{2} \, {d^a}_{bc} \, e^b_{\mu} \, e^c_{\nu}-\frac{1}{6} \, g_{\mu\nu} \, g^{\lambda\rho} \, {d^a}_{bc} \, e^b_{\lambda} \, e^c_{\rho}
\end{equation}
This is symmetric under interchange of $\mu$ and $\nu$. The second term on the righthand side ensures 
that $e^a_{(\mu\nu)}$ is traceless: $g^{\mu\nu} \, e^a_{(\mu\nu)}=0$ so that $e^a_{(\mu\nu)}$ has five independent components with respect to the indices $\mu$, $\nu$. In spin-3 gravity the index $a$ is 
associated with $SL(3,R)$ and runs over $a=1,2,\dots,8$. So the set of generalized vielbeins, ($e^a_{\mu}$ and $e^a_{(\mu\nu)}$), makes up an 8 $\times$ 8 matrix.  

This allows us to define the inverse vielbeins, $E_a^{\mu}$ and $E_a^{(\mu\nu)}$, by the relations,
\begin{eqnarray}
E^{\mu}_a \ e^a_{\nu} &=& \delta^{\mu}_{\nu}, \qquad 
E^{\mu}_a \ e^a_{(\nu\lambda)} = 0,  \nonumber \\
E^{(\mu\nu)}_a \ e^a_{\lambda} &=&0, \qquad 
E^{(\mu\nu)}_a \ e^a_{(\lambda \rho)} = \delta^{\mu}_{\lambda} \, 
\delta^{\nu}_{\rho}+
\delta^{\nu}_{\lambda} \, \delta^{\mu}_{\rho}
-\frac{2}{3} \ g_{\lambda\rho} \ g^{\mu\nu},
\label{Ee}
\end{eqnarray}
and then solve the torsion-free condition 
$\partial_{\mu} \, e^a_{\nu}-\partial_{\nu} \, e^a_{\mu}+{f^a}_{bc} \, \omega^b_{\mu} \, e^c_{\nu}-{f^a}_{bc} \, \omega^b_{\nu} \, e^c_{\mu}=0$ 
and express the spin connection $\omega^a_{\mu}$  in terms of the vielbeins. The result is 
\begin{equation}
\omega_{\mu}^a=\omega_{\mu}^a(e) \equiv \frac{1}{12} \ {f^{ab}}_{c} \ 
E_b^{\lambda} \ \nabla_{\mu} \ e^c_{\lambda}+\frac{1}{24} \ {f^{ab}}_{c} \ 
E_b^{(\lambda\rho)} \ \nabla_{\mu} \ e^c_{(\lambda\rho)}.
\label{omegae}
\end{equation}
Here $\nabla_{\mu}$ is a new covariant derivative.  For this definition,  we need to introduce some notation for indices. Let $M, N, \dots$ denote a set of two types of indices, $\mu$ and $(\nu,\lambda)$. 
Then covariant derivatives of tensors, $V_M$, $V^M$, with this type of indices are defined by
\begin{eqnarray}
\nabla_{\mu} \, V_M &=& \partial_{\mu} \, V_M-\Gamma^N_{\mu M} \, V_N
=\partial_{\mu} \, V_M-\Gamma^{\nu}_{\mu M} \, V_{\nu}-\frac{1}{2} \,\Gamma^{(\nu\lambda)}_{\mu M} \, V_{(\nu\lambda)} , \\
\nabla_{\mu} \, V^M &=& \partial_{\mu} \, V^M+\Gamma^M_{\mu N} \, V^N= \partial_{\mu} \, V^M+\Gamma^M_{\mu \nu} \, V^{\nu}+\frac{1}{2} \, \Gamma^M_{\mu (\nu\lambda)} \, V^{(\nu\lambda)}.
\end{eqnarray}
Note that a factor $\frac{1}{2}$ is associated to the summation over the pair of indices  $(\mu\nu)$. We 
will use this summation convention throughout this paper. There are four types of connections $\Gamma^M_{\mu N}$ according to the types of $M$ and $N$.  These are generalizations of the Christoffel symbol in the ordinary gravity.\footnote{A generalization of Christoffel symbol in higher-spin gauge theories was once considered in \cite{WF}. The direction of the generalization is, however, different from ours.} In \cite{FN} the expression for these connections are determined in terms of the metric-like quantities in such a way that the full covariant derivatives of the generalized vielbeins vanish: $D_{\mu} \, e^a_{\nu}= \nabla_{\mu} e^a_{\nu}+{f^a}_{bc} \, \omega^b_{\mu} \, e^a_{\nu}=0$ and $D_{\mu} \, e^a_{(\nu\lambda)}=
\nabla_{\mu} e^a_{(\nu\rho)}+{f^a}_{bc} \, \omega^b_{\mu} \, e^a_{(\nu\rho)}=0$. 
The (extended) metric compatible with the covariant derivatives $\nabla_{\mu}$ is given by 
\begin{equation}
G_{MN}= e^a_M \, e_{aN}. \label{GMN}
\end{equation} 
This is decomposed into four blocks and three of them are related to the metric and the spin-3 field: the first block is the ordinary metric, $G_{\mu\nu}=g_{\mu\nu}$. Off-diagonal  blocks are    $G_{\mu(\nu\rho)}=G_{(\nu\rho)\mu}=\phi_{\mu\nu\rho}-\frac{1}{3} \, g_{\nu\rho} \, g^{\lambda\sigma} \phi_{\lambda\sigma\mu}$, where $\phi_{\mu\nu\rho}$ is the spin-3 field. The last one $G_{(\mu\nu) (\lambda\rho)}$ is new, but in principle can be expressed in terms of $g_{\mu\nu}$ and $\phi_{\mu\nu\lambda}$, as is displayed in appendix B. However, the covariant derivative $\nabla_{\mu}$ defined above mixes the two types of indices, $\mu$ and $(\mu\nu)$, and the last component $G_{(\mu\nu) (\lambda\rho)}$ is also important. The perturbative expansions of $G_{MN}$ and $G^{MN}$ in powers of $\phi$ are given in appendix B. 

In order to distinguish the ordinary Christoffel symbol from the above new connections $\Gamma^M_{\mu N}$, the former will be denoted as $\hat{\Gamma}^{\mu}_{\nu\lambda}$ throughout this paper. The covariant derivatives and the curvature tensor associated with the Christoffel symbol will be denoted as $\hat{\nabla}_{\mu}$ and 
${\hat{R}^{\lambda}}_{\ \ \rho\mu\nu}$, respectively. 

3D spin-3 gravity is defined by $SL(3,R) \times SL(3,R)$ Chern-Simons action and the field variables are 
the vielbein $e^a_{\mu}$ and the spin connection $\omega^a_{\mu}$. This is a first-order formalism. By substituting $\omega^a_{\mu}(e)$ in (\ref{omegae}) into $\omega^a_{\mu}$ in the Chern-Simons action, an action integral in the second-order formalism was obtained in \cite{FN}. 
In the next section we will express the action integral and several geometric quantities only in terms of the metric and the spin-3 field by using perturbative expansions  in $\phi$. 

\section{Vielbein in terms of metric-like fields}
\hspace{5mm}
As was explained in \cite{FN}, there are various metric-like fields in spin-3 gravity. Among them the metric 
field $g_{\mu\nu}$ and the spin-3 field $\phi_{\mu\nu\lambda}$ are important because 
the others are assumed to be expressible in terms of these.  They are define by\footnote{For the definitions of the basis $t_a$ of $sl(3,R)$, see appendix A.}
\begin{eqnarray}
g_{\mu\nu} &=& \frac{1}{2} \, \mbox{tr} \, e_{\mu} \, e_{\nu}=h_{ab} \, e^a_{\mu} \, e^b_{\nu}, \label{gee} \\
\phi_{\mu\nu\lambda} &=& \frac{1}{4} \,  \mbox{tr} \, e_{\mu} \, \{ e_{\nu}, \, e_{\lambda} \}=\frac{1}{2} \, d_{abc} \, e^a_{\mu} \, e^b_{\nu} \, e^c_{\lambda}
\end{eqnarray}
Here $\{ \ , \ \}$ is an anti-commutator.  Other metric-like fields are similarly defined 
in terms of traces of  products of the vielbeins, so if the above relations were  
solved for $e_{\mu}$, all the metric-like fields would be expressed in terms of 
 $g_{\mu\nu}$ and the spin-3 field $\phi_{\mu\nu\lambda}$. This is what we are up to.
The vielbein $e^a_{\mu}$ has 24 components, and among them 8 of those are gauge 
degrees of freedom for local 'Lorentz rotations'. Up to these gauge transformations, 
the vielbein is expected to be obtained uniquely. 

By using the Killing metric $h_{ab}$ \footnote{See appendix A for our conventions. } the explicit form of relation (\ref{gee}) reads 
\begin{eqnarray}
g_{\mu\nu} = e^2_{\mu} \, e^2_{\nu}-2 \, (e^1_{\mu} \, e^3_{\nu}+e^3_{\mu} \, e^1_{\nu})
+8 \,  (e^4_{\mu} \, e^8_{\nu}+e^8_{\mu} \, e^4_{\nu})-2 \,  (e^5_{\mu} \, e^7_{\nu}+e^7_{\mu} \, e^5_{\nu})+\frac{4}{3} \, e^6_{\mu} \, e^6_{\nu}
\end{eqnarray}
Similar, but more involved eqs for $\phi$ can also be written down explicitly using the symmetric structure constant $d_{abc}$ The indices $\mu$, $\nu$.. run over $r, t, \phi$. 

At present, this attempt  can be fulfilled only perturbatively: we must assume that $\phi_{\mu\nu\lambda}$ is small, and resort to expansions in powers of $\phi$. We will use the following gauge fixing conditions for local frame rotations.
\begin{equation}
e^a_{r}=0,   \qquad (a \neq 2,6)
\end{equation}
 The remaining two are given by
\begin{equation}
e^1_{t}=e^3_{t}, \qquad e^7_t \, e^1_{\phi}+e^5_t \, e^3_{\phi}=0. \label{extragauge}
\end{equation}

First, the eqs for $g_{rr}$ and $\phi_{rrr}$, 
\begin{eqnarray}
g_{rr} &=& (e^2_r)^2+ \frac{4}{3} \, (e^6_r)^2, \label{eqgrr} \\
\phi_{rrr} &=& -\frac{8}{9} \, (e^6_r)^3+2 \, (e^2_r)^2 \, e^6_r, 
\end{eqnarray}
are combined into a cubic eq for $e^6_r$
\begin{equation}
(e^6_r)^3-\frac{9}{16} \, g_{rr} \, e^6_r+\frac{9}{32} \, \phi_{rrr}=0  \label{cubic}
\end{equation}
This eq has one or three real roots according to the sign of the discriminant $D= 
(81/4096) \, \phi_{rrr}^2-(27/256) \, g_{rr}^3$:  for $D < 0$ there are three real roots.  
\begin{equation}
e^6_r= \omega^k \, (-\frac{9}{64} \, \phi_{rrr}+i \, \sqrt{-D})^{1/3}+\omega^{3-k}
 \, (-\frac{9}{64} \, \phi_{rrr}-i \, \sqrt{-D})^{1/3} \qquad (k=0,1,2)
\end{equation}
Here $\omega=e^{2\pi i/3}$, and $\sqrt{-D}$ denotes the positive root.  For $D \geq 0$ there is only a single one. 
\begin{equation}
e^6_r= (-\frac{9}{64} \, \phi_{rrr}+ \sqrt{D})^{1/3}+
 \, (-\frac{9}{64} \, \phi_{rrr}- \sqrt{D})^{1/3} 
\end{equation}
Assuming that $g_{rr} >0$,\footnote{Throughout this paper the world-volume indices take values $\mu, \nu, \dots=r,t,\phi$, and the signature is $(+,-,+)$. } $D$ turns out negative for small $|\phi_{rrr}|$,  and there are three solutions for the inversion problem.   For $\phi=0$, the solutions are $e^6_r=0, \pm \frac{3}{4} \, \sqrt{g_{rr}}$. Because $\phi=0$ corresponds to the ordinary spin-2 gravity, in what follows, we will choose the branch which reduces to $e^6_r=0$ at $\phi=0$.  This solution is smoothly connected to the one for $D \geq 0$ and large $|\phi_{rrr}|$. Then the small $\phi$ expansion for $e^6_r$ is given by 
\begin{equation}
e^6_r= \frac{1}{2} \,  (g_{rr})^{-1} \, \phi_{rrr}+\frac{2}{9} \, (g_{rr})^{-4} \, (\phi_{rrr})^3+\dots \label{sole6r}
\end{equation}
In turn, $e^2_r$ is determined by solving  (\ref{eqgrr}).
\begin{eqnarray}
e^2_r &=& \sqrt{g_{rr}-\frac{4}{3} \, (e^6_r)^2} = \sqrt{g_{rr}}+\dots
\end{eqnarray}
The above eq shows that there is an upper bound for $|\phi_{rrr}|$, given the value of $g_{rr}$: (\ref{cubic}) shows that $\phi_{rrr}$ grows as $(e^6_r)^3$ for large $e^6_r$. 

The eqs for $g_{rt}$, $g_{r\phi}$, $\phi_{rrt}$ and  $\phi_{rr\phi}$ determine $e^6_t$, $e^6_{\phi}$, $e^2_t$ and $e^2_{\phi}$. Results are 
\begin{eqnarray}
e^6_t &=& \frac{\frac{3}{2} \, \phi_{rrt}-2 \, g_{rt} \, e^6_r}{g_{rr}-\frac{16}{3} \, (e^6_r)^2}, 
\\
e^2_r &=& \frac{(g_{rr}-\frac{8}{3} \, (e^6_r)^2) \, g_{rt}-2 \, \phi_{rrt} \, e^6_r}{
g_{rr}-\frac{16}{3} \, (e^6_r)^2}, \\
e^6_{\phi} &=& \frac{\frac{3}{2} \, \phi_{rr\phi}-2 \, g_{r\phi} \, e^6_r}{g_{rr}-\frac{16}{3} \, (e^6_r)^2}, \\
e^2_{\phi} &=& \frac{(g_{rr}-\frac{8}{3} \, (e^6_r)^2) \, g_{r\phi}-2\, \phi_{rr\phi} \, e^6_r}{
(g_{rr}-\frac{16}{3} \, (e^6_r)^2) \, e^2_r}.
\end{eqnarray}
The solution (\ref{sole6r}) is to be substituted into the above.

The remaining eqs for $g_{tt}$, $g_{t\phi}$, $g_{\phi\phi}$, $\phi_{rtt}$, $\phi_{rt\phi}$, $\phi_{r\phi\phi}$, $\phi_{ttt}$, $\phi_{tt\phi}$, $\phi_{t\phi\phi}$ and $\phi_{\phi\phi\phi}$
provide a set of coupled eqs that can at best be solved only by perturbations in powers of $\phi$. 
The redundancy in these eqs can be removed by the gauge fixing (\ref{extragauge}). To leading order ${\cal O}(\phi^0)$ we have 
\begin{eqnarray}
e^1 &=& \frac{1}{2} \, \sqrt{\frac{-g}{g_{rr}} \, g^{\phi\phi}} \, dt+ \frac{1}{2 \, \sqrt{g_{rr} \, g^{\phi\phi}}} \, (-\sqrt{-g} \, g^{t\phi}-\sqrt{g_{rr}}) \, d\phi, \\
e^2 &=& \sqrt{g_{rr}} \, dr+\frac{1}{\sqrt{g_{rr}}} \, g_{rt} \, dt+\frac{1}{\sqrt{g_{rr}}} \, g_{r\phi} \, d\phi, \\
e^3 &=&\frac{1}{2} \, \sqrt{\frac{-g}{g_{rr}} \, g^{\phi\phi}} \, dt+ \frac{1}{2 \, \sqrt{g_{rr} \, g^{\phi\phi}}} \, (-\sqrt{-g} \, g^{t\phi}+\sqrt{g_{rr}}) \, d\phi. 
\end{eqnarray}
Here $g^{\mu\nu}$ is the inverse of $g_{\mu\nu}$ and $g$ is the determinant of $g_{\mu\nu}$. 
Other vielbeins $e^4, \dots, e^8$ begin with ${\cal O}(\phi^1)$.  
In order to compute the action integral to the first non-trivial order, and in terms of the metric-like 
fields, we computed $e^1, e^2 , e^3$ up to 
${\cal O}(\phi^2)$ and $e^4, \dots, e^8$ up to ${\cal O}(\phi^3)$.

\subsection{ Connections and Curvature Tensor }
\hspace{5mm}
By using the above results we can compute all quantities necessary for writing down the action integral. 
These include the generalized cosmological term.\footnote{ $\phi_{\mu}=\phi_{\mu\nu\lambda} \, g^{\nu\lambda}$. Indices are raised and lowered by $g^{\mu\nu}$ and $g_{\mu\nu}$, except for $\xi^M$, $\xi_M$ and $C^M$, $C_M$ in sec.4. In the latter case, $G^{MN}$ and $G_{MN}$ play the role of interchanging the indices.}
\begin{equation}
\tilde{e} = \frac{1}{6} \, \epsilon^{\mu\nu\lambda} \, f_{abc} \, e^a_{\mu} \, e^b_{\nu} \, e^c_{\lambda}
= \sqrt{-g} \, \left( 1+\frac{1}{2} \, \phi_{\mu} \, \phi^{\mu}-\frac{1}{3} \, \phi_{\mu\nu\lambda} \, \phi^{\mu\nu\lambda} +{\cal O}(\phi^4)      \right)  \label{cosmological}
\end{equation}
Here $\epsilon^{\mu\nu\lambda}$ is a unit completely anti-symmetric symbol with $\epsilon^{rt\phi}=+1$. 

To compute Christoffel-like connections $\Gamma^M_{\mu N}$, it is necessary to define tensor $M_{(\mu\nu)( \lambda\rho)}$ and its inverse, $J^{(\mu\nu)( \lambda\rho)}$. These are defined\cite{FN} as 
\begin{eqnarray}
M_{(\mu\nu)( \lambda\rho)} &=& G_{(\mu\nu)( \lambda\rho)}-G_{(\mu\nu)\alpha} \, g^{\alpha\beta} \, G_{\beta( \lambda\rho)}, \\
\frac{1}{2} \, J^{(\mu\nu)(\sigma\kappa)} \, M_{(\sigma\kappa)(\lambda\rho)} &=& P^{\mu\nu}_{\lambda\rho}
\equiv \delta^{\mu}_{\lambda} \, \delta^{\nu}_{\rho} +
\delta^{\mu}_{\rho} \, \delta^{\nu}_{\lambda} -\frac{2}{3}\, g^{\mu\nu} \, g_{\lambda\rho} \label{Proj}
\end{eqnarray}
Here $P_{\mu\nu}^{\lambda\rho}$ is a projector onto a space of symmetric traceless tensors.\footnote{$\frac{1}{2} \, P^{\mu\nu}_{\kappa\sigma} \, P^{\kappa\sigma}_{\lambda\rho}=P^{\mu\nu}_{\lambda\rho}$} These have the 
expansion, $M=M^{(0)}+M^{(2)}+\dots$, where $M^{(n)}={\cal O}(\phi^n)$,  and similarly for $J$.
The 0-th terms are given by 
\begin{eqnarray}
M^{(0)}_{(\mu\nu)( \lambda\rho)} &=& \frac{1}{4} \, g_{\mu\sigma} \, g_{\nu\kappa} \, P^{\sigma\kappa}_{\lambda\rho},
\\
J^{(0)(\mu\nu)(\lambda\rho)} &=& 4 \, g^{\sigma\lambda} \, g^{\kappa\rho} \, P_{\sigma\kappa}^{\mu\nu}
\label{J0}
\end{eqnarray}

One of the four connections, $\Gamma^{(\lambda\rho)}_{\mu\nu}$, is given by 
\begin{equation}
\Gamma^{(\lambda\rho)}_{\mu\nu}= \frac{1}{6} (\hat{\nabla}_{\mu} \, \Phi_{\nu\kappa\sigma}+\hat{\nabla}_{\nu} \,
\Phi_{\mu\kappa\sigma}-\hat{\nabla}_{\kappa} \, \Phi_{\mu\nu\sigma}) \, J^{(\kappa\sigma)(\lambda\rho)}+
\frac{1}{4} \, S_{\mu\nu,\kappa\sigma} \, J^{(\kappa\sigma)(\lambda\rho)}. \label{Gamma21}
\end{equation}
Here $\Phi$ is the traceless part of $\phi$: $\Phi_{\mu\nu\lambda}=\phi_{\mu\nu\lambda}-\frac{1}{5} \, 
(g_{\mu\nu} \, \phi_{\lambda}+g_{\nu\lambda} \, \phi_{\mu}+g_{\lambda\mu} \, \phi_{\nu})$.  $S_{\mu\nu,\kappa\sigma}$ is a tensor obtained by solving a few algebraic equations\footnote{Eqs (3.30) and (3.32) in \cite{FN} } and the formal solution was presented in appendix D of \cite{FN}, where $S_{\mu\nu,\kappa\sigma}$ is given as an infinite sum of $S^{(n)}_{\mu\nu,\kappa\sigma}$, which is ${\cal O}(\phi^{2n+1})$. Up to ${\cal O}(\phi^{2})$, it is given by
\begin{eqnarray}
S_{\mu\nu,\lambda\rho} =S^{(0)}_{\mu\nu,\lambda\rho}
&=&   -\frac{3}{5} \, g_{\mu\nu} \, ({\cal A}_{\lambda\rho}
+{\cal A}_{\rho\lambda})
-\frac{2}{5} \, g_{\lambda\rho} \,  ({\cal A}_{\mu\nu}+{\cal A}_{\nu\mu})   \nonumber \\
&&
+\frac{3}{5} \, g_{\mu\nu} \, g_{\lambda\rho} \, {{\cal A}_{\alpha}}^{\alpha} -\frac{3}{10} \, (g_{\mu\lambda} \, g_{\nu\rho}+g_{\mu\rho} \, g_{\nu\lambda}) \, {\cal A}^{\alpha}_{\alpha} \nonumber \\
&& +\frac{1}{5} \, g_{\mu\lambda} \, (2 \, {\cal A}_{\nu\rho}+
{\cal A}_{\rho\nu}) 
+\frac{1}{5} \, g_{\nu\lambda} \, (2 \, {\cal A}_{\mu\rho}+ 
{\cal A}_{\rho\mu}) \nonumber \\
&& +\frac{1}{5} \, g_{\mu\rho} \, (2 \, {\cal A}_{\nu\lambda}
+ {\cal A}_{\lambda\nu})
+\frac{1}{5} \, g_{\nu\rho} \, (2\, {\cal A}_{\mu\lambda}+ 
{\cal A}_{\lambda\mu}) +{\cal O}(\phi^3), \label{S1}
\end{eqnarray}
where ${\cal A}_{\mu\nu}$ is given by 
\begin{eqnarray}
{\cal A}_{\mu\nu} &=& \hat{\nabla}_{\mu} \, \phi_{\nu}-\frac{5}{9} \, \hat{\nabla}^{\lambda} \, \Phi_{\mu\nu\lambda} +{\cal O}(\phi^3) \nonumber \\
&=& \frac{10}{9} \, \hat{\nabla}_{\mu} \, \phi_{\nu}+\frac{1}{9} \, \hat{\nabla}_{\nu} \, \phi_{\mu}-\frac{5}{9} \, \hat{\nabla}_{\rho} \, {\phi_{\mu\nu}}^{\rho}+\frac{1}{9} \, g_{\mu\nu} \, \hat{\nabla}_{\lambda} \, \phi^{\lambda}+
{\cal O}(\phi^3).
\end{eqnarray}
In eq (D.2) of \cite{FN}, which defines ${\cal A}_{\mu\nu}$, there also appears a term containing $W_{\sigma\kappa}$. This quantity, however, can be shown to be ${\cal O}(\phi^3)$ and does not contribute here. 

By using (\ref{Gamma21}), (\ref{J0}) and (\ref{S1}), we obtain
\begin{eqnarray}
\Gamma^{(\lambda\rho)}_{\mu\nu} &=& -\frac{4}{3} \, \hat{\nabla}^{(\lambda} \, {\phi^{\rho)}}_{\mu\nu}
-\frac{8}{3} \, g_{\mu\nu} \, \hat{\nabla}^{(\lambda} \, \phi^{\rho)}+\frac{8}{3} \, \delta_{(\mu}^{(\lambda} \, \hat{\nabla}_{\nu)} \, \phi^{\rho)}+\frac{8}{3} \, \hat{\nabla}_{(\mu} \, {\phi^{\lambda\rho}}_{\nu)} \nonumber \\
&&+\frac{8}{3} \, \delta^{(\lambda}_{(\mu} \, \hat{\nabla}^{\rho)}\, \phi_{\nu)}-\frac{8}{3} \, g^{\lambda\rho} \, \hat{\nabla}_{(\mu} \, \phi_{\nu)}+\frac{4}{3} \, g_{\mu\nu} \, \hat{\nabla}_{\sigma} \, \phi^{\lambda\rho\sigma}-\frac{8}{3} \, \delta^{(\lambda}_{(\mu} \, \hat{\nabla}_{|\sigma|} \, {{\phi^{\rho)}}_{\nu)}}^{\sigma} \nonumber \\
&& +\frac{4}{3} \, g^{\lambda\rho} \, \hat{\nabla}_{\sigma} \, {\phi_{\mu\nu}}^{\sigma}-\frac{2}{3} \, \delta^{(\lambda}_{\mu} \, \delta^{\rho)}_{\nu} \, \hat{\nabla}_{\sigma} \, \phi^{\sigma}+\frac{2}{3} \, g^{\lambda\rho} \, g_{\mu\nu} \, \hat{\nabla}_{\sigma} \, \phi^{\sigma} +{\cal O}(\phi^3). \label{Gamma21}
\end{eqnarray}
Indices between parentheses are meant to be completely symmetrized, and dividing by the number of terms 
that are needed for symmetrization is understood. 

Then $\Gamma^{\lambda}_{\mu\nu}$ is obtained by using the formula  $\Gamma^{\lambda}_{\mu\nu}=\hat{\Gamma}^{\lambda}_{\mu\nu}-
\frac{1}{2} \, \Gamma^{(\sigma\kappa)}_{\mu\nu} \, {\phi_{\sigma\kappa}}^{\lambda}$. \footnote{eq (3.17) in \cite{FN}}
\begin{eqnarray}
\Gamma^{\lambda}_{\mu\nu} &=& \hat{\Gamma}^{\lambda}_{\mu\nu}-\frac{4}{3} \, \phi^{\lambda \rho\sigma} \, \hat{\nabla}_{(\mu} \, \phi_{\nu)\rho\sigma}+\frac{4}{3} \, \phi^{\lambda} \, \hat{\nabla}_{(\mu} \, \phi_{\nu)}-\frac{4}{3} \, {{\phi^{\lambda}}_{(\mu}}^{\rho} \, \hat{\nabla}_{\nu)} \,  \phi_{\rho} +\frac{1}{3} \, {\phi^{\lambda}}_{\mu\nu} \, \hat{\nabla}_{\rho} \, \phi^{\rho} \nonumber \\
&& -\frac{4}{3} \,  {{\phi^{\lambda}}_{(\mu}}^{\rho} \, \hat{\nabla}_{|\rho|} \, \phi_{\nu)}+\frac{2}{3} \, \phi^{\lambda\rho\sigma} \, \hat{\nabla}_{\rho} \, \phi_{\mu\nu\sigma}-\frac{2}{3} \, \phi^{\lambda} \, \hat{\nabla}_{\rho} \, {\phi_{\mu\nu}}^{\rho}+\frac{4}{3} \, {\phi^{\lambda\rho}}_{(\mu} \, \hat{\nabla}_{|\sigma|}
\, {\phi_{\nu)\rho}}^{\sigma} \nonumber \\
&&+\frac{4}{3} \, g_{\mu\nu} \, \phi^{\lambda \rho\sigma} \, \hat{\nabla}_{\rho} \, \phi_{\sigma}-\frac{2}{3} \, g_{\mu\nu} \, \phi^{\lambda\rho\sigma} \, \hat{\nabla}_{\kappa} \, {\phi_{\rho\sigma}}^{\kappa}-\frac{1}{3} \, g_{\mu\nu} \, \phi^{\lambda} \, \hat{\nabla}_{\rho} \phi^{\rho}+{\cal O}(\phi^4)
\end{eqnarray}

The remaining two, $\Gamma^{\rho}_{\mu (\nu\lambda)}$ and $\Gamma^{(\rho\sigma)}_{\mu (\nu\lambda)}$
are obtained by using eqs (4.12) and (4.13) in \cite{FN}. For computing $\Gamma^{\rho}_{\mu (\nu\lambda)}$ 
to the leading order, which is ${\cal O}(\phi^1)$, we need to evaluate a quantity, 
\begin{eqnarray}
{K^{\rho}}_{(\sigma\kappa)\lambda} & \equiv & {d^a}_{bc} \, E^{\rho}_a \, e^b_{(\sigma\kappa)} \, e^c_{\lambda}
\nonumber \\  
&=&-\frac{1}{3} \, \delta^{\rho}_{\lambda} \, g_{\sigma\kappa}+\delta^{\rho}_{(\sigma} \, g_{\kappa)\lambda}
+{\cal O}(\phi^2)  \label{K1}
\end{eqnarray}
and we obtain 
\begin{eqnarray}
\Gamma^{\rho}_{\mu (\nu\lambda)} &=& -\frac{1}{3} \, \hat{\nabla}^{\rho} \, \phi_{\mu\nu\lambda}
+\frac{1}{3} \, g_{\nu\lambda} \, \hat{\nabla}^{\rho} \, \phi_{\mu}-\frac{1}{3} \, g_{\mu(\nu} \, \hat{\nabla}^{\rho} \, \phi_{\lambda)}+\frac{2}{3} \, \hat{\nabla}_{\mu} \, {\phi^{\rho}}_{\nu\lambda} \nonumber \\
&& -\frac{1}{3} \, \delta^{\rho}_{(\nu} \, \hat{\nabla}_{|\mu|} \, \phi_{\lambda)}+\frac{1}{3} \, \hat{\nabla}_{(\nu} \, {\phi^{\rho}}_{\lambda)\mu}-\frac{1}{3} \, g_{\mu(\nu} \, \hat{\nabla}_{\lambda)} \, 
\phi^{\rho}-\frac{1}{3} \, \delta^{\rho}_{(\nu} \, \hat{\nabla}_{\lambda)} \, \phi_{\mu}\nonumber \\
&& +\frac{2}{3} \, \delta^{\rho}_{\mu} \, \hat{\nabla}_{(\nu} \, \phi_{\lambda)} -\frac{1}{3} \, g_{\nu\lambda} \, 
\hat{\nabla}_{\sigma} \,  {\phi^{\sigma \rho}}_{\mu} +\frac{1}{3} \, g_{\mu(\nu} \, \hat{\nabla}_{|\sigma|} \, 
{\phi^{\sigma\rho}}_{\lambda)}+\frac{1}{3} \, \delta^{\rho}_{(\nu} \, \hat{\nabla}_{|\sigma|} \, {\phi_{\lambda)\mu}}^{\sigma} \nonumber \\
&& -\frac{1}{3} \, \delta^{\rho}_{\mu} \, \hat{\nabla}_{\sigma} \, {\phi_{\nu\lambda}}^{\sigma}+\frac{1}{6} \, 
\delta^{\rho}_{(\nu} \, g_{\lambda)\mu} \, \hat{\nabla}_{\sigma} \, \phi^{\sigma}-\frac{1}{6} \, \delta^{\rho}_{\mu} \, g_{\nu\lambda} \,  \hat{\nabla}_{\sigma} \, \phi^{\sigma}+{\cal O}(\phi^3). \label{Gamma12}
\end{eqnarray}

 For computing $\Gamma^{(\rho\sigma)}_{\mu (\nu\lambda)}$, we also need to evaluate another quantity, 
\begin{eqnarray}
{K^{(\rho\sigma)}}_{(\kappa\eta)\lambda} & \equiv & {d^a}_{bc} \, E^{(\rho\sigma)}_a \, e^b_{(\kappa\eta)} \, e^c_{\lambda}
\nonumber \\  
&=&-\frac{4}{3} \, g_{\lambda(\kappa} \, {\phi^{\rho\sigma}}_{\eta)}  +\frac{4}{3} \, g_{\kappa\eta} \, {\phi^{\rho\sigma}}_{\lambda}+\frac{8}{3} \, \delta^{(\rho}_{\lambda} \, {\phi^{\sigma)}}_{\kappa\eta}
-\frac{8}{3} \, \delta^{(\rho}_{(\kappa} \, {\phi^{\sigma)}}_{\eta)\lambda}-\frac{4}{3} \, g_{\lambda(\eta} \,  \delta^{(\rho}_{\kappa)} \, \phi^{\sigma)}  \nonumber \\
&& -\frac{4}{3} \, \phi_{(\eta} \, \delta^{(\rho}_{\kappa)} \, \delta^{\sigma)}_{\lambda} +\frac{4}{3} \, g^{\rho\sigma} \, g_{\lambda(\kappa} \, \phi_{\eta)}+\frac{4}{3} \, \delta^{\rho}_{(\kappa} \, \delta^{\sigma}_{\eta)} \, \phi_{\lambda}-\frac{8}{9} \, g^{\rho\sigma} \, g_{\kappa\eta} \, \phi_{\lambda}
+{\cal O}(\phi^3). \label{K3}
\end{eqnarray}
By using this tensor, we obtain
\begin{eqnarray}
\Gamma^{(\rho\sigma)}_{\mu (\nu\lambda)} &=& 4 \, \delta^{(\rho}_{(\nu} \, \Gamma^{\sigma)}_{|\mu|\lambda)}-\frac{4}{3} \, g^{\rho\sigma} \, g_{\kappa(\nu} \, \Gamma^{\kappa}_{|\mu|\lambda)}-\frac{2}{9} \, g^{\rho\sigma} \, g_{\nu\lambda} \, 
{\phi_{\kappa\tau}}^{\eta} \Gamma^{(\kappa\tau)}_{\mu\eta}
 +\frac{2}{3} \, g^{\eta(\rho} \, g_{\nu\lambda} \, {\phi^{\sigma)}}_{\kappa\tau} \, \Gamma^{(\kappa\tau)}_{\mu\eta} \nonumber \\
&& -\frac{2}{9}  \, g_{\nu\lambda} \, g^{\eta(\rho} \, \phi^{\sigma)} \Gamma^{(\kappa\tau)}_{\mu\eta}
+\frac{1}{2} \,  {K^{(\rho\sigma)}}_{(\kappa\tau)(\nu} \, \Gamma^{(\kappa\tau)}_{|\mu| \lambda)}  \nonumber \\
&&-\frac{1}{6} \, g_{\nu\lambda} \, g^{\alpha\eta} \, {K^{(\rho\sigma)}}_{(\kappa\tau)\eta} \, \Gamma^{(\kappa\tau)}_{\mu\alpha}+{\cal O}(\phi^4)
\end{eqnarray}
Note that this expression contains other kinds of $\Gamma$'s and they must be substituted. The final result is 
complicated and will not be displayed here. This connection has a non-vanishing trace with respect to the lower paired indices:
$g^{\nu\lambda} \, \Gamma^{(\rho\sigma)}_{\mu(\nu\lambda)}=-2 \, \partial_{\mu} \, g^{\rho\sigma}-\frac{2}{3} \, g^{\rho\sigma} \, g^{\nu\lambda} \, \partial_{\mu} \, g_{\nu\lambda}$. This is because the definition of this connection, $\Gamma^{(\rho\sigma)}_{\mu(\nu\lambda)}=E^{(\rho\sigma)}_a \, (\partial_{\mu} \, e^a_{(\nu\lambda)}-\nabla_{\mu} \, e^a_{(\nu\lambda)})$ contains the derivative $\partial_{\mu}$ of the vielbein. However, contrary to the expectation, the similarly defined connection,   $\Gamma^{\rho}_{\mu(\nu\lambda)}$, does not 
have a trace, $g^{\nu\lambda} \, \Gamma^{\rho}_{\mu(\nu\lambda)}=0$, as can be checked by using (\ref{Gamma12}). 

Finally, we turn to the generalized curvature tensor. This was defined in eq (6.11) of \cite{FN}.
\begin{equation}
{R^M}_{N\mu\nu} \equiv \partial_{\mu} \, \Gamma^M_{\nu N}-\partial_{\nu} \, \Gamma^M_{\mu N}
+\Gamma^M_{\mu K} \, \Gamma^K_{\nu N}-\Gamma^M_{\nu K} \, \Gamma^K_{\mu N}. \label{gRiemann}
\end{equation}
The components can be obtained by substituting the above results, and some of them are presented in appendix D.  
Here we make a brief comment.  Firstly, to the leading order, the component ${R^{\lambda}}_{\rho\mu\nu}$ 
agrees with the ordinary Riemann tensor in 3 dimensions, as it should.
\begin{eqnarray}
{R^{\lambda}}_{\rho\mu\nu} = {\hat{R}^{\lambda}}_{\rho\mu\nu} +{\cal O}(\phi^2),
\end{eqnarray}
where $\displaystyle  {\hat{R}^{\lambda}}_{\ \ \rho\mu\nu} =\partial_{\mu} \, \hat{\Gamma}^{\lambda}_{\nu \rho}-\partial_{\nu} \, \hat{\Gamma}^{\lambda}_{\mu \rho}
+\hat{\Gamma}^{\lambda}_{\mu \kappa} \, \hat{\Gamma}^{\kappa}_{\nu \rho}-\hat{\Gamma}^{\lambda}_{\nu \kappa} \, \hat{\Gamma}^{\kappa}_{\mu \rho}=\delta^{\lambda}_{\mu} \, \hat{R}_{\rho\nu}-\frac{1}{2} \, \hat{R} \, \delta^{\lambda}_{\mu} \, g_{\nu\rho}+\dots$. The second-order terms are too complicated to present here. 

Secondly, two types of components, ${R^{(\rho\sigma)}}_{\lambda\mu\nu}$ and ${R^{(\rho\sigma)}}_{(\lambda\kappa)\mu\nu}$, turn out to contain terms proportional to the ordinary Christoffel symbols. Therefore, these components do not act as tensors under diffeomorphism.  
The reason can be traced to the derivative $\partial_{\mu}$ on $\Gamma^M_{\nu N}$ in (\ref{gRiemann}). 
The traceless condition $g_{\rho\sigma} \, {R^{(\rho\sigma)}}_{N\mu\nu}=0$ is jeopardized by the derivative. 
However, those terms  proportional to the Christoffels are all proportional to $g^{\rho\sigma}$, and  by contracting these components by the projector $P_{\rho\sigma}^{\alpha\beta}$ (\ref{Proj}), we can obtain quantities covariant under diffeomorphisms. 
This means that more appropriate definition of the generalized curvature tensor may be the one, obtained by projecting out some terms by using (\ref{Proj}),
like $
\left[ {R^{(\rho\sigma)}}_{\lambda\mu\nu}\right]_{\text{redefined}} = \frac{1}{2} \, P^{\rho\sigma}_{\alpha\beta} \, {R^{(\alpha\beta)}}_{\lambda\mu\nu}$ 

If the indices of these curvature tensors are, however,  contracted with other tensors, such as ${F_{\mu M}}^N$ (\ref{Fabc}) below,  which also play the role of projector, it is not necessary to perform the above-mentioned redefinition. This is indeed the case for the calculation of the action integral in the next subsection.

\subsection{ Action Integral  for pure spin-3 Gravity}
\hspace{5mm}
As was shown in \cite{FN}, the second-order action for spin-3 gravity obtained by substituting (\ref{omegae}) into the Chern-Simons action is given by 
\begin{eqnarray}
S_{\text{second-order}} &=& \frac{k}{4\pi \ell} \, \int \mbox{tr} \, e \wedge (d\omega(e)+ \omega(e) \wedge \omega(e)+\frac{1}{3 \ell^2} \, e \wedge e), \nonumber \\
&=& \frac{k}{48\pi\ell} \, \int \, d^3x \, \left\{ -
\epsilon^{\mu\nu\lambda} \,  {F_{\mu M}}^N\, {R^M}_{N\nu\lambda}+\frac{24}{\ell^2} \, 
\tilde{e}
\right\}
 \label{CSsecond}
\end{eqnarray} 
Here $k=\ell/4G$ and $G$ is the 3D gravitational constant, and $\ell$ is the cosmological length related to the cosmological constant by $\Lambda=-2/\ell^2$.\footnote{The normalization of the action (4.30) in \cite{FN} is not appropriate. It must be corrected by a factor $1/4$. } $\tilde{e}$ is the cosmological term presented in (\ref{cosmological}). 
The quantity ${F_{\mu M}}^N$ multiplying the generalized curvature is given by
\begin{equation}
{F_{\mu M}}^N \equiv {f^a}_{bc} \, e^c_{\mu} \, e^b_M \, E^N_a.  \label{Fabc}
\end{equation}
Because $e^b_{(\nu\lambda)}$ and $E^{(\nu\lambda)}_a$ are traceless when contracted with $g^{\nu\lambda}$ and  $g_{\nu\lambda}$, respectively, this quantity acts as a projector. So, only the tensor part of the generalized curvature, ${R^M}_{N\nu\lambda}$, contributes to the action integral, and the action integral is invariant under diffeomorphism. 
Therefore the projection in terms of $P^{\mu\nu}_{\lambda\rho}$ mentioned in the previous subsection  is {\em not} necessary in this case. 
To express the action in terms of $g_{\mu\nu}$ and $\phi_{\mu\nu\lambda}$, perturbative expansions for the tensor ${F_{\mu M}}^N$ must be worked out. Some of the components are displayed in appendix E. 

The resulting second-order action up to ${\cal O}(\phi^3)$  is given by 
\begin{equation}
S_{\text{second-order}} = \frac{k}{48\pi} \, \int d^3x \, \sqrt{-g} \, \left(L_0+L_2 +{\cal O}(\phi^4)\right).  \label{Ssecond}
\end{equation}
Here $L_0=12(\hat{R}+ \frac{2}{\ell^2})$ is the Lagrangian for Einstein-Hilbert action. 
An interesting observation is that not only ${R^{\lambda}}_{\rho\mu\nu}$ but also ${R^{(\lambda\rho)}}_{(\kappa\sigma)\mu\nu}$ contribute to $L_0$. This modifies the coefficient 
in front of Einstein-Hilbert action. 
The next term is given by 
\begin{eqnarray}
L_2 &=& -\frac{8}{\ell^2} \, \phi_{\mu\nu\rho} \, \phi^{\mu\nu\rho}+\frac{12}{\ell^2} \, \phi_{\mu} \, \phi^{\mu}+16  \, \phi_{\mu\lambda\rho} \, {\phi_{\nu}}^{\lambda\rho} \, \hat{R}^{\mu\nu}-2 \, \phi_{\mu} \, \phi_{\nu} \, 
\hat{R}^{\mu\nu}-8 \, \phi_{\mu\nu\lambda} \, \phi^{\lambda} \, \hat{R}^{\mu\nu} \nonumber \\ &&-4 \, \phi_{\mu\nu\lambda} \, \phi^{\mu\nu\lambda} \, \hat{R}+2 \, \phi_{\mu} \, \phi^{\mu} \, \hat{R}
-2 \, \phi_{\mu} \, \hat{\nabla}^{\mu} \, \hat{\nabla}_{\nu} \, \phi^{\nu}+24 \, \phi^{\mu\nu\lambda} \, \hat{\nabla}_{\nu} \, \hat{\nabla}_{\lambda} \, \phi_{\mu} \nonumber \\
&& -16 \,  \phi^{\mu\nu\lambda} \, \hat{\nabla}_{\lambda} \, \hat{\nabla}_{\sigma} \, {\phi_{\mu\nu}}^{\sigma}+4\, \hat{\nabla}_{\mu}\, \phi_{\nu} \, \hat{\nabla}^{\nu} \, \phi^{\mu}+4 \, \hat{\nabla}_{\mu}\, \phi_{\nu} \, \hat{\nabla}^{\mu} \, \phi^{\nu}-16 \, \hat{\nabla}_{\mu} \, \phi^{\mu\nu\lambda} \, \hat{\nabla}_{\rho} \, {\phi^{\rho}}_{\nu\lambda} \nonumber \\
&& +16 \, \hat{\nabla}^{\mu} \, \phi^{\nu} \, \hat{\nabla}_{\lambda}\, {\phi_{\mu\nu}}^{\lambda}-2 \, 
(\hat{\nabla}_{\mu} \, \phi^{\mu})^2+2 \, \phi^{\mu} \, \hat{\nabla}_{\nu} \, \hat{\nabla}_{\mu} \, \phi^{\nu}
-8 \, \phi^{\mu\nu\lambda} \, \hat{\nabla}_{\rho} \, \hat{\nabla}_{\lambda} \, {\phi_{\mu\nu}}^{\rho} \nonumber \\
&& +8 \, \phi^{\mu\nu\lambda} \, \hat{\nabla}^{\rho} \, \hat{\nabla}_{\rho} \, \phi_{\mu\nu\lambda}-4 \, 
\hat{\nabla}_{\mu} \, \phi_{\nu\lambda\rho} \, \hat{\nabla}^{\rho} \, \phi^{\nu\lambda\mu}+\frac{20}{3} \, 
\hat{\nabla}_{\mu} \, \phi_{\nu\lambda\rho} \, \hat{\nabla}^{\mu} \, \phi^{\nu\lambda\rho}.
\end{eqnarray}
This Lagrangian can be rewritten into the Fronsdal form by partial integration.
\begin{eqnarray}
L_2 &=& \frac{4}{3} \, \left[ \phi^{\mu\nu\rho} \, ({\cal F}_{\mu\nu\rho}-\frac{3}{2} \, g_{\mu\nu} \, {\cal F}_{\rho}) -\frac{3}{2} \, \hat{R} \, \phi_{\mu\nu\rho} \, \phi^{\mu\nu\rho}+\frac{9}{2} \, \hat{R}_{\rho\sigma} \,{ \phi^{\rho}}_{\mu\nu} \, \phi^{\sigma\mu\nu}-\frac{9}{4} \, \hat{R}_{\rho\sigma} \, \phi^{\rho} \, \phi^{\sigma}
\right.  \nonumber \\  && \left.-\frac{6}{\ell^2} \, \phi_{\mu\nu\rho} \, \phi^{\mu\nu\rho}+\frac{9}{\ell^2} \, \phi_{\mu} \, \phi^{\mu} \right]
+\frac{1}{\sqrt{-g}} \, \partial_{\rho} \, \left[ \sqrt{-g} \, Q^{\rho} \right] \label{L21}
\end{eqnarray}
${\cal F}_{\mu\nu\rho}$ is the Fronsdal tensor
\begin{eqnarray}
{\cal F}_{\mu\nu\rho} &=& \hat{\nabla}^{\lambda} \, \hat{\nabla}_{\lambda} \, \phi_{\mu\nu\rho} 
-\frac{3}{2} \, \hat{\nabla}^{\lambda} \, \hat{\nabla}_{(\mu} \, \phi_{\nu\rho)\lambda}-\frac{3}{2} \, \hat{\nabla}_{(\mu} \, \hat{\nabla}^{\lambda} \, \phi_{\nu\rho)\lambda}+3 \, \hat{\nabla}_{(\mu} \, \hat{\nabla}_{\nu} \, \phi_{\rho)},
\end{eqnarray}
 and ${\cal F}_{\mu}={\cal F}_{\mu\nu\lambda} \, g^{\nu\lambda}$.  $Q^{\rho}$ in the surface term is given by 
\begin{eqnarray}
Q^{\rho} &=& \frac{20}{3} \, \phi_{\mu\nu\lambda} \, \hat{\nabla}^{\rho} \, \phi^{\mu\nu\lambda} +4 \, \phi_{\mu} \, \hat{\nabla}^{\rho} \, \phi^{\mu}+4 \, \phi_{\mu} \, \hat{\nabla}^{\mu} \, \phi^{\rho}-2 \, \phi^{\rho} \, \hat{\nabla}_{\mu} \, \phi^{\mu}-4 \, \phi_{\mu\nu\lambda} \, \hat{\nabla}^{\lambda} \, \phi^{\mu\nu\rho}
\nonumber \\ &&-16 \, \phi^{\mu\nu\rho} \, \hat{\nabla}^{\lambda} \, \phi_{\mu\nu\lambda}+16 \, \phi^{\mu\nu\rho} \, \hat{\nabla}_{\nu} \, \phi_{\mu}+4 \, \phi^{\mu\nu\rho} \, \hat{\nabla}_{\nu} \, \phi_{\mu}-4 \, \phi^{\mu} \, \hat{\nabla}_{\nu} \, {\phi_{\mu}}^{\nu\rho}.  
\end{eqnarray}
In this way the second-order action can be divided into three integrals:
\begin{equation}
S_{\text{second order}} = S_{\text{EH}}+S_{\text{free Fronsdal}}+S_{\text{boundary}}
\end{equation}
The bulk part of (\ref{L21}) is the linearized spin-3 Fronsdal action with $\phi$ mass terms, and agrees with the result of \cite{Campometric}. \footnote{Our $\phi$ and the spin-3 field $\varphi$ in \cite{Campometric} are related as $\phi_{\mu\nu\rho}= 3 \, \varphi_{\mu\nu\rho}$. }  $S_{\text{boundary}}=\int d^2x \, \sqrt{-g} \, Q^r$ connects 
the CS action in the metric-like formalism and the Einstein-Fronsdal type action in the metric-like formalism.

\subsection{Transformations of the Metric-Like Quantities}
\hspace{5mm}
The transformation properties of $g_{\mu\nu}$ and $\phi_{\mu\nu\rho}$ were studied in \cite{FN}. 
It was shown that $g_{\mu\nu}$ transforms as
\begin{equation}
\delta \, g_{\mu\nu} = \hat{\nabla}_{\mu} \, \xi_{\nu}+\hat{\nabla}_{\nu} \, \xi_{\mu}-\Gamma^{(\lambda\rho)}_{\mu\nu} \, \zeta_{(\lambda\rho)}=\nabla_{\mu} \, \xi_{\nu}+\nabla_{\nu}\, \xi_{\mu}. \label{delgmn}
\end{equation}
$\xi_{\mu}$ and $\zeta_{\mu\nu}=\xi_{(\mu\nu)}-G_{(\mu\nu)\rho} \, g^{\rho\sigma} \, \xi_{\sigma}$ are parameters of diffeomorphism and spin-3 transformation.\footnote{$\xi_M=\frac{1}{2} \, \text{tr} \, (\Lambda_-  e_M)$. See eqs (5.4) and (5.5) of \cite{FN}. 
$\zeta_{\mu\nu}$ satisfies $\zeta_{\mu\nu} \, g^{\mu\nu}=0$.} 
So, $g_{\mu\nu}$ behaves as a tensor under diffeomorphism.  By using (\ref{Gamma21}), the transformation rule under spin-3 transformation is also derived. 
\begin{eqnarray}
\delta_{\zeta} \, g_{\mu\nu} &=& -\frac{8}{3} \, \zeta^{\lambda\rho} \, \hat{\nabla}_{(\mu} \, \phi_{\nu)\lambda\rho}-\frac{8}{3} \, {\zeta_{(\mu}}^{\lambda} \, \hat{\nabla}_{\nu)} \, \phi_{\lambda}
-\frac{8}{3} \, {\zeta_{(\mu}}^{\lambda} \, \hat{\nabla}_{|\lambda|} \, \phi_{\nu)} \nonumber \\
&&+\frac{4}{3} \, \zeta^{\lambda\rho} \, \hat{\nabla}_{\lambda} \, \phi_{\mu\nu\rho}+\frac{8}{3} \, 
{\zeta_{(\mu}}^{\lambda} \, \hat{\nabla}^{\rho} \, \phi_{\nu)\lambda\rho}+\frac{8}{3} \, g_{\mu\nu} \, \zeta^{\lambda\rho} \, \hat{\nabla}_{\lambda} \, \phi_{\rho} \nonumber \\ &&
+\frac{2}{3} \, \zeta_{\mu\nu} \, \hat{\nabla}_{\lambda} \, \phi^{\lambda}-\frac{4}{3} \, g_{\mu\nu} \, \zeta^{\lambda\rho} \, \hat{\nabla}^{\sigma} \, \phi_{\lambda\rho\sigma}+{\cal O}(\phi^3) \label{delgmn2}
\end{eqnarray}
This result agrees with that of \cite{Campometric}. 

The case of $\phi_{\mu\nu\rho} =\frac{1}{4} \, \mbox{tr} \, e_{\mu} \, \{e_{\nu}, \, e_{\rho} \}$ is more complicated. Under diffeomorphism it was shown \cite{FN} that $\phi$ transforms as \footnote{There is a typo in eq (5.13) of \cite{FN}. The coefficient $1/5$ at the top of the last line of (\ref{phixi}) is missing. }
\begin{eqnarray}
\delta_{\xi} \, \phi_{\mu\nu\lambda} &=& 
\xi^{\sigma} \, \hat{\nabla}_{\sigma} \, \phi_{\mu\nu\lambda}
+ \hat{\nabla}_{\mu} \, \xi^{\sigma} \, \phi_{\sigma\nu\lambda}+
\hat{\nabla}_{\lambda} \,  \xi^{\sigma} \, \phi_{\sigma\nu\mu}+
\hat{\nabla}_{\nu} \, \xi^{\sigma} \, \phi_{\sigma\mu\lambda} \nonumber \\
&&+\frac{1}{5}  \, g_{\mu\nu} \, \xi^{\sigma} \, \left(
 \hat{\nabla}_{\lambda} \,  {\phi_{\sigma\kappa}}^{\kappa}-
\hat{\nabla}_{\sigma} \,  {\phi_{\lambda\kappa}}^{\kappa}
\right) 
+\frac{1}{5}  \, g_{\nu\lambda} \, \xi^{\sigma} \, \left(
 \hat{\nabla}_{\mu} \,  {\phi_{\sigma\kappa}}^{\kappa}-
\hat{\nabla}_{\sigma} \,  {\phi_{\mu\kappa}}^{\kappa}
\right) \nonumber \\ && 
+\frac{1}{5}  \, g_{\mu\lambda} \, \xi^{\sigma} \, \left(
 \hat{\nabla}_{\nu} \,  {\phi_{\sigma\kappa}}^{\kappa}-
\hat{\nabla}_{\sigma} \,  {\phi_{\nu\kappa}}^{\kappa}
\right) \nonumber \\
&& +\left\{ \frac{1}{5} \, g^{\alpha\beta} \,( S_{\mu\alpha,\nu\beta} \, \xi_{\lambda}+
 S_{\mu\alpha,\lambda\beta} \, \xi_{\nu})-\xi^{\alpha} \, S_{\mu\nu,\lambda\alpha}+ \text{cyclic permutations of} \  \mu,\nu, \lambda \right\}. \nonumber \\ &&
\label{phixi}
\end{eqnarray}
The appearance of this expression is quite different from the transformation rule of a rank 3 tensor. Thus the transformation rule of  $\phi_{\mu\nu\rho} =\frac{1}{4} \, \mbox{tr} \, e_{\mu} \, \{e_{\nu}, \, e_{\rho} \}$ is a nontrivial issue, although the tensorial property of the counterpart, $g_{\mu\nu} =\frac{1}{2} \, \mbox{tr} \, e_{\mu} \, e_{\nu}$,  is easy to justify. We have checked that $\phi_{\mu\nu\rho}$ behaves as  a tensor up to ${\cal O}(\phi^4)$:
\begin{equation}
\delta_{\xi} \, \phi_{\mu\nu\rho} = \xi_{\lambda} \, \hat{\nabla}^{\lambda} \, \phi_ {\mu\nu\rho}+
3 \, {\phi_{(\mu\nu}}^{\lambda} \, \hat{\nabla}_{\rho)} \, \xi_{\lambda}+{\cal O}(\phi^5). \label{delphi}
\end{equation}
Since our standing point is to derive the transformation properties of the metric-like fields by starting from the transformation rules of the frame-like fields under $SL(3,R) \times SL(3,R)$, this fact is not evident. All order proof is not yet obtained. 

Finally, we computed the spin-3 transformation of $\phi$ by using eqs (5.15) and (5.3) of \cite{FN}. 
\begin{eqnarray}
\delta_{\zeta} \, \phi_{\mu\nu\lambda} &=& \partial_{\mu} \, \zeta_{\nu\lambda}-\frac{1}{2} \, \Gamma^{(\rho\sigma)}_{\mu(\nu\lambda)} \, \zeta_{\rho\sigma}+ (\text{cyclic permutations of } \mu,\nu,\lambda) \nonumber \\
&=& 3 \, \hat{\nabla}_{(\mu} \, \zeta_{\nu\lambda)}+{\cal O}(\phi^2) \label{delphi2}
\end{eqnarray}
The 0-th order term of this transformation was first obtained in \cite{Campometric}. We also computed 
the next ${\cal O}(\phi^2)$ terms. These are non-vanishing. However, they are complicated to display here. 
The action integral (\ref{Ssecond}) is invariant under the above ${\cal O}(\phi^0)$ transformations. This was first shown by \cite{Campometric}.


\section{Matter Coupled to Spin-3 Gravity}
\hspace{5mm}
In this section we will couple a $(B, C)$ system composed of 0-form $C$ and 2-form $B=\frac{1}{2} \, B_{\mu\nu}\, dx^{\mu} \wedge dx^{\nu}$ to the spin-3 gravity topologically. These are 3 $\times$ 3 matrices 
and can be expanded into the basis $\{t_0, t_a\}$ ($t_0=\bm{1}$ is an identity matrix.)
\begin{eqnarray}
C &=& C^A \, t_A= C^0 \, t_0+C^a \, t_a, \label{C} \\
B &=& B^A \, t_A = B^0 \, t_0+B^a \, t_a \label{B}
\end{eqnarray}
The action integral is given by 
\begin{equation}
S_{\text{matter}} = \int \mbox{tr} \, B \wedge (dC+AC-C\bar{A}). \label{topological 
action}
\end{equation}
This action is invariant under $SL(3,R) \times SL(3,R)$ gauge transformation: 
\begin{eqnarray}
A & \rightarrow & A'=U^{-1} \, dU+U^{-1} \, A \, U, \qquad 
\bar{A}  \rightarrow  \bar{A}'=\bar{U}^{-1} \, d\bar{U}+\bar{U}^{-1} \, \bar{A} \, \bar{U},  \nonumber \\
C & \rightarrow & C'=U^{-1} \, C \, \bar{U}, \qquad B \rightarrow B'=
\bar{U}^{-1} \, B \, U. \label{SL3Rmatter}
\end{eqnarray}
Here $U$ and $\bar{U}$ are 3 $\times$ 3 matrices corresponding to the first and second $SL(3,R)$, respectively.  As we will see in (\ref{bC1})-(\ref{bC2}), the 0-th components $B^0$ and $C^0$ in (\ref{C})-(\ref{B}), proportional to $\bm{1}$,  are necessary. By using the vielbein and the spin connection\footnote{$A=\omega+\ell^{-1} \, e$ and $\bar{A}=\omega-\ell^{-1} \, e$.}, action integral (\ref{topological action}) in terms of the components reads 
\begin{eqnarray}
S_{\text{matter}} &= &\int d^3x \, \epsilon^{\mu\nu\lambda} \, 
\mbox{tr} \, \frac{1}{2} \, 
B_{\mu\nu} \, (\partial_{\lambda} \, C+A_{\lambda} \, C-C \, \bar{A}_{\lambda}) \nonumber \\
&=& \int d^3x \, \epsilon^{\mu\nu\lambda} \, B_{A\mu\nu} \, (\partial_{\lambda} \, C^A
+{f^A}_{bc} \, \omega^b_{\lambda} \, C^c+\frac{1}{\ell} \, {d^A}_{bC} \, e^b_{\lambda} \, 
C^C) \label{topological action2}
\end{eqnarray}
Here $A$, $B$, $C$,.. run over $0,1,2,\dots,8$, while $a$, $b$, $c$,  ...  over $1,2,\dots, 8$.  These indices are raised and lowered by $h_{AB}$ and and its inverse $h^{AB}$. The 
Killing metric $h_{ab}$ for $sl(3,R)$ is defined in appendix A and $h_{00}=
3/2$ and $h_{a0}=h_{0a}=0$. Hence $B_0=\frac{3}{2} \, B^0$. ${f^A}_{BC}$ vanishes, if at least one of $A$, $B$, $C$ are 0. ${f^a}_{bc}$ is the structure constant for $SL(3,R)$ and  ${d^a}_{bc}$ the 
invariant symmetric tensor.  ${d^0}_{ab} =(4/3) \, h_{ab} $, ${d^a}_{b0}=2 \, \delta^a_b  $. 
Let us note that this action does not contain the metric tensor like the Chern-Simons action, hence it is topological: it is invariant under general coordinate transformations.  This symmetry is independent of the above gauge symmetry.

\subsection{Infinitesimal Gauge Transformations}
\hspace{5mm}
Gauge transformation (\ref{SL3Rmatter}) in infinitesimal form will be studied now. Let us write 
$U$ and $\bar{U}$ as $U=e^{\Lambda}\approx 1+\Lambda$, $\bar{U}=e^{\bar{\Lambda}} \approx 1+\bar{\Lambda}$. The transformations are classified into two sets;
(a) local Lorentz-like transformation $\Lambda=\bar{\Lambda}=\Lambda_+$; 
(b) local translation (diffeo+spin-3 transformation) $\Lambda=-\bar{\Lambda}=\Lambda_-$.  

\begin{description}
\item [(a)] Under local Lorentz-like transformation, $e$ and $\omega$ transform as 
$\delta e=[e, \Lambda_+]$ and $\delta \omega=d\Lambda_++[\omega,\Lambda_+]$. 
A transformation rule for matter is 
\begin{eqnarray}
\delta \, C&=&-\Lambda_+ \, C+C \, \Lambda_+=[C, \Lambda_+], \\
\delta \, B&=& B \, \Lambda_+-\Lambda_+ \, B=[B, \Lambda_+]. 
\end{eqnarray}
Writing $\Lambda_+=t_a \, \Lambda^a_+$, the transformation of the components is 
\begin{eqnarray}
\delta \, C^0 &=& 0, \qquad \delta \, C^a={f^a}_{bc} \, C^b \, \Lambda^c_+, \\
\delta \, B^0 &=& 0, \qquad \delta \, B^a={f^a}_{bc} \, B^b \, \Lambda^c_+. 
\end{eqnarray}
So the 0-th components of the matter are singlets under local Lorentz-like transformation. 

\item [(b)] Under local translation, $e$ and $\omega$ transform as 
\begin{eqnarray}
\delta \, e &=& \ell (d\Lambda_-+[\omega, \Lambda_-]) \equiv \ell \, D^{(L)} \, \Lambda_-, \label{localtranse}\\
\delta  \, \omega &=& (1/\ell) \, [e, \Lambda_-]. \label{localtransomega}
\end{eqnarray} 
The matter fields transform as 
\begin{eqnarray}
\delta \, C &=& -\Lambda_- \, C-C \, \Lambda_-=-\{\Lambda_-, C\}, \label{localtransC} \\
\delta \, B &=& \{\Lambda_-, B \} \label{localtransB}
\end{eqnarray}
In the transformation of $C^a$ the 0-th component inevitably appear, and thus we
are forced to introduce them from the beginning; 
\begin{eqnarray}
\delta \, C^0 &=& -{d^0}_{bc} \, \Lambda_-^b \, C^c= -\frac{4}{3} \, \Lambda^a_- \, C_a, \label{bC1}\\ 
\delta \, C^a &=& -{d^a}_{bC} \, \Lambda^b_- \, C^C= -{d^a}_{bc} \, \Lambda^b_- \, C^c-2 \, \Lambda^a_- \, C^0
\label{bC2}
\end{eqnarray}
\begin{eqnarray}
\delta \, B^0 &=& {d^0}_{bc} \, B^c= \frac{4}{3} \, \Lambda^a_- \, B_a, \label{bB1} \\ 
\delta \, B^a &=& {d^a}_{bC} \, \Lambda^b_- \, B^C= {d^a}_{bc} \, \Lambda^b_- \, B^c+2 \, \Lambda^a_- \, B^0
\label{dB2}
\end{eqnarray}
Later, we will show that $C^0=(1/3) \, \mbox{tr} \, C$ is a scalar under ordinary diffeomorphism, but transforms non-trivially under spin-3 transformation. 
\end{description}

\subsection{Extra Local Symmetry}
\hspace{5mm}
In addition to the $SL(3,R) \times SL(3,R)$ gauge symmetry (\ref{SL3Rmatter}), 
the total action $S_{\text{tot}}=S_{CS}+S_{\text{matter}}$ has the following local symmetry.
\begin{eqnarray}
\delta_{\Xi} \, A & = & -\frac{8\pi}{k} \, \left(C \, \Xi-\frac{1}{3} \, \text{tr} \, (C\, \Xi)\right),  \nonumber\\
\delta_{\Xi} \, \bar{A} &=& -\frac{8\pi}{k} \, \left(\Xi \, C-\frac{1}{3} \, \text{tr} \, (C\, \Xi)\right), \nonumber \\
\delta_{\Xi} \, B &=& d \, \Xi+\Xi \wedge A+ \bar{A} \wedge \Xi, \nonumber \\
\delta_{\Xi} \, C &=& 0.  \label{extralocalsymmetry}
\end{eqnarray}
Here $A, \, \bar{A}=\omega \pm \frac{1}{\ell} \, e$ are Chern-Simons gauge connections, and $\Xi=\Xi_{\mu} \, dx^{\mu}$ is 
a one-form gauge function which is also a $3 \times 3$ matrix.  The trace terms on the righthand sides of $\delta_{\Xi} \, A$ and $\delta_{\Xi} \, \bar{A}$ are introduced to ensure the tracelessness of $A$ and $\bar{A}$.\footnote{By introducing a diagonal U(1) gauge field $A^0=\bar{A}^0$, which corresponds to $t_0=\bm{1}$, one could avoid subtracting the trace terms. 
This additional gauge field $A^0$ would cancel out in the action integral altogether. } This symmetry can be proved by direct calculation. 

It can be shown that when $\Xi$ is written as 
\begin{eqnarray}
\Xi=d\, \Sigma+\bar{A} \, \Sigma-\Sigma \, A  \label{XiSigma}
\end{eqnarray}
for some zero-form $\Sigma$, the above transformation reduces to a gauge transformation  (\ref{SL3Rmatter})  for  only $A$ and $\bar{A}$, up to the equation of motion for $C$.
\begin{eqnarray}
\delta_{\Sigma} \, A & = & -\frac{8\pi}{k} \, \left\{ d(C  \Sigma)+A  (C \Sigma)-(C \Sigma)  A \right\}+\frac{8\pi}{k} \, (dC+AC-C\bar{A}) \,  \Sigma-\frac{1}{3} \, \times \text{trace}, \nonumber \\
\delta_{\Sigma} \, \bar{A} &=& -\frac{8\pi}{k} \, \left\{d(\Sigma  C)-\bar{A}  (\Sigma  C) +(\Sigma  C)  \bar{A}  \right\}+\frac{8\pi}{k} \, \Sigma \, (dC+AC-C\bar{A})-\frac{1}{3} \, \times \text{trace}, \nonumber \\
\delta_{\Sigma} \, B &=& \bar{F} \, \Sigma-\Sigma \, F, \nonumber \\
\delta_{\Sigma} \, C &=& 0. \label{AbarA}
\end{eqnarray}
Here $F=dA+A\wedge A$ and $\bar{F}=d\bar{A}+\bar{A} \wedge \bar{A}$, and `trace' is the trace of the preceding terms. Note that the transformation of $B$ (and $C$) does not coincide with the gauge transformation (\ref{SL3Rmatter}).  By performing additional infinitesimal gauge transformation (\ref{SL3Rmatter}) with $U\sim 1+\frac{8\pi}{k} \, C \, \Sigma$ and $\bar{U}\sim 1+\frac{8\pi}{k} \, \Sigma \, C$, we obtain 
\begin{eqnarray}
\delta'_{\Sigma} \, A & = & \frac{8\pi}{k} \, (dC+AC-C\bar{A}) \,  \Sigma-\frac{1}{3} \times \text{trace}, \nonumber \\
\delta'_{\Sigma} \, \bar{A} &=& \frac{8\pi}{k} \, \Sigma \, (dC+AC-C\bar{A})-\frac{1}{3} \times \text{trace}, \nonumber \\
\delta'_{\Sigma} \, B &=& \{\bar{F}+\frac{8\pi}{k} \, (BC-\frac{1}{3} \, \text{tr}CB)\} \, \Sigma-\Sigma \, 
\{F+\frac{8\pi}{k} \, (CB-\frac{1}{3} \, \text{tr}CB))\}, \nonumber \\  
\delta'_{\Sigma} \, C &=& 0. \label{reducible}
\end{eqnarray}
The righthand sides are all equations of motion. Hence, this combined transformation is trivial on-shell, and so one component of $\Xi_{\mu}$ is not effective.  As a result, the transformation (\ref{extralocalsymmetry}) with (\ref{XiSigma}) reduces on shell to SL(3,R) $\times $ SL(3,R) gauge transformation with gauge parameters $\Lambda= -\frac{8\pi}{k} \, C \Sigma$ and $\bar{\Lambda}= -\frac{8\pi}{k} \,  \Sigma C$.

\subsection{Hamiltonian Analysis and Scalars of 3d Vasiliev Theory }
\hspace{5mm}
As the total action $S_{\text{tot}}=S_{CS}+S_{\text{matter}}$ is constructed as the integral of products of forms 
without an explicit metric, it is topological. Let us first identify the physical degrees of freedom. This is performed by Hamiltonian methods. The extra 
local symmetry just mentioned above leads to an interesting result. Since the action is first-order in derivatives, and is constructed as the integral of products of forms without an explicit metric, it is already in a form of Hamiltonian. 
We will show that there is no propagating `field' degrees of freedom in the $SL(3,R) \times SL(3,R)$ model. This is due to the equation of motion for $C$ (\ref{eqmoC}) below, 
which determines $C$ in terms of the gauge fields on the Cauchy surface. Only the degree of freedom at a single point on the Cauchy surface remains. This analysis will, however,  also be applicable to higher-spin gravity based on $HS[\lambda] \times HS[\lambda]$ CS theory+ BC matter system. 
Because in this case $C$ and $B$ have an infinite number of internal degrees of freedom, the modes at a single point would turn into propagating degrees of freedom.  

The momentum $P$ conjugate to $C$ is given by $P\equiv  \frac{1}{2} \, \epsilon^{ ij} \, B_{ij}$. Here $i,j=\phi, r$.
The momenta conjugate to 
$A_i$ and $\bar{A}_i$ are  $\pi_A^i \equiv \frac{k}{16\pi} \, \epsilon^{ij} \, A_j$ and $\pi_{\bar{A}}^i \equiv \frac{-k}{16\pi} \, 
\epsilon^{ij} \, 
\bar{A}_j$, respectively. The momentum $\Pi_{B}^i$ conjugate to $B_{ti}$ does not exist. Then the primary constraint 
$\Pi_{B}^i \approx 0$ generates  a secondary constraint $\chi_i \equiv \partial_i \, C+A_i \, C-C \, \bar{A}_i \approx 0$. 
These constraints agree with the spatial components $i=\phi, \rho$ of the equations of motion for $C$. 
Similarly, the momenta  $\pi_A$ and $\pi_{\bar{A}}$ conjugate to $A_t$ and $\bar{A}_t$, respectively, obey $\pi_{A} 
\approx \pi_{\bar{A}} \approx 0$.  These lead to secondary constraints, $\psi \equiv \epsilon^{ij} \, (F_{ij}+\frac{8\pi}{k} \, C \, B_{ij}-\frac{8\pi}{3k} \, \text{tr}(CB_{ij})) \approx 0$ and 
$\bar{\psi} \equiv \epsilon^{ij} \, (\bar{F}_{ij}+\frac{8\pi}{k} \, B_{ij} \, C-\frac{8\pi}{3k} \, \text{tr}(CB_{ij})) \approx 0$. 
The Hamiltonian is a sum of Lagrange multipliers times these constraint functions. 

Constraints $\pi_A \approx \pi_{\bar{A}} \approx \Pi_B^i \approx 0$ are first-class. This means that $A_t$, $\bar{A}_t$, $B_{ti}$, as well as $\pi_A$, $\pi_{\bar{A}}$, $\Pi_B^i$, are unphysical. So we can fix gauge  $B_{t\phi}=B_{tr}=0$ by using the symmetry (\ref{extralocalsymmetry}). The two constraints $\psi$ and $\bar{\psi}$ generate $SL(3,R) \times SL(3,R)$ gauge transformations, and are first-class. The  constraint $\chi_i$, the generator of the local transformation (\ref{extralocalsymmetry}),  transforms covariantly under these gauge  transformations, hence $\chi_i \approx 0$ is also first-class. Now, the gauge fields $A_i$ are eliminated by $\psi \approx 0$ and an appropriate gauge fixing, and are non-propagating in the bulk. So are $\bar{A}_i$. As for $\chi_i$, the role of these constraints is not to eliminate $C$, but to determine the derivatives of the field $C$ in the spatial directions, $\phi$, $r$. Corresponding to $\chi_i$, we thus propose to fix gauge by the conditions $\tilde{\chi}_i \equiv \partial_i \, P-P \, A_i+\bar{A}_i \, P \approx 0$.\footnote{This gauge fixing will be done by using diffeomorphism. Due to the reducibility (\ref{reducible}),  variations of $\tilde{\chi}_i$ under transformation (\ref{AbarA}) vanish, hence the matrix, whose components are Poisson brackets of $\chi_i$ and $\tilde{\chi}_j$, is degenerate. However, the Poisson brackets of $\chi_i$, $\tilde{\chi}_i$ and the gauge fixing functions for $A_i$, $\bar{A}_i$ are non-vanishing, and the constraints $\chi_i$, $\tilde{\chi}_i   \approx 0$ become second-class.} 
When these differential equations are solved, the integration constants for the solutions $C$ and $P$, will be subject to Dirac bracket relations. In this way,  almost all degrees of freedom of $C$ and $P$ will be eliminated. 
Combining the gauge fixing $\tilde{\chi}_i \approx 0$ with the equation of motion for $P$\footnote{Recall that we set $B_{t\phi}=B_{tr}=0$ .}, 
we obtain the set of  equations for $P$, $\partial_{\mu} \, P-P \, A_{\mu}+\bar{A}_{\mu} \, P=0$. This provides the counterpart of the set of equations for $C$, $\partial_{\mu} \, C+C \, A_{\mu}-\bar{A}_{\mu} \, C=0$. Hence $C$ and $P= B_{r \phi}$ are strongly constrained on the Cauchy surface by $\chi_i \approx 0$ and $\tilde{\chi}_i \approx 0$. 

We will now turn to Lagrangian formulation. For the moment, we will consider only the matter action $S_{\text{matter}}$, and fix the 
Chern-Simons gauge connections $A$, $\bar{A}$ to some background fields.  We will discuss that the fields $C$ and $B_{t\phi}$ can
 be identified 
with the scalar fields $C$ and $\tilde{C}$ of 3d free Vasiliev theory\cite{Vasiliev3d}\cite{AKP2}. In Vasiliev theory the gauge connections are solutions to the flatness equations $F=\bar{F}=0$, and they cannot be treated as independent dynamical field variables. The equations of motion for our matter system  are given by 
\begin{eqnarray}
dC +A \, C-C \, \bar{A} &=& 0, \label{eqmoC} \\
dB-B \wedge A+ \bar{A} \wedge B &=& 0. 
\end{eqnarray}
The equation for $B$ resembles that for $\tilde{C}$ \cite{AKP2}, $d\tilde{C}-\tilde{C} \wedge A+ \bar{A} \wedge \tilde{C} = 0$, but three components of $B$ are mixed up;
$\partial_{t} \, B_{\phi r}+\partial_{\phi} \, B_{rt}+\partial_r \, B_{t\phi}+\cdots=0$.
When the Chern-Simons gauge fields $A$, $\bar{A}$ are flat background connections corresponding to such as AdS$_3$ space, the symmetry (\ref{extralocalsymmetry}) cannot be used. In this case, however, the Chern-Simons gauge fields are pure gauges, and at least locally, they can be removed by gauge transformation (\ref{SL3Rmatter}). Now, the action integral for the gauge-transformed matters $C'$, $B'$ (with $A'=\bar{A}'=0$), $S_{\text{matter}}=\int \text{tr} B' \wedge d \, C'$, has a local symmetry 
\begin{equation}
\delta \, B' = d \, \Xi ,\quad \delta \, C'=0,
\end{equation}
where $\Xi=\Xi_{\mu} \, dx^{\mu}$ is a one-form gauge function. This symmetry is similar to  (\ref{extralocalsymmetry}), but 
the gauge connections $A'=\bar{A}'=0$ are fixed. 
Although there are three components $\Xi_{\mu}$, the one-form which can be written as  $\Xi=d\Sigma$ is redundant, 
and only two of the three components of $B'_{\mu\nu}$ can be gauged away.  
Let us set $B'_{rt}=B'_{r\phi}=0$. Then the equation of motion for $B'_{t\phi}$ is given by 
\begin{eqnarray}
\partial_{r} \, B'_{t\phi }=0. \label{Bprime}
\end{eqnarray}

Let us recall the construction of the solution to the equation of motion for $C$ presented in\cite{HKP}.
Because the equation of motion for $C$ is the same as in \cite{HKP} the solution is the same. The equation of motion for $C'$ is given by $\partial_{\mu} \, C'=0$ and the solution is $C'=$ constant matrix. By choosing an  appropriate constant matrix $C'$ and performing the inverse gauge transformation $C=U \, C' \, \bar{U}^{-1}=b^{-1} \, e^{-\Lambda_0} \, C' \, e^{\bar{\Lambda_0}} \, b^{-1}$,  which changes the spacetime into AdS$_3$, the covariantly constant solution for $C$ can be obtained as in  \cite{HKP}\cite{AKP2}. We do not repeat the calculation here. In this case $\text{tr} \, C$ can be shown to satisfy Klein-Gordon equation in AdS$_3$, with mass $m^2=8/\ell^2$, in exactly the same manner as in \cite{HKP}\cite{AKP2}. 
Our matter field $C$ corresponds to the scalar in the 3d Vasiliev theory with the deformation parameter $\lambda=3$. 

On the other hand the solution to (\ref{Bprime}) is of a form, $B'_{t\phi}=f(t,\phi)$ where $f(t,\phi)$ is an arbitrary 3 $\times$ 3 matrix function which depends on $t$, $\phi$.  Although the equation of motion for $B'_{t\phi}$ is less restrictive than that for $C'$, the delta-function boundary condition\cite{KP2} on the inversely-gauge transformed field $B_{t\phi}=\bar{U} \, B'_{t\phi} \, U^{-1}$ of form
\begin{eqnarray}
\text{tr} \, B_{t\phi} \sim e^{a \,r} \, \delta^{(2)}(x-x') +\cdots \qquad (\text{as} \ r \rightarrow \infty)
\end{eqnarray}
can determine $B'_{t\phi}$ and $B_{t\phi}$, and the value $a=-(1+\lambda)=-4$. When $B'_{t\phi}=$ 
constant matrix, $\text{tr} \, B_{t\phi}$ also satisfies Klein-Gordon equation with mass $m^2=8/\ell^2$.
Hence, by identifying $\tilde{C}=B_{t\phi}$, this model may be taken as a Lagrangian formulation of 
(free) 3d Vasiliev scalar fields. This model is dual to a non-unitary CFT on the boundary, because $\lambda=3$, but by 
replacing in the action $S_{\text{tot}}$ the matrices $A_{\mu}$, $\bar{A}_{\mu}$, $C$, and $B_{\mu\nu}$ by functions and 
replacing the matrix multiplication by the lone-star product\cite{lonestar}\cite{AKP2} , we can also 
realize the unitary model with $0 \leq \lambda \leq 1$. 
 
Let us now consider the full action including the Chern-Simons action, and treat $A$ and $\bar{A}$ as dynamical variables. 
When the Chern-Simons gauge fields are not background fields, classically they must follow the equations of motion
\begin{eqnarray}
F=dA+A \wedge A&=& -\frac{8\pi}{k} \,\left(CB-\frac{1}{3} \, \text{tr}(CB)\right),  \nonumber \\
\bar{F}=d\bar{A}+\bar{A} \wedge \bar{A}&=& -\frac{8\pi}{k} \, \left(BC-\frac{1}{3} \, \text{tr}(BC)\right), 
\label{nonflatF}
\end{eqnarray}
and generally, they are not flat. In this case the gauge fields cannot be gauged away, and the equations of 
motion for $C$ and $B$ cannot be solved by the above method. However, by treating $r$ as a `time coordinate',  
using the extra local symmetry (\ref{extralocalsymmetry}) and  using the Hamiltonian analysis at the beginning of this 
subsection, we may 
set the components of $B_{\mu\nu}$ except for $B_{t\phi}$ (instead of $B_{r\phi}$) to zero,  and impose the gauge fixing conditions
\begin{eqnarray}
\tilde{\chi}_{\mu}\equiv  \partial_{\mu} \, B_{t \phi}+\bar{A}_{\mu} \,  B_{t \phi}-B_{t \phi} \, A_{\mu} \approx 0. \qquad (\mu=t, \phi)
\label{B tildeC eom}
\end{eqnarray}
The equation of motion for $B_{t\phi}$ is (\ref{B tildeC eom}) with $\mu=r$. 
Hence $B_{t \phi}$ will play the role similar to that  of $\tilde{C}$ in the 3d Vasiliev theory, even when the gauge connections are not fixed backgrounds.  It will be interesting to study this model from the point of view of AdS/CFT correspondence. 
Here we will not attempt this. 

In the following, it will be shown that a torsion appears in the spin connection, and the interaction terms 
for $C$ and $B$ are introduced into the action integral in the metric-like formulation.

\subsection{Spin-3 Gravity with Torsion}
\hspace{5mm}
Now we will eliminate the spin connection $\omega^a_{\mu}$ to obtain the metric-like theory. 
If we used the solution $\omega^a_{\mu}(e)$, (\ref{omegae}), then the transformation rules of $g_{\mu\nu}$ and 
$\phi_{\mu\nu\lambda}$ would be 
the same as in the pure spin-3 gravity theory.  The transformation rules of $C$ under 
diffeomorphism\footnote{These are given by (\ref{trC1}) and (\ref{trC2}) below with $\Delta \Gamma$ set to zero.} would also coincide with the usual rules for 
tensors. However,  the spin-3 transformation rules for $C$, (\ref{Cspin3-1}) and (\ref{Cspin3-2}) presented below with $\Delta \Gamma=0$, would not leave the action invariant. 
At present it is not clear if it is possible to modify the action and the transformation rules to recover the symmetry. 

For this reason we will solve the equation of motion for the total action with respect to $\omega^a_{\mu}$. 
When the matter fields are coupled to the pure spin-3 gravity, a torsion is introduced. 
To see this,  let us consider our total action in the Palatini formalism. 
\begin{equation}
S_{\text{tot}}[e,\omega,B,C] = \frac{k}{4\pi \ell} \, \int \text{tr} \, \left(e\wedge R(\omega)+\frac{1}{3\ell^2} \, e \wedge e \right)
+\int \text{tr} \, B\wedge  \left(dC+\frac{1}{\ell} \, \{e, \, C \}+ [\omega, C] \right)
\label{totalaction}
\end{equation}
Here $R(\omega)=d\omega+\omega \wedge \omega$. The first integral is the Chern-Simons action.\cite{AT}\cite{WittenCS}\cite{Campoleoni} The equation of motion obtained by variation with respect to $\omega$ is no longer a torsion-free condition:
\begin{equation}
\partial_{\mu} \, e^a_{\nu}+{f^a}_{bc} \, \omega^b_{\mu} \, e^c_{\nu}-(\mu \leftrightarrow \nu) = -T_{\mu\nu}. \label{torsionfulleq}
\end{equation}
Here $T_{\mu\nu}=T^M_{\mu\nu} \, e_M$ is a torsion tensor defined by
\begin{equation}
T_{\mu\nu} = -\frac{4\pi \ell}{k} \, [B_{\mu\nu}, C]. \label{torsiondef}
\end{equation}
Eq (\ref{torsionfulleq}) defines a new spin connection $\tilde{\omega}^a_{\mu}(e,B,C)$ and a new set of 
connections $\tilde{\Gamma}^M_{\mu N}$. The equation 
\begin{equation}
\partial_{\mu} \, e^a_{\nu}+{f^a}_{bc} \, \tilde{\omega}^b_{\mu}\ e^c_{\nu}=\tilde{\Gamma}^M_{\mu\nu} \, e_M
\label{tildeomegaGamma}
\end{equation}
with $(\tilde{\Gamma}^M_{\mu\nu}-\tilde{\Gamma}^M_{\nu\mu}) \, e_M=-T_{\mu\nu}$ 
can be solved for $\tilde{\Gamma}^M_{\mu\nu}$, by the same procedure as in sec.3 of \cite{FN} and then 
$\tilde{\Gamma}^M_{\mu(\nu\lambda)}$ is determined in terms of them. The result can be written in the form 
\begin{equation}
\tilde{\Gamma}^M_{\mu N}= \Gamma^M_{\mu N}+\Delta \, \Gamma^M_{\mu N}, \label{DeltaGamma}
\end{equation}
where $\Gamma^M_{\mu N} $ is the connection for the pure spin-3 gravity. For the component, $\tilde{\Gamma}^{(\kappa\sigma)}_{\mu\nu}$,  the difference is given by 
\begin{eqnarray}
\Delta \, \Gamma^{(\kappa\sigma)}_{\mu\nu} &=& -\frac{1}{2} \, T^{(\kappa\sigma)}_{\mu\nu} +\frac{1}{8} \, J^{(\kappa\sigma)(\lambda\rho)} \, \left[  T_{\mu\lambda,(\nu\rho)} + \ T_{\mu\rho,(\nu\lambda)} - 2 g_{\lambda\rho} \ T_{\mu}{}^{\sigma}{}_{,(\nu\sigma)} \right. \nonumber \\ 
&& +   g_{\nu\rho} \
T_{\mu}{}^{\sigma}{}_{,(\lambda\sigma)} + \
g_{\nu\lambda} \
T_{\mu}{}^{\sigma}{}_{,(\rho\sigma)} + \
T_{\nu\lambda,(\mu\rho)} + \
T_{\nu\rho,(\mu\lambda)} - 2 g_{\lambda\rho} \
T_{\nu}{}^{\sigma}{}_{,(\mu\sigma)}                                  \nonumber \\ 
&& + \  g_{\mu\rho} \    T_{\nu}{}^{\sigma}{}_{,(\lambda\sigma)}   
 +
g_{\mu\lambda} \
T_{\nu}{}^{\sigma}{}_{,(\rho\sigma)} + \
g_{\nu\rho} \         
T_{\lambda}{}^{\sigma}{}_{,(\mu\sigma)} + \
g_{\mu\rho} \
T_{\lambda}{}^{\sigma}{}_{,(\nu\sigma)}  \nonumber \\ 
&& \left.  - 2 \
g_{\mu\nu} \
T_{\lambda}{}^{\sigma}{}_{,(\rho\sigma)} + \
g_{\nu\lambda} \
T_{\rho}{}^{\sigma}{}_{,(\mu\sigma)} + \
g_{\mu\lambda} \
T_{\rho}{}^{\sigma}{}_{,(\nu\sigma)} - 2 \
g_{\mu\nu} \
T_{\rho}{}^{\sigma}{}_{,(\lambda\sigma)}       \right]   +{\cal O}(\phi^1) \label{delGamma0} \nonumber \\ && 
\end{eqnarray}
The next order contribution, $\Delta \, [\Gamma^{(\tau\eta)}_{\mu\nu}]_1 \ (={\cal O}(\phi^1))$ is displayed in appendix H. The difference of another connection,
$\Delta \, \Gamma^{\rho}_{\mu\nu}$, is related to $\Delta \, \Gamma^{(\rho\sigma)}_{\mu\nu}$ by 
\begin{equation}
\Delta \,  \Gamma^{M}_{\mu\nu} \, G_{M \lambda }=\Delta \,  \Gamma^{\rho}_{\mu\nu} \, G_{\rho \lambda }+\frac{1}{2} \, \Delta \,  \Gamma^{(\rho\sigma)}_{\mu\nu} \, G_{(\rho\sigma)\lambda } =- \frac{1}{2} \, (T_{\mu\nu,\lambda}+T_{\lambda\mu,\nu}+T_{\lambda\nu,\mu}), \label{TTT}
\end{equation}
In the above equations, we used a notation.
\begin{equation}
T_{\mu\nu,M}= T^N_{\mu\nu} \, G_{NM}
\end{equation}
Expression (\ref{TTT}) satisfies $\Delta \Gamma^M_{\mu\nu} \, G_{M \lambda}+\Delta \Gamma^M_{\mu\lambda} \, G_{M\nu}=0$ 
due to the anti-symmetry of $T_{\mu\nu}$. 
In (\ref{delGamma0}) the indices of $T_{\mu\nu,(\lambda\rho)}$ are raised by $g^{\kappa\sigma}$. 
Let us note that $\Delta \,  \Gamma^{M}_{\mu\nu}$ is not symmetric under interchange of $\mu$ and $\nu$, due to the torsion. 
The other two connections $\tilde{\Gamma}^M_{\mu(\nu\lambda)}$ are obtained by replacing $\Gamma^M_{\mu N}$ by $\tilde{\Gamma}^M_{\mu N}$ in eqs (4.12) and (4.13) of \cite{FN}. 

The spin connection $\tilde{\omega}^a_{\mu}$ which satisfies (\ref{tildeomegaGamma}) is now given by 
\begin{equation}
\tilde{\omega}_{\mu}^a(e,B,C) \equiv \frac{1}{12} \ {f^{ab}}_{c} \ 
E_b^M \ \tilde{\nabla}_{\mu} \ e^c_M.
\label{tildeomegae}
\end{equation}
Here $\tilde{\nabla}_{\mu}$ is the covariant derivative associated with $\tilde{\Gamma}^M_{\mu N}$. 
$G_{MN}$ is compatible with $\tilde{\nabla}_{\mu}$. 
The generalized curvature tensor which corresponds to the above new connection is defined by 
\begin{eqnarray}
{\tilde{R}^M}{}_{ N\mu\nu}& = & \partial_{\mu} \, \tilde{\Gamma}^M_{\nu N}-\partial_{\nu} \, \tilde{\Gamma}^M_{\mu N}
+\tilde{\Gamma}^M_{\mu K} \, \tilde{\Gamma}^K_{\nu N}-\tilde{\Gamma}^M_{\nu K} \, \tilde{\Gamma}^K_{\mu N} \nonumber \\
&=& {R^M}_{N\mu\nu}+\nabla_{\mu} \, \Delta \, \Gamma^M_{\nu N}-\nabla_{\nu} \, \Delta \, \Gamma^M_{\mu N}
+\Delta \, \Gamma^M_{\mu K} \, \Delta \, \Gamma^K_{\nu N}-\Delta \, \Gamma^M_{\nu K} \, \Delta \, \Gamma^K_{\mu N}. 
\label{tildeR}
\end{eqnarray}
In the second line of the above equation, the covariant derivative $\nabla_{\mu} \, \Delta \, \Gamma^M_{\nu K}$ is used here for brevity with tacit understanding that the non-existing component $\Delta \, \Gamma^M_{(\mu\nu)N}=0$. 
After substitution of (\ref{tildeomegae}) into (\ref{totalaction}), we obtain the total action.
\begin{eqnarray}
\tilde{S}_{\text{tot}} \equiv   S_{\text{tot}}[e,\tilde{\omega},B,C]   &=& \frac{k}{48\pi \ell} \, \int d^3x \, \left(-\epsilon^{\mu\nu\lambda} \, {F_{\mu M}}^N \, {\tilde{R}^M}{}_{N \nu \lambda}+\frac{24}{\ell^2} \, \tilde{e}\right) \nonumber \\
&&+ \int \text{tr} \, B \wedge \left(d C+\frac{1}{\ell} \, \{ e, \, C \}+ [\tilde{\omega}, \, C]\right)
\end{eqnarray}
When the generalized curvature tensor (\ref{tildeR}) is substituted into the above equation,  
those terms linear in $\nabla_{\nu} \, \Delta \, \Gamma^M_{\lambda N}$ turn out total derivative 
ones, and can be dropped. This is because the metric-like quantity ${F_{\mu M}}^N$ is made of the vielbeins, and covariantly constant: $\nabla_{\nu} \, {F_{\mu M}}^N=D_{\nu} \, {F_{\mu M}}^N=0$.\footnote{$D_{\nu}$ is the full covariant derivative.} Finally, only the quadratic terms remain, and the total action is given by
\begin{eqnarray}
\tilde{S}_{\text{tot}} &=& S_{\text{second-order}}+\frac{k}{48\pi \ell} \, \int d^3x \, \left(-2 \, \epsilon^{\mu\nu\lambda} \, {F_{\mu M}}^N \, \Delta \, \Gamma^M_{\nu K} \, \Delta \, \Gamma^K_{\lambda N}\right) \nonumber \\
&&+ \int \text{tr} \, B \wedge \left(d C+\frac{1}{\ell} \, \{ e, \, C \}+ [\tilde{\omega}, \, C]\right) \label{Stilde}
\end{eqnarray}
The first term in the first line is the pure spin-3 gravity action (\ref{CSsecond}).  
The last term in the first line is quadratic in the torsion tensor and depends on $B$ and $C$.

\subsection{Proof of Local Translation Invariance of the Action}
\hspace{5mm}
In order to prove that $\tilde{S}_{\text{tot}}=S_{\text{tot}}[e,\tilde{\omega},B,C]$ is invariant under the local translation
(\ref{localtranse}), (\ref{localtransC})-(\ref{localtransB}), we will use the 1.5 order formalism. Under variation of the fields, $\tilde{S}_{\text{tot}}$ 
changes as follows. 
\begin{eqnarray}
\delta \, \tilde{S}_{\text{tot}} = \frac{\delta \, S_{\text{tot}}}{\delta e} \, \delta \, e+\frac{\delta \, S_{\text{tot}}}{\delta B} \, \delta \, B+\frac{\delta \, S_{\text{tot}}}{\delta C} \, \delta \, C+\frac{\delta \, S_{\text{tot}}}{\delta \omega} \, \delta \, \tilde{\omega}[e,B,C]
\end{eqnarray}
After the functional differentiations, we set $\omega=\tilde{\omega}$ on the righthand side. 
The last variation $\delta \, \tilde{\omega}[e,B,C]$ is computed according to the dependence of the solution $\omega=\tilde{\omega}$ on 
the other fields, $e_{\mu}$, $B$ and $C$. This is actually very complicated, but because $\tilde{\omega}$ solves the equation of motion, the 
functional derivative multiplying this variation vanishes at $\omega=\tilde{\omega}$. This means that when calculating the variation, we can 
keep $\tilde{\omega}$ fixed. 

For the pure spin-3 gravity part of the action, we only need to vary the vielbein. The $e \wedge R(\tilde{\omega})$ 
part is invariant up to total derivative terms, because of (\ref{localtranse}) and the Bianchi identity, $d \, R(\omega)+
\omega \wedge R(\omega)-R(\omega) \wedge \omega=0$. The variation of the generalized cosmological term (\ref{cosmological}) is a total derivative:
$\delta \, \tilde{e}= \partial_{\mu} \, (\frac{1}{2} \, \epsilon^{\mu\nu\lambda} \, f_{abc} \, \Lambda^a_- \, e^b_{\nu} \, e^c_{\lambda})$. 

For the matter part, if the spin connection is kept fixed,  we obtain after simple calculation, the variation of the matter Lagrangian, 
\begin{eqnarray}
\epsilon^{\mu\nu\lambda} \,  f_{abc} \, e^b_{\mu} \, \Lambda^c_-  \, [B_{\nu\lambda}, \, C]^a=\frac{k}{4\pi \ell} \, \epsilon^{\mu\nu\lambda}  \, f_{abc} \, e^b_{\mu} \, \Lambda^c_-  \, T^a_{\nu\lambda}.
\end{eqnarray}
 Here the definition of the torsion (\ref{torsiondef}) is used. Due to (\ref{torsionfulleq}), this is a total derivative like the variation of the generalized cosmological term mentioned above. Therefore the invariance of $S_{\text{tot}}[e,\tilde{\omega},B,C]$ is proved.

\subsection{World-volume Components of the Matter}
\hspace{5mm}
The matter fields, $C^a$ and $B^a$, introduced above have an internal index $a$ and 
transform non-trivially under local Lorentz-like transformation (a). By contracting these 
fields with the generalized vielbein $e^a_{\mu}$ and $e^a_{(\mu\nu)}$, we will obtain fields 
which are neutral to local Lorentz-like transformation. The new fields are $C^0$,  
$C^M= E^M_a \, C^a$ and $B^0$, $B_M= e_M^a \, B^a$. 
$C^a$ and $C^M$ are in one-to-one correspondence with each other: $C^a=C^M \, e^a_M$. 

We now introduce a Lorentz covariant derivative for $C^a$. 
\begin{eqnarray}
\tilde{D}^{(L)}_{\mu} \, C^a &=& \partial_{\mu} \, C^a+{f^a}_{bc} \, \tilde{\omega}^b_{\mu} \, C^c, 
\nonumber \\
\tilde{D}^{(L)}_{\mu} \, C^0 &=& \partial_{\mu} \, C^0. 
\end{eqnarray}
Similar definition is made for $\tilde{D}^{(L)}_{\mu} \, B^a_{\nu\lambda}$ and $\tilde{D}^{(L)}_{\mu} \, B^0_{\nu\lambda}$. 
By replacing $\tilde{D}^{(L)}_{\mu}$ on $C^a$ by the full covariant derivative $\tilde{D}_{\mu}$ and 
using $C^a=C^M \, e^a_M$, recalling that $e^a_M$ is covariantly constant, we obtain 
\begin{equation}
\tilde{D}^{(L)}_{\lambda} \, C^a = \tilde{D}_{\lambda} \, (C^M \, e^a_M)=e^a_M \, \tilde{\nabla}_{\lambda} \, C^M.
\end{equation}
Here $\tilde{\nabla}_{\mu}$ is the covariant derivative corresponding to $\tilde{\Gamma}^M_{\mu N}$. 

By assembling the results of the above replacements, matter action (\ref{topological action2}) takes the form.  
\begin{eqnarray}
S_{\text{matter}} &=& \int d^3x \, \epsilon^{\mu\nu\lambda} \, \left[
B_{N\mu\nu} \, E_a^N \, \left(e^a_M \, \tilde{\nabla}_{\lambda}\, C^M + \frac{1}{\ell} \, {d^a}_{bc} e^b_{\lambda} \, C^c+\frac{2}{\ell} \, e^a_{\lambda} \, C^0 \right) \right. \nonumber \\  && \left. +
B_{0 \mu\nu} \, \left(\partial_{\lambda} \, C^0+\frac{4}{3\ell} \, e^b_{\lambda} \, C_b\right)\right].
\end{eqnarray}
We now rewrite $B$ as 
\begin{eqnarray}
{B_N}^{\lambda}  &\equiv& \frac{1}{\sqrt{-g}} \, \epsilon^{\mu\nu\lambda}\, B_{N\mu\nu}
= \varepsilon^{\mu\nu\lambda}\, B_{N\mu\nu},  \label{BNl} \\
{B_0}^{\lambda}  &\equiv& \frac{1}{\sqrt{-g}} \, \epsilon^{\mu\nu\lambda}\, B_{0\mu\nu}= \varepsilon^{\mu\nu\lambda}\, B_{0\mu\nu}.
\label{B0l}
\end{eqnarray}
Here  $\varepsilon^{\mu\nu\lambda}$ is a completely anti-symmetric tensor, and $g=\text{det} \, g_{\mu\nu}$ is the determinant of the metric tensor. We have 
\begin{eqnarray}
S_{\text{matter}} &=& \int d^3x \, \sqrt{-g} \, \left[
{B_N}^{\lambda} \,  \left( \tilde{\nabla}_{\lambda}\, C^N + \frac{1}{\ell} \, {K^N}_{M\lambda}  \, C^M
\right)+\frac{2}{\ell} \, {B_{\lambda}}^{\lambda} \, C^0  \right.      \nonumber \\  
&& \left. +
{B_0}^{\lambda} \, \left(\partial_{\lambda} \, C^0+\frac{4}{3\ell} \,  C_{\lambda}  \right)\right] \nonumber \\
&=& \int d^3x \, \sqrt{-g} \, \left[
{B_N}^{\lambda} \,  \left( \nabla_{\lambda}\, C^N + \frac{1}{\ell} \, {K^N}_{M\lambda}  \, C^M
\right)+\frac{2}{\ell} \, {B_{\lambda}}^{\lambda} \, C^0  \right.      \nonumber \\  
&& \left. +
{B_0}^{\lambda} \, \left(\partial_{\lambda} \, C^0+\frac{4}{3\ell} \,  C_{\lambda}  \right)  +{B_N}^{\lambda} \, C^M \, \Delta \Gamma^N_{\lambda M}\right] 
\label{MatterMetriclike}
\end{eqnarray}
Here the indices $M$, $N$ of $C$ are raised and lowered in terms of $G_{MN}$ and $G^{MN}$: $C_{\lambda}=G_{\lambda N} \, C^N$.\footnote{
Since $C^0$ is not invariant under spin-3 transformation, a suitable covariant derivative for $C^0$ needs to be devised like $\nabla_{\mu} \, \rho^a$ defined in eq(4.8) of \cite{FN}.  This will not be attempted in this paper. } After the second equality, $\tilde{\nabla}_{\lambda}$ is replaced by $\nabla_{\lambda}$ and a new term including $\Delta \Gamma^N_{\lambda M}$ appeared due to the difference of the connections. This is a fourth order interaction 
$(BC)^2$ of the matter fields. 
Also ${K^N}_{ML}$ is a metric-like quantity defined by 
\begin{equation}
{K^N}_{ML} \equiv {d^a}_{bc} \, E^N_a \, e^b_M \, e^c_{L} \label{tensorK}
\end{equation}
This quantity has the following expansion in powers of $\phi$. 
\begin{eqnarray}
{K^{\mu}}_{\nu \lambda} &=& \frac{2}{3} \, g_{\nu\lambda} \, \phi^{\mu}+{\cal O}(\phi^3), \\
{K^{(\mu\nu)}}_{\lambda\rho} &=& 2 \, \left(\delta^{\mu}_{\lambda} \, \delta^{\nu}_{\rho}+\delta^{\nu}_{\lambda} \, \delta^{\mu}_{\rho}-\frac{2}{3} \, g^{\mu\nu} \, g_{\lambda\rho}\right)+{\cal O}(\phi^2)
\end{eqnarray}
${K^{\mu}}_{(\nu\lambda)\rho}$ and ${K^{\mu\nu}}_{(\lambda\rho)\kappa}$ are already displayed in (\ref{K1}) and (\ref{K3}).  The other components are presented in appendix C.  In appendix F, expansion of the part of the matter action, which does not depend on the torsion,  is presented in powers of $\phi$ up to ${\cal O}(\phi^1)$.     Those parts which comes from $\Delta \Gamma^N_{\mu M}$ turn out too complicated to write down.           

To summarize, the total action is given by 
\begin{eqnarray}
\tilde{S}_{\text{tot}} &=& S_{\text{second-order}} \nonumber \\ && +\int d^3x \, \sqrt{-g} \, \left[
{B_N}^{\lambda} \,  \left( \nabla_{\lambda}\, C^N + \frac{1}{\ell} \, {K^N}_{M\lambda}  \, C^M
\right)+\frac{2}{\ell} \, {B_{\lambda}}^{\lambda} \, C^0   +{B_0}^{\lambda} \, \left(\partial_{\lambda} \, C^0 +\frac{4}{3\ell} \,  C_{\lambda}  \right) \right] \nonumber \\ 
&& 
 +\int d^3x \, \left[\, \sqrt{-g} {B_N}^{\lambda} \, C^M \, \Delta \Gamma^N_{\lambda M}
-\frac{k}{24\pi \ell} \, \epsilon^{\mu\nu\lambda} \, {F_{\mu M}}^N \, \Delta \, \Gamma^M_{\nu K} \, \Delta \, \Gamma^K_{\lambda N}\right]
\end{eqnarray}
Those terms in the third line come from the torsion and represent matter interactions.

\subsection{Symmetry of the Matter-Coupled Theory}
\hspace{5mm}
The transformation rules for the metric-like fields  $g_{\mu\nu}$ and $\phi_{\mu\nu\lambda}$ under the local translation (b) are modified from (\ref{delgmn})-(\ref{delphi2}), because of the torsion terms. The new rules are obtained by replacing $\nabla_{\mu}$ and $\Gamma^M_{\mu N}$ by  $\tilde{\nabla}_{\mu}$ and $\tilde{\Gamma}^M_{\mu N}$ in the above equations, respectively.  As a result, the local translations of $g_{\mu\nu}$ and $\phi_{\mu\nu\lambda}$ depend on the matter fields. It might be puzzling that even the diffeomorphism of the metric-like quantities depend on the matter fields through the torsion tensor. 
For example, 
\begin{eqnarray}
\delta \, g_{\mu\nu} &=&  \tilde{\nabla}_{\mu} \, \xi_{\nu}+\tilde{\nabla}_{\nu} \, \xi_{\mu} \nonumber \\
&=& \nabla_{\mu} \, \xi_{\nu}+\nabla_{\nu} \, \xi_{\mu}- \, \Delta \Gamma^M_{\mu\nu} \, \xi_M
- \, \Delta \Gamma^M_{\nu\mu} \, \xi_M. 
\label{diffeo-like}
\end{eqnarray}
For diffeomorphism, $\xi^{(\mu\nu)}=0$ and $\xi_M=G_{M \lambda} \, \xi^{\lambda}$.  There are extra terms in (\ref{diffeo-like}) 
which should not appear in general coordinate transformation. However, as we mentioned before, our theory is originally a topological theory in the frame-like approach, and is diffeomorphism invariant, when the vielbein is transformed as a covariant vector. As a result, $\tilde{S}_{\text{tot}}$ in the metric-like 
approach is also invariant under diffeomorphism,  when  $g_{\mu\nu}$ and $\phi_{\mu\nu\lambda}$ as well as $C^0$, $C^{\mu}$, $C^{(\mu\nu)}$ and ${B_0}^{\lambda}$, ${B_{\mu}}^{\lambda}$ and ${B_{(\mu\nu)}}^{\lambda}$ transform as tensors in the usual way. 
The difference between the diffeomorphism and the diffeomorphism-like transformation (\ref{diffeo-like}) should also be the symmetry transformation of the action integral. Therefore, the symmetry of the matter-coupled theory in the metric-like formalism seems to be  larger than that of the pure spin-3 gravity! As we saw in sec.3, in the pure spin-3 case a part of local translation agrees with diffeomorphism. To study this symmetry in more details, we need to compute the symmetry algebra. 
This will not, however,  be performed here.

To conclude that the symmetry in the metric-like formalism becomes larger after coupling to the matter fields, 
let us study if it is possible to redefine the transformation parameters $\xi_M$ so that transformation (\ref{diffeo-like}) takes the
ordinary form (\ref{delgmn}). The diffeomorphism for $g_{\mu\nu}$, (\ref{diffeo-like}), is rewritten as
\begin{eqnarray}
\delta_{\text{diffeo}} \, g_{\mu\nu} =  (g_{\nu\lambda} \, \partial_{\mu} \, \xi^{\lambda}  +g_{\mu\lambda} \, \partial_{\nu} \, \xi^{\lambda}  +\xi^{\lambda} \, \partial_{\lambda} \, g_{\mu\nu} )
-( \Delta\Gamma^M_{\mu\nu}
+ \, \Delta \Gamma^M_{\nu\mu}) \, G_{M\lambda} \, \xi^{\lambda}.  \label{diffg}
\end{eqnarray}
The spin-3 transformation for $g_{\mu\nu}$ is given by\footnote{Note that  $\xi^{\lambda}= g^{\lambda\rho} \, \xi_{\rho}$ and $\zeta_{(\lambda\rho)}=\xi_{(\lambda\rho)}-g_{(\lambda\rho)\sigma} \, \xi^{\sigma}$. }
\begin{eqnarray}
\delta_{\text{spin-3}} \, g_{\mu\nu} =  -\Gamma_{\mu\nu}^{(\rho\sigma)} \, \zeta_{(\rho\sigma)}
-\frac{1}{2} \, (\Delta \, \Gamma_{\mu\nu}^{ (\rho\sigma)}+\Delta \, 
\Gamma_{\nu\mu}^{ (\rho\sigma)}) \, \zeta_{(\rho\sigma)}.  \label{spin3g}
\end{eqnarray}
Let us consider the case, where $C$ and $B$ are small and it is possible to perform perturbation in $BC$ (or  the torsion $[B,C]$). 
The first terms of (\ref{diffg}) and (\ref{spin3g}) are ${\cal O}((BC)^0)$ and the second terms are ${\cal O}((BC)^1)$ 
due to $\Delta \Gamma$. If it is possible to choose  $\zeta_{(\rho\sigma)}$ as a function of $\xi^{\lambda}$ in such 
a way that the transformation (\ref{spin3g}) cancels the second term of  (\ref{diffg}),  we may simply redefine the combined transformation as a new diffeomorphism $\delta'_{\text{diffeo}} \, g_{\mu\nu}$. Similarly, if it is possible to choose   $\xi^{\lambda}$
 as a function of $\zeta_{(\rho\sigma)}$ in such a way that the transformation (\ref{diffg}) 
cancels the second term of  (\ref{spin3g}), the combined transformation is a new spin-3 transformation $\delta'_{\text{spin-3}} \, g_{\mu\nu}$. This procedure is equivalent to investigate if it is possible to modify the relation among the transformation functions of the  local translation,  $\Lambda_-^a$, $\xi_{\mu}$ and $\xi_{(\mu\nu)}$,  from  $\Lambda_-^a =\, E^{a\mu} \, \xi_{\mu}+\frac{1}{2} \,  E^{a(\mu\nu)} \, \xi_{(\mu\nu)}$ to a new one, in such a way that the transformation of the metric tensor $\delta \, g_{\mu\nu}= \delta e^{a}_{\mu} \, e_{a}^{\nu}+e_{a\mu} \, \delta e^a_{\nu}$ 
with $\delta e^a_{\mu}= \partial_{\mu} \, \Lambda^a_-+{f^a}_{bc} \, \tilde{\omega}^b_{\mu}(e,B,C) \, \Lambda^c_-$, expressed in terms of the metric-like quantities take the `pure-spin-3-gravity form'  for diffeomorphism and spin-3 transformation.

The above redefinition of the diffeomorphism and spin-3 transformation is, however, not possible, because  $\Gamma_{\mu(\rho\sigma)}^{\lambda}$, as given in (\ref{Gamma12}),  is first order in $\phi_{\mu\nu\lambda}$. At least within perturbation in $\phi_{\mu\nu\lambda}$, (\ref{spin3g}) cannot be solved for $\zeta_{(\rho\sigma)}$. The spin-3 algebra which stems from local translation is modified due to matter coupling.

To summarize, the diffeomorphism symmetry in the metric-like formalism which stems from the fact that the vielbein is a vector of diffeomorphism and  a part of the local translation symmetry coincides only in the pure spin-3 gravity. When matters are coupled to spin-3 gravity, part of the local translation in the frame-like formalism is converted into a symmetry transformation in the metric-like formalism which is different from diffeomorphism.  So the matter-coupled theory is a 3d gravity with new spin-2 and spin-3 gauge symmetry. 

We now turn to the transformation rule of $C^A$ under the local translation (b). Instead of the gauge parameter 
$\Lambda_-$, let us introduce new functions $\xi_M= \ell \, \Lambda^a_- \, e_{aM}$ and $\xi^M= \ell \, \Lambda_-^a \, E_a^M$. Then (\ref{bC1}) is rewritten as 
\begin{equation}
\delta \, C^0 = -\frac{4}{3\ell } \, \xi_M \, C^M.
\end{equation}
This does not look like a diffeomorphism of matter fields. However, if $\xi^{(\mu\nu)}=0$, the eq of motion (\ref{eomC0}) can be 
used to show that 
\begin{eqnarray}
 \xi^{\mu} \ \partial_{\mu} \, C^0 &=& (\ell \, E^{\mu}_a \, \Lambda_-^a )\, (-\frac{4}{3\ell} \, e_{b\mu} \, C^b)=
-\frac{4}{3} \, \Lambda^a_- \, (h_{ab}-\frac{1}{2} \,  E_a^{(\mu\nu)} \, e_{b(\mu\nu)} )\, C^b \nonumber \\
&=& -\frac{4}{3} \, \Lambda^a_- \, C_a+ \frac{2}{3} \, \xi^{(\mu\nu)} \,e_{b(\mu\nu)} \, C^b =
-\frac{4}{3\ell} \, \xi^M \, C_M =\delta \, C^0
\end{eqnarray}
At the second equality of the first line, a relation $E^M_a \, e^b_M=\delta_a^b$, which is a counterpart of (\ref{Ee}), is used. 
Therefore $\delta \, C^0$ is the ordinary diffeomorphism for $C^0$. On the other hand, if $\xi^{\mu}=0$, 
we have 
\begin{eqnarray}
\delta \, C^0&= &-\frac{2}{3\ell} \, \xi^{(\mu\nu)} \, C_{(\mu\nu)}=-\frac{2}{3\ell} \, \xi^{(\mu\nu)} \, 
G_{(\mu\nu)M} \, C^M \nonumber \\
&=& -\frac{1}{6\ell} \, \xi^{(\mu\nu)} \, C^{(\lambda\rho)} \, g_{\mu\lambda} \, g_{\nu\rho}
-\frac{2}{3\ell} \, \xi^{(\mu\nu)} \, C^{\lambda} \, \phi_{\mu\nu\lambda}
+{\cal O}(\phi^2) 
\end{eqnarray}
$C^0$ mixes with $C^M$ under spin-3 transformation. 

The transformation of $C^M$ is more involved. Let us recall the transformations of the vielbeins. The vielbein transforms as $\delta \, e^a_{\mu}= \ell \, \tilde{D}^{(L)}_{\mu}\, \Lambda^a_-$. 
The transformation of the additional one, $e^a_{(\mu\nu)}= \frac{1}{4} \, {d^a}_{bc} \, e^b_{\lambda} \, 
e^c_{\rho} \, P^{\lambda\rho}_{\mu\nu}$ is given by 
\begin{equation}
\delta \, e^a_{(\mu\nu)} = \frac{1}{4} \, {d^a}_{bc} \, e^b_{\lambda} \, e^c_{\rho} \, \delta \, P^{\lambda\rho}_{\mu\nu} +
\frac{1}{2} \, {d^a}_{bc} \, E^{bM} \, e^c_{\rho} \, \tilde{\nabla}_{\lambda} \, \xi_M \, P^{\lambda\rho}_{\mu\nu}.
\end{equation}
Then by using  (\ref{bC2}) and (\ref{delgmn}), we have 
\begin{eqnarray}
\delta \, C^M=\delta \, (E^M_a \, C^a) &=& -\frac{1}{\ell} \, {K^M}_{NL} \, C^L \, \xi^N-C^{\mu} \, {P^M}_N \,  \tilde{\nabla}_{\mu} \, \xi^M-\frac{1}{2} \, C^{(\mu\nu)}
\, {K^{M}}_{N\mu} \, \tilde{\nabla}_{\nu} \, \xi^N\nonumber \\ 
&&+\frac{1}{6} \, C^{(\mu\nu)} \, {K^M}_{\lambda\rho} \, g^{\lambda\rho} \, \tilde{\nabla}_{\mu} \, \xi_{\nu}-\frac{2}{\ell} \, C^0 \, \xi^M +C^N \, \delta ({P^M}_N).
\end{eqnarray}
Here $\xi_M$ and $\xi^N$ are related as $\xi_M= G_{MN} \, \xi^N$. If $M=(\mu\nu)$ and $N=(\lambda\rho)$, ${P^M}_N=P^{\mu\nu}_{\lambda\rho}$. If  $M=\mu$ and $N=\nu$, ${P^M}_N={\delta^{\mu}}_{\nu}$. Otherwise, ${P^M}_N=0$. 

Because the matter system is topological, it may be allowed to use the equations of motion to rewrite the above transformation. 
The equations of motion for $C^0$ and $C^a$ derived from (\ref{topological action2}) are as follows.
\begin{eqnarray}
& \partial_{\mu} \, C^0+\frac{4}{3\ell} \, e_{a\mu} \, C^a=0, \label{eomC0} \\
& \tilde{D}^{(L)}_{\mu} \, C^a+\frac{1}{\ell} \, {d^a}_{bc} \, e^b_{\mu} \, C^c+\frac{2}{\ell} \, e^a_{\mu} \, C^0=0 \label{eomCa}
\end{eqnarray}
Those for $B$'s are
\begin{eqnarray}
& \partial_{[\mu} \, B^0_{\nu\lambda]}-\frac{4}{3\ell} \, e_{a[\mu} \, B^a_{\nu\lambda]}   =0, \\
& \tilde{D}^{(L)}_{[\mu} \, B^a_{\nu\lambda]}-\frac{1}{\ell} \, {d^a}_{bc} \, e^b_{[\mu} \, B^c_{\nu\lambda]}
-\frac{2}{\ell} \, e^a_{[\mu} \, B^0_{\nu\lambda]}=0.  
\end{eqnarray}
Here $[ \ , \  ]$ stands for complete anti-symmetrization of the indices in between. 

Eq (\ref{eomCa}) leads to the relations.
\begin{eqnarray}
\tilde{\nabla}_{\mu} \, C^{\nu} &=& -\frac{1}{\ell} \, {K^{\nu}}_{\mu M} \, C^M-\frac{2}{\ell} \, \delta^{\nu}_{\mu} \, C^0, \label{eom1} \\
\tilde{\nabla}_{\mu} \, C^{(\nu\lambda)} &=& -\frac{1}{\ell} \, {K^{(\nu\lambda)}}_{\mu M} \, C^M+\frac{1}{3}\, g^{\nu\lambda} \, g_{\rho\sigma} \, \tilde{\nabla}_{\mu} \, C^{(\rho\sigma)}. \label{eom2}
\end{eqnarray}
Note that the second term on the righthand side of the last equation does not vanish, because 
the two indices of $g_{\rho\sigma}$ is contracted with the single index $M=(\rho\sigma)$ of $C^M$. This is not the ordinary rule of contraction of indices, and so $g_{\rho\sigma}$ cannot go through $\tilde{\nabla}_{\mu}$. 

When $\xi^{(\mu\nu)}=0$, by using (\ref{eom1}) and (\ref{eom2}), we obtain the transformations,\footnote{Relation with $\xi_{\mu}$, $\zeta_{\mu\nu}$ in subsec.3.3 is, $\xi_{\mu}=g_{\mu\nu} \, \xi^{\nu}$ in the case of diffeomorphism ($\xi^{(\mu\nu)}=0$), and $\zeta_{\mu\nu}=\frac{1}{2} \, M_{(\mu\nu)(\lambda\rho)} \, \xi^{(\lambda\rho)}$ in the case of spin-3 transformation ($\xi^{\mu}=0$).}
\begin{eqnarray}
\delta \, C^{\mu} &=& \xi^{\rho} \, \tilde{\nabla}_{\rho} \, C^{\mu}-C^{\rho} \, \tilde{\nabla}_{\rho} \, \xi^{\mu}-\frac{1}{2} \, C^{(\rho\sigma)} \, {K^{\mu}}_{M\rho} \, \tilde{\nabla}_{\sigma} \, \xi^M 
+\frac{1}{6} \, C^{(\rho\sigma)} \, {K^{\mu}}_{\kappa\tau} \, g^{\kappa\tau} \, \tilde{\nabla}_{\rho} \, \xi_{\sigma},    \nonumber \\ 
& =& \xi^{\rho} \, \partial_{\rho} \, C^{\mu} -C^{\rho} \, \partial_{\rho} \, \xi^{\mu} 
+\frac{1}{2} \, C^{(\rho\sigma)} \, {K^{\mu}}_{M\rho} \, 
\Delta \, \Gamma^M_{\sigma \kappa} \, \xi^{\kappa} 
-\frac{1}{6} \, C^{(\rho\sigma)} \, {K^{\mu}}_{\kappa\tau} \, g^{\kappa\tau} \, \Delta \, \Gamma^M_{\rho\sigma} \, G_{M\lambda} \, \xi^{\lambda} \nonumber \\  &&
+\xi^{\rho} \, \Delta\Gamma^{\mu}_{\rho M} \, C^M-C^{\rho} \, \Delta\Gamma^{\mu}_{\rho\lambda} \, \xi^{\lambda}+{\cal O}(\phi^2), \label{trC1} \\  
\delta \, C^{(\mu\nu)} &=& \xi^{\rho} \, \tilde{\nabla}_{\rho} \, C^{(\mu\nu)}-C^{\rho} \, \tilde{\nabla}_{\rho} \, \xi^{(\mu\nu)}+\frac{1}{3} \, g^{\mu\nu} \, g_{\rho\sigma} \, C^{\lambda} \, \tilde{\nabla}_{\lambda} \, \xi^{(\rho\sigma)}-\frac{1}{2} \, C^{(\lambda\rho)} \,{K^{(\mu\nu)}}_{M \lambda} \, \tilde{\nabla}_{\rho} \, \xi^M  \nonumber \\ 
&&+\frac{1}{6} \, C^{(\lambda\rho)} \, {K^{(\mu\nu)}}_{\sigma\kappa} \, g^{\sigma\kappa} \, \tilde{\nabla}_{\lambda} \, \xi_{\rho} -\frac{2}{3} \, g^{\mu\nu} \, C^{(\lambda\rho)} \, \tilde{\nabla}_{\lambda} \, \xi_{\rho}-\frac{1}{3} \, g^{\mu\nu} \, g_{\kappa\sigma} \, \xi^{\rho} \, \tilde{\nabla}_{\rho} \, C^{(\kappa\sigma)} \nonumber \\
&  =&
\xi^{\rho} \, \partial_{\rho} \, C^{(\mu\nu)}-C^{(\mu\rho)} \, \partial_{\rho} \, \xi^{\nu} -C^{(\rho\nu)} \, \partial_{\rho} \, \xi^{\mu}
+\xi^{\rho} \, \Delta \, \Gamma^{(\mu\nu)}_{\rho M} \, C^M-C^{\rho} \, \Delta \, \Gamma^{(\mu\nu)}_{\rho \lambda} \, \xi^{\lambda} \nonumber \\ 
&&+\frac{1}{3} \, g^{\mu\nu} \, g_{\rho\sigma} \, C^{\lambda} \, \Delta \, \Gamma^{(\rho\sigma)}_{\lambda\kappa} \, \xi^{\kappa}
-\frac{1}{2} \, C^{(\lambda\rho)} \,{K^{(\mu\nu)}}_{M \lambda} \, \Delta \, \Gamma^M_{\rho\sigma} \, \xi^{\sigma}  \nonumber \\ 
&&-\frac{1}{6} \, C^{(\lambda\rho)} \, {K^{(\mu\nu)}}_{\sigma\kappa} \, g^{\sigma\kappa} \, \Delta \, \Gamma^M_{\lambda\rho} \, G_{M\tau} \, \xi^{\tau} +\frac{2}{3} \, g^{\mu\nu} \, C^{(\lambda\rho)} \, \Delta \Gamma^M_{\lambda\rho} \, G_{M\tau} \, \xi^{\tau} \nonumber \\ &&-\frac{1}{3} \, g^{\mu\nu} \, g_{\kappa\sigma} \, \xi^{\rho} \, \Delta\, \Gamma_{\rho M}^{(\kappa\sigma)} \, C^{M} 
+{\cal O}(\phi^2). \label{trC2}
\end{eqnarray}
In both equations, those terms containing $\phi$ canceled in a non-trivial way except in the torsion terms containing $\Delta \, \Gamma$'s. Except for those terms with $\Delta \Gamma$, these transformations are those for a vector and a rank-2 tensor.  This is the same situation as for the diffeomorphism of $\phi_{\mu\nu\rho}$ discussed at the end of the subsec.3.4. 
If there is no torsion, the above transformation coincides with diffeomorphism. 
The torsion terms are cubic in the matter fields. 
Due to the torsion terms, the local translation of $C^{\mu}$ and $C^{(\mu\nu)}$ with $\xi^ {(\lambda\rho)}=0$ does not coincide with the diffeomorphism for  a vector and a rank-2 tensor. The matter action (\ref{Mactionexp}), however, is clearly invariant under the diffeomorphism.\footnote{ $T^M_{\mu\nu}$ and $\Delta \Gamma^N_{\mu M}$ behave as tensors under diffeomorphism.} Therefore $\tilde{S}_{\text{tot}}$ is also invariant under the usual diffeomorphism, although it is different from the local translation (b). 
As we discussed at the beginning of this subsection, the symmetry of the spin-3 gravity coupled to matter in the metric-like formalism is larger than that of the pure spin-3 gravity. 
Let us also note that in the above equation,  although $\xi^{(\mu\nu)}=0$, quantities such as $\tilde{\nabla}_{\lambda} \, \xi^{(\rho\sigma)}= \tilde{\Gamma}^{(\rho\sigma)}_{\lambda\kappa} \, \xi^{\kappa}$ do not vanish.

On the other hand, when $\xi^{\mu}=0$, we have the spin-3 transformation.
\begin{eqnarray}
\delta \, C^{\mu} &=& -\frac{1}{2\ell} \, {K^{\mu}}_{(\nu\rho)M} \, C^M \, \xi^{(\nu\rho)} -C^{\nu} \, \tilde{\nabla}_{\nu} \, \xi^{\mu}-\frac{1}{2} \, 
C^{(\lambda\rho)} \, {K^{\mu}}_{M \lambda} \, \tilde{\nabla}_{\rho} \, \xi^M \nonumber \\
&&+\frac{1}{6} \, C^{(\lambda\rho)} \, {K^{\mu}}_{\nu\sigma} \, g^{\nu\sigma} \, \tilde{\nabla}_{\lambda} \, \xi_{\rho} \nonumber \\
&=& \left[\delta \, C^{\mu}\right]_0+\left[\delta \, C^{\mu}\right]_1
-\frac{1}{2} \, C^{\nu} \, \Delta \, \Gamma^{\mu}_{\nu (\lambda\sigma)} \, \xi^{(\lambda\sigma)}-\frac{1}{4} \, 
C^{(\lambda\rho)} \, {K^{\mu}}_{M \lambda} \, \Delta \, \Gamma^M_{\rho(\sigma\kappa)} \, \xi^{(\sigma\kappa)} \nonumber \\ &&
-\frac{1}{6} \, C^{(\lambda\rho)} \, {K^{\mu}}_{\nu\sigma} \, g^{\nu\sigma} \, \Delta \, \Gamma^M_{\lambda\rho} \, \xi_{M} 
+ {\cal O}(\phi^2),  \label{Cspin3-1}
\end{eqnarray}
\begin{eqnarray}
\delta \, C^{(\mu\nu)} &=& -\frac{1}{2\ell} \, {K^{(\mu\nu)}}_{(\lambda\rho)M} \, C^M \, \xi^{(\lambda\rho)}-\frac{2}{\ell} \, C^0 \, \xi^{(\mu\nu)}-C^{\rho} \,\tilde{\nabla}_{\rho} \, \xi^{(\mu\nu)}\nonumber \\ &&-\frac{1}{2} \, C^{(\lambda\rho)} \, 
{K^{(\mu\nu)}}_{M\lambda} \, \tilde{\nabla}_{\rho} \, \xi^M 
+\frac{1}{6} \, C^{(\lambda\rho)} \, {K^{(\mu\nu)}}_{\sigma\kappa} \, g^{\sigma\kappa} \, \tilde{\nabla}_{\lambda} \, \xi_{\rho}
\nonumber \\
&&-\frac{2}{3} \, g^{\mu\nu} \, C^{(\lambda\rho)} \, \tilde{\nabla}_{\rho} \, \xi_{\lambda}+\frac{1}{3} \, g^{\mu\nu} \, g_{\sigma\kappa} \, C^{\lambda} \, \tilde{\nabla}_{\lambda} \, \xi^{(\sigma\kappa)} \nonumber \\
&=& \left[\delta \, C^{(\mu\nu)}\right]_0+\left[\delta \, C^{(\mu\nu)}\right]_1
-\, \frac{1}{2} \, C^{\rho} \,\Delta \, \Gamma^{(\mu\nu)}_{\rho(\sigma\kappa)} \, \xi^{(\sigma\kappa)}\nonumber \\ &&-\frac{1}{4} \, C^{(\lambda\rho)} \, 
{K^{(\mu\nu)}}_{M\lambda} \, \Delta \, \Gamma^M_{\rho (\sigma\kappa)} \, \xi^{(\sigma\kappa)} 
-\frac{1}{6} \, C^{(\lambda\rho)} \, {K^{(\mu\nu)}}_{\sigma\kappa} \, g^{\sigma\kappa} \, \Delta \, \Gamma^M_{\lambda\rho} \, \xi_M
\nonumber \\
&&+\frac{2}{3} \, g^{\mu\nu} \, C^{(\lambda\rho)} \, \Delta \, \Gamma^M_{\rho\lambda} \, \xi_M+\frac{1}{6} \, g^{\mu\nu} \, g_{\sigma\kappa} \, C^{\lambda} \, \Delta \, \Gamma^{(\sigma\kappa)}_{\lambda (\tau\eta)} \, \xi^{(\tau\eta)}
+ {\cal O}(\phi^2). \label{Cspin3-2}
\end{eqnarray}
$[\delta \, C^M]_n$ is of ${\cal O}(\phi^n)$, and does not depend on $\Delta \Gamma^M_{\mu N}$.
The $\phi$ expansions of these terms are presented in appendix G.  The explicit forms of the terms coming 
from the torsion is not worked out here explicitly due mainly to the page size.

Transformation of the field $B$ can also be worked out. As for diffeomorphism, because the matter action (\ref{Mactionexp}) is manifestly invariant, ${B^0}^{\mu}$ and  ${B_M}^{\mu}={{B}_{\nu}}^{\mu}, \, {{B}_{(\nu\lambda)}}^{\mu}$ must also transform as tensors with additional terms containing $\Delta \Gamma$.
As for the spin-3 transformation, analysis similar to the $C$ fields lead to the following transformations (with $\xi^{\mu}=0$).
\begin{eqnarray}
\delta \, {B_0}^{\lambda} &=& \frac{1}{\ell} \, \xi^{(\mu\nu)} \, {B_{(\mu\nu)}}^{\lambda}-\frac{1}{2} \, g^{\rho\tau} \, G_{\tau(\kappa\sigma)} \, \tilde{\nabla}_{\rho} \, \xi^{(\kappa\sigma)} \, {B_0}^{\lambda},  \\
\delta \, {B_\rho}^{\lambda} &=& -\frac{1}{2} \, g^{\alpha\beta} \, G_{\beta(\kappa\sigma)} \, \tilde{\nabla}_{\alpha} \, \xi^{(\kappa\sigma)} \, {B_{\rho}}^{\lambda}+\frac{1}{2} \, {B_{\tau}}^{\lambda} \, \tilde{\Gamma}^{\tau}_{\rho(\kappa\sigma)} \, \xi^{(\kappa\sigma)}+\frac{1}{2} \, {B_{(\kappa\sigma)}}^{\lambda} \, \tilde{\nabla}_{\rho} \, \xi^{(\kappa\sigma)} \nonumber \\
&& + \frac{1}{2\ell} \, \xi^{(\kappa\sigma)} \, {K_{(\kappa\sigma)\rho}}^{\tau} \, {B_{\tau}}^{\lambda}+ \frac{1}{4\ell} \, \xi^{(\kappa\sigma)} \, {K_{(\kappa\sigma)\rho}}^{(\alpha\beta)} \, {B_{(\alpha\beta)}}^{\lambda} \nonumber \\ && +\frac{2}{3\ell} \, G_{\rho(\kappa\sigma)} \, \xi^{(\kappa\sigma)} \, {B_0}^{\lambda}, 
\end{eqnarray}
\begin{eqnarray}
\delta \, {B_{(\rho\sigma)}}^{\lambda} &=&-\frac{1}{2} \, g^{\alpha\beta} \, G_{\beta(\kappa\eta)} \, \tilde{\nabla}_{\alpha} \, \xi^{(\kappa\eta)} \, {B_{(\rho\sigma)}}^{\lambda}-\frac{1}{6} \, {K^M}_{\kappa\eta} \, {B_{M}}^{\lambda} \, g^{\kappa\eta} \, (G_{\sigma N} \, \tilde{\nabla}_{\rho} \, \xi^N+G_{\rho N} \, \tilde{\nabla}_{\sigma} \, \xi^N)\nonumber \\
&&+\frac{1}{3} \, {K^M}_{\kappa\eta} \, {B_M}^{\lambda} \, g_{\rho\sigma} \, g^{\kappa\alpha} \, g^{\eta\beta} \, G_{\beta N} \, \tilde{\nabla}_{\alpha} \, \xi^N+\frac{1}{2} \, {K^N}_{M\kappa} \, \tilde{\nabla}_{\eta} \, \xi^M \, P^{\kappa\eta}_{\rho\sigma} \, {B_N}^{\lambda}\nonumber \\
&& +\frac{1}{\ell} \, {K_{(\rho\sigma)M}}^N \, \xi^M \, {{B_N}}^{\lambda}+\frac{2}{3\ell} \, G_{(\rho\sigma)(\kappa\tau)} \, \xi^{(\kappa\tau)} \, {B_0}^{\lambda}  
\end{eqnarray}

\section{Summary}
\hspace{5mm}
We expressed the generalized connections $\Gamma^M_{\mu N}$  and curvature tensors ${R^M}_{N\mu \nu}$, which 
were introduced in our previous work\cite{FN}, in terms of the metric $g_{\mu\nu}$ and the spin-3 field 
$\phi_{\mu\nu\lambda}$ explicitly by means of perturbative expansions in $\phi$.  The matter coupling to 0-form 
$C$ and 2-form $B$ fields of the spin-3 gravity is introduced in the action formalism, firstly in the frame-like 
approach, and then translated into the metric-like approach. 
We eliminated the spin connection $\omega^a_{\mu}$ by  solving the equation of motion for the total action to obtain 
the solution $\omega^a_{\mu}=\tilde{\omega}^a_{\mu}(e,B,C)$. This spin connection has a  torsion and this leads to an 
action which contains interaction terms of the form $(BC)^2$ introduced due to the torsion. We found that the symmetry 
in the metric-like formalism is enhanced, when the  matter fields are coupled to the spin-3 gravity.

The construction of the matter coupling presented in this paper can be applied to other 3D 
higher-spin gravity theories based on $SL(N,R) \times SL(N,R)$ and $hs[\lambda] \times hs[\lambda]$ Chern-Simons 
theories, and a similar conclusion is expected. It will also be possible to introduce topological matter composed of two 
1-forms, $C_{\mu}$ and $B_{\mu}$ in a similar way. 

Finally, we will discuss on AdS/CFT correspondence\cite{ads1}\cite{ads2}\cite{ads3}. 
Since we obtained a matter theory interacting with 3d spin-3 gravity, it will be natural to study AdS/CFT correspondence for this model. 
In AdS/CFT correspondence, $C^0=(1/3)\, \text{tr}C$ will serve as a source for a scalar operator $O$ on the boundary, as in \cite{AKP2} for 3d Vasiliev theory. There will be another operator $\tilde{O}$ which corresponds to $B^0_{t\phi}$. 
Usually, in  AdS/CFT correspondence, the solutions to the equations of motion are substituted into the action integral. 
The equations of motion for the gauge fields, $A,\bar{A}=\omega \pm \frac{1}{\ell} \, e$, are given by (\ref{nonflatF})
and in general, these gauge fields are not flat connections, and we cannot use the method of \cite{HKP} to find solutions. 
These equations describe the back reaction of the matter to the gravity, and it was shown in this paper that there is 
non-vanishing torsion.  One can show that  (\ref{nonflatF}) is consistent with the matter equations of motion by using the 
Bianchi identities,  $dF+A \wedge F-F \wedge A=0$ and similar equation for $\bar{F}$. We need to study solutions to the full set of the equations of motion. Especially, we need to investigate if there exist asymptotically AdS$_3$ solutions, and if $\text{tr} \, C$ and $\text{tr} \, B_{t\phi}$  satisfy Klein Gordon equation in a spacetime with torsion, {\em etc}. Then, as in the case of AdS/CFT duality for spinors, we might choose $\int_{\partial M} \text{tr} \, (B\wedge C)$ as a boundary action integral.\cite{Hen}\footnote{$\partial M$ is the boundary of the spacetime.} When $A$ and $\bar{A}$ are fixed flat backgrounds, this method will not work, because the solutions $C$ and $B$ are covariantly constant and the value of the boundary action will be too trivial to provide a two-point function, even though the Lagrangian formulation is available. When $A$ and $\bar{A}$ are dynamical variables and non-flat,  it may be possible to extract information about the correlation functions in the boundary CFT from the boundary action. 

In this paper it is found that the action integral can be rewritten in terms of the world-volume components of $C$ and $B$, and these 
components transform in a non-trivial way under the spin-3 transformation.  Then those components of $C$ and $B$ different from  
$C^0=\text{tr} \, C$ and $B^0=\text{tr} \, B$ might take part in the AdS/CFT correspondence. 
What is the role of the other components of $C$ and $B$ in the gravity/CFT correspondence? It is known that in spin-3 
gravity, there exists $W_3$ current in the CFT on the boundary\cite{Campoleoni}\cite{HR}  and in the CFT 
with $W$ symmetry, the OPEs of the primary field $O$ and the $W_3$ current contain new fields,  $\hat{W}_{-1} \,  O$ 
and $\hat{W}_{-2} \,  O$, which are not simply related to $O$ by just differentiations. \cite{Watts}
\begin{equation}
W(z) \, O(z')= \frac{w}{(z-z')^3}  \, O(z')+\frac{1}{(z-z')^2} \, \hat{W}_{-1} \, O(z') +\frac{1}{z-z'} \, \hat{W}_{-2} \, O(z') +\dots
\end{equation}
 $C^{\mu}$ and $C^{(\mu\nu)}$ might be the sources for $\hat{W}_{-1} \,  O$ and $\hat{W}_{-2} \,  O$, and their anti-chiral counterparts, where $W_n$ is the expansion mode of the $W_3$ current in the boundary CFT. 
It remains to be studied if these OPE's can be observed in the AdS/CFT correspondence.


\newpage
\setcounter{section}{0}
\renewcommand{\thesection}{\Alph{section}}

\section{Notations for sl(3,R) algebra}
\hspace{5mm}
In this appendix notations related to $sl(3,R)$ algebra are summarized. 

Let the generators $L_i \ (i=-1,0,1)$, $W_n \ (n=-2, \ldots 2)$ satisfy 
an $sl(3, R)$ algebra.
\begin{eqnarray}
\ [L_i,L_j] &=& (i-j) \, L_{i+j}, \qquad 
\ [L_i,W_n] = (2i-n) \, W_{i+n}, \nonumber \\
\ [W_m,W_n] &=& -\frac{1}{3} (m-n) \, \{2m^2+2n^2-mn-8\} \, L_{m+n}
\end{eqnarray}
We use the same three-dimensional representation as in \cite{Campoleoni} with 
the parameter $\sigma=-1$.
\begin{eqnarray}
L_1 &=& \left(\begin{array}{ccc}
        0  & 0& 0 \\
        1 & 0 & 0 \\
        0 & 1 & 0  \end{array}\right), \qquad 
L_0  = \left(\begin{array}{ccc}
        1  & 0& 0 \\
        0 & 0 & 0 \\
        0 & 0 & -1 \end{array} \right), \qquad 
L_{-1}  = \left(\begin{array}{ccc}
        0  & -2& 0 \\
        0 & 0 & -2 \\
        0 & 0 & 0  \end{array}\right), \nonumber \\
W_2 &=& \left(\begin{array}{ccc}
        0  & 0& 0 \\
        0 & 0 & 0 \\
        2 & 0 & 0  \end{array}\right), \qquad 
W_1  = \left(\begin{array}{ccc}
        0  & 0& 0 \\
        1 & 0 & 0 \\
        0 & -1 & 0 \end{array} \right), \qquad 
W_0  = \frac{2}{3} \, \left(\begin{array}{ccc}
        1  & 0& 0 \\
        0 & -2 & 0 \\
        0 & 0 & 1  \end{array}\right), \nonumber \\
W_{-1} &=& \left(\begin{array}{ccc}
        0  & -2 & 0 \\
        0 & 0 & 2 \\
        0 & 0 & 0 \end{array} \right), \qquad 
W_{-2}  = \left(\begin{array}{ccc}
        0  & 0& 8 \\
        0 & 0 & 0 \\
        0 & 0 & 0  \end{array}\right)
\label{generators}
\end{eqnarray}
Non-vanishing norms of these matrices are given by 
\begin{eqnarray}
\mbox{tr} \ (L_0)^2=2, \quad \mbox{tr} \ (L_{-1}L_1)=-4, \quad \mbox{tr} \ 
 (W_0)^2=\frac{8}{3}, \quad \mbox{tr} \ (W_1W_{-1})=-4, \quad \mbox{tr} \
 (W_2W_{-2})=16.
\end{eqnarray}
These generators will also be collectively denoted as $t_a, (a=1, \ldots,8)$,
\begin{eqnarray}
t_1&=&L_1, \quad t_2=L_0, \quad t_3= L_{-1}, \nonumber \\
t_4 &=&W_2, \quad t_5=W_1, \quad t_6= W_0, \quad t_7= W_{-1}, \quad  
t_8= W_{-2}. 
\end{eqnarray}
The structure constants ${f_{ab}}^c$ are defined by 
\begin{eqnarray}
 [ t_a, t_b]= {f_{ab}}^c \, t_c.
\end{eqnarray}
The Killing metric $h_{ab}$ for the local frame is defined by 
\begin{eqnarray}
h_{ab}=\frac{1}{2} \, \mbox{tr} \, (t_a t_b)
\end{eqnarray}
Its nonzero components are given by $h_{22}=1, \ h_{13}=h_{31}=-2, 
h_{48}=h_{84}=8, \ h_{57}=h_{75}=-2, \ h_{66}=4/3$. 
This metric tensor has a signature $(3,5)$. 
Indices of the local frame are raised and lowered by $h_{ab}$ and its inverse 
$h^{ab}$. Then $f_{abc} \equiv {f_{ab}}^d \, h_{dc}$ 
is completely anti-symmetric in the three indices. It can be shown that 
$f_{abc}$ and $h_{ab}$ are related by
\begin{equation}
h_{ab}= -\frac{1}{12} \, {f_a}^{cd} \, f_{bcd}. \label{hff}
\end{equation}
The structure constants are given by
\begin{eqnarray}
&& f_{123} = -2, \ f_{158}=8, \ f_{167}=-4, \ f_{248}=-16, \nonumber \\
&& f_{257} = 2, \ f_{347}=8, \ f_{356}=-4
\end{eqnarray}

The invariant tensor ${d_{ab}}^c$ is defined by 
\begin{eqnarray}
 \{t_a, t_b \} = {d_{ab}}^c \, t_c+ {d_{ab}}^0 \, t_0,
\end{eqnarray}
where $t_0=\bm{I}$ is an identity matrix. The constant with the lowered 
index $d_{abc}= {d_{ab}}^d \, h_{dc}$ is completely symmetric in all the 
indices. 
These constants are given by
\begin{eqnarray}
&& d_{127}=d_{235}=-2, \quad  d_{136}= d_{226}=d_{567}=\frac{4}{3},  \quad 
d_{118}= d_{334}=8, \nonumber \\
&& d_{468}=\frac{32}{3}, \quad d_{477}= d_{558}=-8, \quad  d_{666}=-\frac{16}{9}, 
\qquad {d_{ab}}^0=\frac{4}{3} \, h_{ab} 
\end{eqnarray}

\section{Metric $G_{MN}=e^a_M \, e_{aN}$ and its inverse $G^{MN}$}
\hspace{5mm}
\begin{eqnarray}
G_{\mu\nu} &=& g_{\mu\nu}, \\
G_{\mu(\nu\lambda)} &=& G_{(\nu\lambda)\mu}=\phi_{\mu\nu\lambda} - \frac{1}{3} \, g_{\nu\lambda} \, \phi_{\mu}, \\
G_{(\mu\nu)(\lambda\rho)} &=& M_{(\mu\nu)(\lambda\rho)}+(\phi_{\mu\nu\sigma}-\frac{1}{3} \, g_{\mu\nu} \, \phi_{\sigma}) \, g^{\sigma\kappa} \, (\phi_{\kappa\lambda\rho}-\frac{1}{3} \, g_{\lambda\rho} \, \phi_{\kappa}) \nonumber \\
&=& \frac{1}{4} \, g_{\mu\rho} \, g_{\nu\lambda} + 
\frac{1}{4} \, g_{\mu\lambda} \, g_{\nu\rho} -  
\frac{1}{6} \, g_{\mu\nu} \, g_{\lambda\rho}+{\cal O}(\phi^2), \\
G^{\mu\nu} &=& g^{\mu\nu} + 2 \, \phi^{\mu\lambda\rho} \, \phi^{\nu}{}_{\lambda\rho} -  \frac{2}{3} \,
\phi^{\mu} \,\phi^{\nu}+{\cal O}(\phi^4), \\
G^{\mu(\nu\lambda)} &=&G^{(\nu\lambda)\mu}=-4 \phi^{\mu\nu\lambda} + \frac{4}{3} \
g^{\nu\lambda} \phi^{\mu}+{\cal O}(\phi^3), \\
G^{(\mu\nu)(\lambda\rho)} &=& J^{(\mu\nu)(\lambda\rho)}=4 \, g^{\mu\rho} \, g^{\nu\lambda} + 4 \,
g^{\mu\lambda} \, g^{\nu\rho} -  \frac{8}{3} \, g^{\mu\nu} \, g^{\lambda\rho}+{\cal O}(\phi^2)
\end{eqnarray}

\section{Metric-like Tensor ${K^M}_{NK}={d^a}_{bc} \, E^M_a \, e^b_N \, e^c_K$}
\hspace{5mm}
\begin{eqnarray}
K^\mu{}_{\nu\lambda} = \frac{2}{3} g_{\nu\lambda} \ 
\phi^{\mu} + {\cal O} (\phi^3),
\end{eqnarray}
\begin{eqnarray}
K^{(\mu\nu)}{}_{\lambda\rho} &=& 4 \delta^{(\mu}_{\rho} \delta^{\nu)}_{\lambda}  -  \
\frac{4}{3} g^{\mu\nu} g_{\lambda\rho}
- \frac{8}{9} g_{\lambda\rho} \
\phi^{\mu\sigma\kappa} \
\phi^{\nu}{}_{\sigma\kappa} -  \frac{4}{9} \
g_{\lambda\rho} \phi^{\mu} \
\phi^{\nu}										\nonumber \\
&~& -  \frac{8}{9} \
g_{\lambda\rho} \phi^{\mu\nu\sigma} \
\phi_{\sigma} + \frac{8}{27} \
g^{\mu\nu} g_{\lambda\rho} \
\phi_{\sigma\kappa\tau} \phi^{\sigma\kappa\tau} \
+ \frac{4}{9} g^{\mu\nu} g_{\lambda\rho} \
\phi^{\kappa} \
\phi_{\kappa} + {\cal O} (\phi^4),
\end{eqnarray}
\begin{eqnarray}
K^\mu{}_{(\nu\lambda)\rho} &=& - \frac{1}{3} \delta^{\mu}_{\rho} g_{\nu\lambda} \
+ \delta^{\mu}_{(\nu} g_{\lambda)\rho} 		
- \frac{2}{3} \phi^{\mu}{}_{\rho}{}^{\sigma} \
\phi_{\nu\lambda\sigma} + \frac{2}{3} \  
\phi^{\mu}{}_{(\nu}{}^{\sigma} \
\phi_{\lambda)\rho\sigma} 
 + \frac{2}{3} \
g_{\rho(\nu}  \phi_{\lambda)\sigma\kappa} \ \phi^{\mu\sigma\kappa} 
  \nonumber \\ &&
+ \frac{2}{3} \
\phi^{\mu}{}_{\rho(\nu} \
\phi_{\lambda)} 
 -  \frac{1}{3} \
g_{\rho(\nu} \phi^{\mu} \
\phi_{\lambda)} - \frac{1}{3} \
\delta^{\mu}{}_{\rho} \
\phi_{\nu} \
\phi_{\lambda}							
 -  \frac{4}{9} \
g_{\nu\lambda} \phi^{\mu\sigma\kappa} \
\phi_{\rho\sigma\kappa}
-  \frac{2}{3} \
\phi^{\mu}{}_{\nu\lambda} \
\phi_{\rho} 
\nonumber \\ &&
+ \frac{2}{9} \
g_{\nu\lambda} \phi^{\mu} \
\phi_{\rho}  + \frac{1}{3} \
\delta^{\mu}_{(\nu} \
\phi_{\lambda)} \
\phi_{\rho}							
+ \frac{2}{9} \
g_{\nu\lambda} \phi^{\mu}{}_{\rho}{}^{\sigma} \
\phi_{\sigma} + \frac{2}{3} \
\delta^{\mu}{}_{\rho} \
\phi_{\nu\lambda}{}^{\sigma} \
\phi_{\sigma}-  \frac{2}{3} \
\delta^{\mu}_{(\nu} \
\phi_{\lambda)\rho}{}^{\sigma} \
\phi_{\sigma}							 \nonumber \\
&& + \frac{1}{27} \
\delta^{\mu}_{\rho} g_{\nu\lambda} \
\phi_{\sigma\kappa\tau} \phi^{\sigma\kappa\tau} \
-  \frac{1}{9} \
\delta^{\mu}_{(\nu} g_{\lambda)\rho} \
\phi_{\sigma\kappa\tau} \phi^{\sigma\kappa\tau} \
-  \frac{1}{9} \delta^{\mu}_{\rho} g_{\nu\lambda} \
\phi^{\kappa} \
\phi_{\kappa} + {\cal O} (\phi^4),
\end{eqnarray}
\begin{eqnarray}
 K^{(\mu\nu)}{}_{(\lambda\rho)\sigma}& =& - \frac{4}{3} g_{\sigma(\lambda} \
\phi^{\mu\nu}{}_{\rho)} + \
\frac{4}{3} g_{\lambda\rho} \
\phi^{\mu\nu}{}_{\sigma} + \frac{8}{3} \
\delta^{(\mu}_{\sigma} \
\phi^{\nu)}{}_{\lambda\rho} -  \frac{8}{3} \
\delta^{(\mu}_{(\lambda} \
\phi^{\nu)}{}_{\rho)\sigma}						
-  \frac{4}{3} \
\delta^{(\mu}_{(\lambda} g_{\rho)\sigma} \
\phi^{\nu)} \nonumber \\ &&-  \frac{4}{3} \
\delta^{(\mu}_{\sigma} \delta^{\nu)}_{(\lambda} \
\phi_{\rho)} + \frac{4}{3} \
g^{\mu\nu} g_{\sigma(\lambda} \
\phi_{\rho)} + \frac{4}{3} \
\delta^{\mu}_{(\lambda} \delta^{\nu}_{\rho)} \
\phi_{\sigma}  -  \frac{8}{9} \
g^{\mu\nu} g_{\lambda\rho} \
\phi_{\sigma} \nonumber \\ &&+ {\cal O} (\phi^3),
\end{eqnarray}
\begin{eqnarray}
 K^\mu{}_{(\nu\lambda)(\rho\sigma)} &= &- \frac{2}{3} g_{\rho\sigma} \
\phi^{\mu}{}_{\nu\lambda} + \frac{2}{3} \
g_{\lambda(\rho} \phi^{\mu}{}_{\sigma)\nu} + \frac{2}{3} \
g_{\nu(\rho} \phi^{\mu}{}_{\sigma)\lambda} -  \frac{2}{3} \
g_{\nu\lambda} \phi^{\mu}{}_{\rho\sigma} -  \
\frac{1}{3} g_{\nu(\rho} g_{\sigma)\lambda} \
\phi^{\mu}  
 \nonumber \\
&& 
+ \frac{1}{3} \
g_{\nu\lambda} g_{\rho\sigma} \
\phi^{\mu}								+ \frac{2}{3} \
\delta^{\mu}{}_{(\rho} \phi_{\sigma)\nu\lambda}  + \frac{2}{3} \
\delta^{\mu}_{(\nu} \phi_{\lambda)\rho\sigma} -  \frac{1}{3} \
\delta^{\mu}_{(\rho} g_{\sigma)(\nu} \
\phi_{\lambda)}  -  \frac{1}{3} \
\delta^{\mu}_{(\nu} g_{\lambda)(\rho} \phi_{\sigma)} \nonumber \\ &&+ {\cal O} (\phi^3), \\
 K^{(\mu\nu)}{}_{(\lambda\rho)(\sigma\kappa)} &= &\frac{4}{3} \delta^{\mu}_{(\kappa} \
\delta^{\nu}_{\sigma)} g_{\lambda\rho}  + \
\frac{4}{3} \delta^{\mu}_{(\lambda} \
\delta^{\nu}_{\rho)} g_{\sigma\kappa} -  \delta^{\mu}_{(\kappa} \
\delta^{\nu}_{\rho)} g_{\lambda\sigma} -  \delta^{\mu}_{(\sigma} \
\delta^{\nu}_{\rho)} g_{\lambda\kappa}								  -  \delta^{\mu}_{(\kappa} \
\delta^{\nu}_{\lambda)} g_{\rho\sigma}  \nonumber \\
&&- \
\delta^{\mu}_{(\sigma} \delta^{\nu}_{\lambda)} \
g_{\rho\kappa} + \frac{4}{3} g^{\mu\nu} \
g_{\lambda(\kappa} g_{\sigma)\rho} -  \frac{8}{9} g^{\mu\nu} \
g_{\lambda\rho} g_{\sigma\kappa} + {\cal O} (\phi^2).
\end{eqnarray}

\section{Generalized Curvature ${R^M}_{N\mu\nu}$}
\hspace{5mm}
\begin{eqnarray}
{R^{(\mu\nu)}}_{(\lambda\rho)\sigma\kappa} &=& 4 \delta^{(\mu}_{(\lambda} {\hat{R}^{\nu)}}_{\ \ \rho)\sigma\kappa} + 2 g^{\mu\nu} \Bigl( \Bigr. \frac{2}{3} g_{\tau(\lambda|} \hat{\Gamma}^\tau_{\kappa\eta} \hat{\Gamma}^\eta_{|\rho)\sigma} - \frac{2}{3} g_{\tau(\lambda|} \hat{\Gamma}^\tau_{\sigma\eta}\hat{\Gamma}^\eta_{|\rho)\kappa} \nonumber \\
&~& + \frac{4}{9} g_{\tau(\lambda} \hat{\Gamma}^\tau_{\rho)\kappa} \hat{\Gamma}^\eta_{\sigma \eta} - \frac{4}{9} g_{\tau(\lambda} \hat{\Gamma}^\tau_{\rho)\sigma} \hat{\Gamma}^\eta_{\kappa\eta} \Bigl. \Bigr) + {\cal O} (\phi^2), 
\end{eqnarray}
In the above component, the ${\cal O}(\phi^2)$ terms are not displayed. 
\begin{eqnarray}
R^\mu{}_{(\nu\lambda)\rho\sigma} &=& -\frac{1}{3} \hat{\nabla}_\rho \hat{\nabla}^\mu \phi_{\nu\lambda\sigma} - \frac{1}{3} g_{\sigma(\nu|} \hat{\nabla}_{\rho} \hat{\nabla}^\mu \phi_{|\lambda)} + \frac{1}{3} \hat{\nabla}_\rho \hat{\nabla}_{(\nu} \phi^\mu{}_{\lambda)\sigma} - \frac{1}{3} g_{\sigma(\nu|} \hat{\nabla}_{\rho} \hat{\nabla}_{|\lambda)} \phi^\mu \nonumber \\
&~& + \frac{2}{3} \delta^\mu_\sigma \hat{\nabla}_\rho \hat{\nabla}_{(\nu} \phi_{\lambda)} - \frac{1}{3} \delta^\mu_{(\nu|} \hat{\nabla}_{\rho} \hat{\nabla}_{|\lambda)} \phi_\sigma + \frac{2}{3} \hat{\nabla}_\rho \hat{\nabla}_\sigma \phi^\mu{}_{\nu\lambda} - \frac{1}{3} \delta^\mu_{(\nu|} \hat{\nabla}_{\rho} \hat{\nabla}_{\sigma} \phi_{|\lambda)}  \nonumber \\ 
&~& + \frac{1}{3} g_{\sigma(\nu|} \hat{\nabla}_\rho \hat{\nabla}_\kappa \phi_{|\lambda)}{}^{\kappa\mu} -\frac{1}{3}\delta^\mu_\sigma \hat{\nabla}_\rho \hat{\nabla}_\kappa \phi^\kappa{}_{\nu\lambda} + \frac{1}{3} \delta^\mu_{(\nu|} \hat{\nabla}_\rho \hat{\nabla}_\kappa \phi_{|\lambda)\sigma}{}^\kappa \nonumber \\
&~& +\frac{1}{6} \delta^\mu_{(\nu} g_{\lambda)\sigma} \hat{\nabla}_\rho \hat{\nabla}_\kappa \phi^\kappa + \frac{1}{3} \hat{\nabla}_\sigma \hat{\nabla}^\mu \phi_{\nu\lambda\rho} +\frac{1}{3} g_{\rho(\nu|} \hat{\nabla}_\sigma \hat{\nabla}^\mu \phi_{|\lambda)} - \frac{1}{3} \hat{\nabla}_\sigma \hat{\nabla}_{(\nu} \phi^\mu{}_{\lambda)\rho}  \nonumber \\
&~& + \frac{1}{3} g_{\rho(\nu|} \hat{\nabla}_\sigma \hat{\nabla}_{|\lambda)} \phi^\mu -\frac{2}{3} \delta^\mu_\rho \hat{\nabla}_\sigma \hat{\nabla}_{(\nu} \phi_{\lambda)} +\frac{1}{3} \delta^\mu_{(\nu|} \hat{\nabla}_\sigma \hat{\nabla}_{|\lambda)} \phi_\rho -\frac{2}{3}\hat{\nabla}_\sigma \hat{\nabla}_\rho \phi^\mu{}_{\nu\lambda} \nonumber \\
 &~& + \frac{1}{3} \delta^\mu_{(\nu|} \hat{\nabla}_\sigma \hat{\nabla}_\rho \phi_{|\lambda)} -\frac{1}{3} g_{\rho(\nu|}\hat{\nabla}_\sigma \hat{\nabla}_\kappa \phi^{\mu\kappa}{}_{|\lambda)} + \frac{1}{3} \delta^\mu_\rho \hat{\nabla}_\sigma \hat{\nabla}^\kappa \phi_{\kappa\nu\lambda} - \frac{1}{3} \delta^\mu_{(\nu|} \hat{\nabla}_\sigma \hat{\nabla}^\kappa \phi_{|\lambda) \kappa\rho} \nonumber \\
 &~& -\frac{1}{6} \delta^\mu_{(\nu} g_{\lambda)\rho} \hat{\nabla}_\sigma \hat{\nabla}^\kappa \phi_\kappa + g_{\nu\lambda} \Bigl( \Bigr. \frac{1}{3} \hat{\nabla}_\rho \hat{\nabla}^\mu \phi_\sigma - \frac{1}{3} \hat{\nabla}_\rho \hat{\nabla}_\kappa \phi^{\kappa\mu}{}_\sigma - \frac{1}{6} \delta^\mu_\sigma \hat{\nabla}_\rho \hat{\nabla}^\kappa \phi_\kappa \nonumber \\
 &~&  - \frac{1}{3} \hat{\nabla}_\sigma \hat{\nabla}^\mu \phi_\rho + \frac{1}{3}\hat{\nabla}_\sigma \hat{\nabla}_\kappa \phi^{\kappa\mu}{}_\rho + \frac{1}{6} \delta^\mu_\rho \hat{\nabla}_\sigma \hat{\nabla}^\kappa \phi_\kappa \Bigl. \Bigr) + {\cal O} (\phi^3),
\end{eqnarray}

\begin{eqnarray}
R^{(\mu\nu)}{}_{\lambda\rho\sigma} &=& -\frac{4}{3} \hat{\nabla}_\rho \hat{\nabla}^{(\mu} \phi^{\nu)}{}_{\lambda\sigma} - \frac{8}{3} g_{\lambda\sigma} \hat{\nabla}_\rho \hat{\nabla}^{(\mu} \phi^{\nu)} + \frac{4}{3} \delta^{(\mu}_\sigma \hat{\nabla}_\rho \hat{\nabla}^{\nu)} \phi_\lambda + \frac{4}{3} \delta^{(\mu}_\lambda \hat{\nabla}_\rho \hat{\nabla}^{\nu)} \phi_\sigma \nonumber \\
&~& + \frac{4}{3} \hat{\nabla}_\rho \hat{\nabla}_\lambda \phi^{\mu\nu}{}_{\sigma} +\frac{4}{3} \delta^{(\mu}_\sigma \hat{\nabla}_\rho \hat{\nabla}_\lambda \phi^{\nu)} + \frac{4}{3} \hat{\nabla}_\rho \hat{\nabla}_\sigma \phi^{\mu\nu}{}_{\lambda} + \frac{4}{3} \delta^{(\mu}_{\lambda} \hat{\nabla}_\rho \hat{\nabla}_\sigma \phi^{\nu)} \nonumber \\
&~& + \frac{4}{3} g_{\lambda\sigma} \hat{\nabla}_\rho \hat{\nabla}_\kappa \phi^{\kappa\mu\nu} - \frac{4}{3} \delta^{(\mu}_\sigma \hat{\nabla}_\rho \hat{\nabla}_\kappa \phi^{\nu)\kappa}{}_{\lambda} -\frac{4}{3} \delta^{(\mu}_\lambda \hat{\nabla}_\rho \hat{\nabla}_\kappa \phi^{\nu)\kappa}{}_{\sigma} -\frac{2}{3} \delta^{(\mu}_\sigma \delta^{\nu)}_\lambda \hat{\nabla}_\rho \hat{\nabla}^\kappa \phi_\kappa \nonumber \\
&~& + g^{\mu\nu} \Bigl( \Bigr. -\frac{4}{3}\hat{\nabla}_\rho\hat{\nabla}_\lambda\phi_\sigma - \frac{4}{3} \hat{\nabla}_\rho \hat{\nabla}_\sigma \phi_\lambda +\frac{4}{3} \hat{\nabla}_\rho \hat{\nabla}^\kappa \phi_{\kappa\lambda\sigma} + \frac{2}{3} g_{\lambda\sigma} \hat{\nabla}_\rho \hat{\nabla}^\kappa \phi_\kappa \nonumber \\
&~& + \frac{8}{9} \hat{\Gamma}^\kappa_{\sigma\tau} \hat{\nabla}_\lambda \phi_{\rho\kappa}{}^\tau + \frac{8}{9} \hat{\Gamma}^{\kappa}_{\rho\kappa}\hat{\nabla}_\lambda \phi_\sigma + \frac{4}{9} \hat{\Gamma}^\kappa_{\sigma\tau} g_{\kappa\rho} \hat{\nabla}_\lambda \phi^\tau +\frac{8}{9} \hat{\Gamma}^\kappa_{\sigma\tau} \hat{\nabla}_\rho \phi_{\lambda\kappa}{}^\tau \nonumber \\
&~& -\frac{8}{9} \hat{\Gamma}^\kappa_{\sigma\kappa} \hat{\nabla}_\rho \phi_\lambda + \frac{4}{9} \hat{\Gamma}^\kappa_{\sigma\tau} g_{\kappa\lambda} \hat{\nabla}_\rho \phi^\tau + \frac{4}{9} \hat{\Gamma}^\kappa_{\lambda\sigma} \hat{\nabla}_\rho \phi_\kappa - \frac{4}{9} \hat{\Gamma}^\kappa_{\sigma\tau} \hat{\nabla}_\kappa \phi^\tau{}_{\lambda\rho} \nonumber \\
&~& -\frac{4}{9} \hat{\Gamma}^\kappa_{\rho\tau} g_{\kappa\sigma} \hat{\nabla}^\tau \phi_\lambda - \frac{4}{9}  \hat{\Gamma}^\kappa_{\rho\tau} g_{\kappa\lambda} \hat{\nabla}^\tau \phi_\sigma-\frac{4}{9} \hat{\Gamma}^\kappa_{\lambda\rho} \hat{\nabla}_\kappa \phi_\sigma - \frac{8}{9} \hat{\Gamma}^\kappa_{\sigma\tau} g_{\lambda\rho} \hat{\nabla}_\kappa\phi^\tau \nonumber \\
&~& -\frac{4}{9} \hat{\Gamma}^\kappa_{\sigma\tau} \hat{\nabla}^\tau \phi_{\kappa\lambda\rho} - \frac{8}{9} \hat{\Gamma}^\kappa_{\rho\kappa} \hat{\nabla}^\tau \phi_{\tau\lambda\sigma} - \frac{4}{9} \hat{\Gamma}^\kappa_{\sigma\tau} g_{\kappa\rho} \hat{\nabla}^\eta \phi_{\eta\lambda}{}^\tau - \frac{4}{9} \hat{\Gamma}^\kappa_{\sigma\tau} g_{\kappa\lambda} \hat{\nabla}^\eta \phi_{\eta \rho}{}^{\tau} \nonumber \\
&~& +\frac{4}{9} \hat{\Gamma}^\kappa_{\lambda\rho} \hat{\nabla}^\tau \phi_{\kappa\tau\sigma} + \frac{8}{9} \hat{\Gamma}^\kappa_{\rho\tau} g_{\lambda\sigma} \hat{\nabla}^\tau \phi_\kappa + \frac{2}{9} \hat{\Gamma}^\kappa_{\lambda\rho} g_{\kappa\sigma} \hat{\nabla}^\tau \phi_\tau + \frac{8}{9} \hat{\Gamma}^\kappa_{\sigma\tau}g_{\lambda\rho} \hat{\nabla}^\eta \phi_{\eta\kappa}{}^\tau \nonumber \\ 
&~& + \frac{4}{9} \hat{\Gamma}^\kappa_{\sigma\kappa} g_{\lambda\rho} \hat{\nabla}^\tau \phi_\tau \Bigl. \Bigr) - \left( \rho \leftrightarrow \sigma \ \ \text{interchanged} \right)  + {\cal O} (\phi^3).
\end{eqnarray}
The remaining component $R^{(\mu\nu)}{}_{(\lambda\rho)\sigma\kappa}$ is too complicated and not presented.

\section{Metric-Like Tensor ${F_{\mu M}}^{N}={f^a}_{bc} \, e^c_{\mu} \, e^b_M \, E^N_a$}
\hspace{5mm}
In the following equations 
\begin{equation}
\varepsilon^{\mu\nu\rho} =\frac{1}{\sqrt{-g}} \, \epsilon^{\mu\nu\rho},  
\end{equation}
where $\epsilon^{rt\phi}=+1 $. The indices are raised and lowered in terms of $g_{\mu\nu}$ and $g^{\mu\nu}$. 
\begin{eqnarray}
F_{\mu\nu}{}^\lambda &=& \varepsilon_{\mu\nu}{}^{\lambda} + \frac{2}{3} \,  \varepsilon^{\lambda\rho\sigma} \
\phi_{\mu\rho}{}^{\kappa} \
\phi_{\nu\sigma\kappa} + \frac{4}{3} \,
\varepsilon_{\mu\nu}{}^{\rho} \
\phi^{\lambda\sigma\kappa} \phi_{\rho\sigma\kappa} \
-  \frac{4}{3} \, \varepsilon_{\mu\nu}{}^{\rho} \
\phi^{\lambda}{}_{\rho}{}^{\sigma} \
\phi_{\sigma} \nonumber \\
&& + \frac{1}{6} \,
\varepsilon_{\mu\nu}{}^{\lambda} \
\phi^{\sigma} \
\phi_{\sigma}+ {\cal O} (\phi^4), \\
F_{\mu(\nu\lambda)}{}^\rho &=& \varepsilon_{\lambda}{}^{\rho\sigma} \
\phi_{\mu\nu\sigma} + \
\varepsilon_{\nu}{}^{\rho\sigma} \
\phi_{\mu\lambda\sigma} - 2 \
\varepsilon_{\mu}{}^{\rho\sigma} \
\phi_{\nu\lambda\sigma} -  \frac{1}{2} \,
\delta_{\lambda}{}^{\rho} \
\varepsilon_{\mu\nu}{}^{\sigma} \
\phi_{\sigma} -  \frac{1}{2} \,
\delta_{\nu}{}^{\rho} \
\varepsilon_{\mu\lambda}{}^{\sigma} \
\phi_{\sigma} \nonumber \\
&~&- \frac{1}{2} \,
\varepsilon_{\lambda}{}^{\rho\sigma} g_{\mu\nu} \
\phi_{\sigma}-  \frac{1}{2} \,
\varepsilon_{\nu}{}^{\rho\sigma} g_{\mu\lambda} \
\phi_{\sigma} + \frac{4}{3} \,
\varepsilon_{\mu}{}^{\rho\sigma} g_{\nu\lambda} \
\phi_{\sigma}+ {\cal O} (\phi^3), \\
F_{\mu\nu}{}^{(\lambda\rho)} &= &2 \varepsilon_{\nu}{}^{\rho\sigma} \
\phi_{\mu}{}^{\lambda}{}_{\sigma} + 2 \
\varepsilon_{\nu}{}^{\lambda\sigma} \
\phi_{\mu}{}^{\rho}{}_{\sigma} - 2 \
\varepsilon_{\mu}{}^{\rho\sigma} \
\phi_{\nu}{}^{\lambda}{}_{\sigma} - 2 \
\varepsilon_{\mu}{}^{\lambda\sigma} \
\phi_{\nu}{}^{\rho}{}_{\sigma}+ {\cal O} (\phi^3), \\
F_{\mu(\nu\lambda)}{}^{(\rho\sigma)} &=& \delta_{\lambda}{}^{\sigma} \
\varepsilon_{\mu\nu}{}^{\rho} + \
\delta_{\lambda}{}^{\rho} \
\varepsilon_{\mu\nu}{}^{\sigma} + \
\delta_{\nu}{}^{\sigma} \
\varepsilon_{\mu\lambda}{}^{\rho} + \
\delta_{\nu}{}^{\rho} \
\varepsilon_{\mu\lambda}{}^{\sigma} + {\cal O} (\phi^2).
\end{eqnarray}
\begin{eqnarray}
F_{(\mu\nu)(\lambda\rho)}{}^\sigma = 
\frac{1}{4} \varepsilon_{\nu\rho}{}^{\sigma} \
g_{\mu\lambda} + \frac{1}{4} \
\varepsilon_{\nu\lambda}{}^{\sigma} g_{\mu\rho} + \
\frac{1}{4} \varepsilon_{\mu\rho}{}^{\sigma} \
g_{\nu\lambda} + \frac{1}{4} \
\varepsilon_{\mu\lambda}{}^{\sigma} g_{\nu\rho}+{\cal O}(\phi^2)
\end{eqnarray}
\begin{eqnarray}
F_{(\mu\nu)(\lambda\rho)}{}^{(\kappa\sigma)} 
&=&\ \frac{4}{3} \delta_{(\lambda|}^{(\kappa} \varepsilon_{\mu}{}^{\sigma)\tau} \phi_{\nu|\rho)\tau}
 + \frac{4}{3} \delta_{(\lambda|}^{(\kappa} \varepsilon_{\nu}{}^{\sigma)\tau} \phi_{\mu|\rho)\tau}
-  \frac{4}{3} \delta_{\mu}^{(\kappa} \varepsilon_{(\lambda|}{}^{\sigma)\tau} \phi_{\nu|\rho)\tau}  \nonumber \\
&~& 
-  \ \frac{4}{3} \delta_{\nu}^{(\kappa} \varepsilon_{(\lambda|}{}^{\sigma)\tau} \phi_{\mu|\rho)\tau}							
- \frac{8}{3} \delta_{(\lambda}^{(\kappa} \varepsilon_{\rho)\tau(\mu} \phi_{\nu)}{}^{\sigma)\tau} 
 -  \frac{8}{3} \varepsilon_{(\lambda}{}^{(\kappa|\tau} g_{\rho)(\mu} \ \phi_{\nu)}{}^{|\sigma)}{}_{\tau}\nonumber \\
&~& 
+  \frac{8}{3} \delta_{(\mu}^{(\kappa} \varepsilon_{\nu)\tau(\lambda} \ \phi_{\rho)}{}^{\sigma)\tau} 
 + \frac{8}{3} \varepsilon_{(\mu}{}^{\tau(\kappa} g_{\nu)(\lambda} \ \phi_{\rho)}{}^{\sigma)}{}_{\tau}  				
 - 2 \ \varepsilon_{\mu(\lambda}{}^{\tau} g_{\rho)\nu} \phi^{\kappa\sigma}{}_{\tau}  \nonumber \\
&~&
- 2 \varepsilon_{\nu(\lambda}{}^{\tau} g_{\rho)\mu} \phi^{\kappa\sigma}{}_{\tau}
- 2 \ \delta_{\mu}^{(\kappa} \delta_{(\lambda}^{\sigma)} \varepsilon_{\rho)\nu}{}^{\tau} \  \phi_{\tau}
- 2 \ \delta_{\nu}^{(\kappa} \delta_{(\lambda}^{\sigma)} \varepsilon_{\rho)\mu}{}^{\tau} \ \phi_{\tau} 		\nonumber \\
&~&
 - 2 \ \delta_{(\lambda|}^{(\kappa} \varepsilon_{\mu}{}^{\sigma)\tau} g_{\nu|\rho)} \phi_{\tau} 
-  2 \ \delta_{(\lambda|}^{(\kappa} \varepsilon_{\nu}{}^{\sigma)\tau} g_{\mu|\rho)} \ \phi_{\tau} 
+  2 \delta_{\mu}^{(\kappa} \ \varepsilon_{(\lambda}{}^{\sigma)\tau} g_{\rho)\nu} \phi_{\tau} \nonumber \\ &&
+2  \ \delta_{\nu}^{(\kappa} \varepsilon_{(\lambda}{}^{\sigma)\tau} g_{\rho)\mu} \phi_{\tau} 		
 + \frac{4}{3} \ \delta_{(\lambda}^{(\kappa} \varepsilon_{\rho)}{}^{\sigma)\tau} g_{\mu\nu} \phi_{\tau}
 -  \ \frac{4}{3} \delta_{(\mu}^{(\kappa} \varepsilon_{\nu)}{}^{\sigma)\tau} g_{\lambda\rho} \ \phi_{\tau}   \nonumber \\ &&
+ \frac{4}{3} \ \varepsilon_{\mu(\lambda}{}^{\tau} g_{\rho)\nu} g^{\kappa\sigma} \phi_{\tau}
+ \frac{4}{3} \ \varepsilon_{\nu(\lambda}{}^{\tau} g_{\rho)\mu} g^{\kappa\sigma} \phi_{\tau}+{\cal O}(\phi^3)
\end{eqnarray}

\section{Matter action up to ${\cal O}(\phi^1)$}
\hspace{5mm}
The matter action is composed of the torsion part $S_{\text{matter}}^{(\text{torsion})}$, which come from $\Delta \Gamma^N_{\mu M}$, and the 
other part, which are expanded according to the power of $\phi$. 
\begin{equation}
S_{\text{ matter}} = S_{\text{matter}}^{(0)}+S_{\text{matter}}^{(1)}+\cdots+S_{\text{matter}}^{(\text{torsion})}, \label{Mactionexp}
\end{equation}
Here only the part which do not depend on the torsion is presented up to ${\cal O}(\phi^1)$. 
\begin{eqnarray}
S_{\text{matter}}^{(0)} &=& \int d^3 x\, \sqrt{-g} \, \left[ {B_{\mu}}^{\lambda} (\hat{\nabla}_{\lambda} \, 
C^{\mu}+\frac{1}{2\ell} \, g_{\lambda\rho} \, C^{(\mu\rho)}) +\frac{1}{2} \, {B_{(\mu\rho)}}^{\lambda} \, ( \hat{\nabla}_{\lambda} \, C^{(\mu\rho)}+\frac{4}{\ell} \, {\delta^{\rho}}_{\lambda}C^{\mu})\right. \nonumber \\ && \left. +\frac{2}{\ell} \, {B_{\lambda}}^{\lambda} \, C^0
+{B_0}^{\lambda} \, (\partial_{\lambda} \, C^0+\frac{4}{3\ell} \, g_{\lambda\nu} \, C^{\nu}) \right],
\end{eqnarray}
\begin{eqnarray}
S^{(1)}_{\mathrm{matter}} &=& \int d^3x \, \sqrt{-g} \, 
\frac{2}{3\ell} \, B_0{}^\mu \, C^{(\nu\lambda)}  \, \phi_{\mu\nu\lambda} 
+ \int d^3x \sqrt{-g} \, \frac{2}{3\ell} \,  B_\mu{}^{\lambda} \,  g_{\lambda\rho} \,  C^\rho \, \phi^\mu
\nonumber \\
&~& + \int d^3x \sqrt{-g} \,
B_\mu{}^\lambda C^{(\rho\sigma)} \Bigl[ \Bigr. -\frac{1}{6} \hat{\nabla}^\mu \phi_{\lambda\rho\sigma} - \frac{1}{6} g_{\lambda\rho} \hat{\nabla}^\mu \phi_\sigma + \frac{1}{3}\hat{\nabla}_\lambda \phi^\mu{}_{\rho\sigma} \nonumber \\
&~& - \delta^\mu_\rho \hat{\nabla}_\lambda \phi_\sigma - \frac{1}{6} g_{\lambda\rho}\hat{\nabla}_\sigma\phi^\mu - \frac{1}{6} \delta^\mu_\rho \hat{\nabla}_\sigma \phi_\lambda + \frac{1}{3} \delta^\mu_\lambda \hat{\nabla}_\rho \phi_\sigma + \frac{1}{6} \hat{\nabla}_\rho \phi^\mu{}_{\lambda\sigma} \nonumber \\
&~& + \frac{1}{6} g_{\lambda\rho} \hat{\nabla}^\kappa \phi^\mu{}_{\sigma\kappa} + \frac{1}{6} \delta^\mu_\rho \hat{\nabla}^\kappa \phi_{\lambda\sigma\kappa} - \frac{1}{6} \delta^\mu_\lambda \hat{\nabla}^\kappa \phi_{\kappa\rho\sigma} + \frac{1}{12}\delta^\mu_\rho g_{\lambda\sigma} \hat{\nabla}^\kappa \phi_\kappa \Bigl. \Bigr] \nonumber \\
&~& + \int d^3x \sqrt{-g} \,
B_{(\mu\nu)}{}^\lambda C^{(\rho\sigma)} \Bigl[ \Bigr. -\frac{1}{3\ell} g_{\lambda\rho}\phi^{\mu\nu}{}_\sigma -\frac{2}{3\ell} \delta^\mu_\rho \phi^\nu{}_{\lambda\sigma} + \frac{2}{3\ell} \delta^\mu_\lambda \phi^\nu{}_{\rho\sigma} \nonumber \\
&~& -\frac{1}{3\ell} \delta^\mu_\rho g_{\lambda\sigma} \phi^\nu + \frac{1}{3\ell}\delta^\mu_\rho \delta^\nu_\sigma \phi_\lambda - \frac{1}{3\ell} \delta^\mu_\lambda \delta^\nu_\rho \phi_\sigma \Bigl. \Bigr] \nonumber \\
&~& + \int d^3x \sqrt{-g} \,
B_{(\mu\nu)}{}^\lambda C^\rho \Bigl[ \Bigr. \frac{2}{3} \hat{\nabla}_\rho \phi^{\mu\nu}{}_\lambda + \frac{2}{3} \delta^\mu_\lambda \hat{\nabla}_\rho \phi^\nu + \frac{2}{3}\delta^\mu_\rho \hat{\nabla}^\nu\phi_\lambda \nonumber \\
&~& -\frac{2}{3} \hat{\nabla}^\mu \phi^\nu{}_{\rho\lambda} + \frac{2}{3} \delta^\mu_\lambda \hat{\nabla}^\nu \phi_\rho + \frac{2}{3} \delta^\mu_\rho \hat{\nabla}_\lambda \phi^\nu - \frac{4}{3} g_{\lambda \rho} \hat{\nabla}^\mu \phi^\nu + \frac{2}{3} \hat{\nabla}_\lambda \phi^{\mu\nu}{}_\rho \nonumber \\
 &~& - \frac{2}{3} \delta^\mu_\lambda \hat{\nabla}^\kappa {{\phi_{\kappa}}^{\nu}}_{\rho} -\frac{2}{3} \delta^\mu_\rho \hat{\nabla}^\kappa \phi^\nu{}_{\kappa\lambda} +\frac{2}{3} g_{\lambda\rho} \hat{\nabla}_\kappa \phi^{\kappa\mu\nu} - \frac{1}{3} \delta^\nu_\lambda \delta^\mu_\rho \hat{\nabla}^\kappa \phi_\kappa \Bigl. \Bigr].
\end{eqnarray}

\section{Spin-3 transformations of $C^{\mu}$ and $C^{(\mu\nu)}$}
\hspace{5mm}
Here only the parts of $\delta \, C^M$, (\ref{Cspin3-1}) and (\ref{Cspin3-2}),  which do not depend on the torsion are presented up to ${\cal O}(\phi^1)$. 
\begin{eqnarray}
\left[\delta \, C^{\mu}\right]_0 &=& -\frac{1}{2\ell} \, \xi^{(\mu\nu)} \, g_{\nu\rho} \, C^{\rho}-\frac{1}{4} \, g_{\lambda \nu} \,  C^{(\nu\rho)} \, \hat{\nabla}_{\rho} \, \xi^{(\mu\lambda)},
\end{eqnarray}
\begin{eqnarray}
\left[\delta \, C^{\mu}\right]_1 &=& \frac{1}{12\ell} \, C^{(\nu\lambda)} \, \phi_{\nu} \, \xi^{(\mu\rho)} \, g_{\rho\lambda}-\frac{1}{6\ell} \, C^{(\nu\lambda)} \, \phi_{\nu\lambda\rho} \, \xi^{(\mu\rho)}+\frac{1}{12\ell} \, C^{(\nu\lambda)} \, \phi^{\mu} \, \xi^{(\rho\sigma)} \, g_{\nu\rho} \, g_{\lambda\sigma}\nonumber \\ 
&&+\frac{1}{12\ell} \, C^{(\mu\nu)} \, \phi_{\lambda} \, \xi^{(\lambda\rho)} \, g_{\nu\rho}-\frac{1}{3\ell} \, C^{(\nu\lambda)} \, {\phi^{\mu}}_{\nu\sigma} \, \xi^{(\sigma\kappa)} \, g_{\kappa\lambda}-\frac{1}{6\ell} \, C^{(\mu\nu)} \, \phi_{\nu\lambda\rho} \, \xi^{(\lambda\rho)} \nonumber \\
&& +\frac{1}{6} \, C^{\nu} \, \xi^{(\lambda\rho)} \, \hat{\nabla}^{\mu} \, \phi_{\nu\lambda\rho}+\frac{1}{6} \, C^{\nu} \, g_{\nu\lambda} \, \xi^{(\lambda\rho)} \, \hat{\nabla}^{\mu} \, \phi_{\rho}-\frac{1}{3} \, C^{\nu} \, \xi^{(\lambda\rho)} \, \hat{\nabla}_{\nu} \, {\phi^{\mu}}_{\lambda\rho} \nonumber \\ &&
+\frac{1}{6} \, C^{\nu} \, \xi^{(\mu\lambda)} \, \hat{\nabla}_{\nu} \, \phi_{\lambda}+\frac{1}{6} \, C^{\nu} \, g_{\nu\lambda} \, \xi^{(\lambda\rho)} \, \hat{\nabla}_{\rho} \, \phi^{\mu}+\frac{1}{6} \, C^{\nu} \, \xi^{(\mu\lambda)} \, \hat{\nabla}_{\lambda} \, \phi_{\nu} \nonumber \\ &&
-\frac{1}{3} \, C^{\mu} \, \xi^{(\nu\lambda)} \, \hat{\nabla}_{\lambda} \, \phi_{\nu}-\frac{1}{6} \, C^{\nu} \, \xi^{(\lambda\rho)} \, \hat{\nabla}_{\rho} \, {{\phi^{\mu}}_{\nu\lambda}}-\frac{1}{6} \, C^{\nu} \, g_{\nu\lambda} \, \xi^{(\lambda\rho)} \, \hat{\nabla}_{\sigma} \, {{\phi^{\mu}}_{\rho}}^{\sigma} \nonumber \\ &&
-\frac{1}{6} \, C^{\nu}\, \xi^{(\mu\lambda)} \, \hat{\nabla}_{\rho} \, {\phi_{\nu\lambda}}^{\rho}+\frac{1}{6} \, C^{\mu} \, \xi^{(\nu\lambda)} \, \hat{\nabla}_{\rho} \, {\phi_{\nu\lambda}}^{\rho}-\frac{1}{12} \, C^{\nu} \, g_{\nu\lambda} \, \xi^{(\mu\lambda)} \, \hat{\nabla}_{\rho} \, \phi^{\rho},
\end{eqnarray}
\begin{eqnarray}
\left[\delta \, C^{(\mu\nu)}\right]_0 &=& -\frac{2}{\ell} \, C^0 \, \xi^{(\mu\nu)}+\frac{1}{2\ell} \, C^{(\mu\rho)} \, g_{\rho\lambda} \, \xi^{(\nu\lambda)}+\frac{1}{2\ell} \, C^{(\nu\rho)} \, g_{\rho\lambda} \, \xi^{(\mu\lambda)}  
\nonumber \\ &&
-\frac{1}{3\ell}\, g^{\mu\nu} \, C^{(\lambda\rho)} \, \xi^{(\sigma\kappa)} \, g_{\lambda\sigma} \, g_{\rho\kappa} -C^{\lambda} \, \hat{\nabla}_{\lambda} \, \xi^{(\mu\nu)},
\end{eqnarray}
\begin{eqnarray}
\left[\delta \, C^{(\mu\nu)}\right]_1 &=& -\frac{1}{3\ell}\, \xi^{(\mu\nu)} \, C^{\lambda} \, \phi_{\lambda}+\frac{1}{3\ell} \, \xi^{(\mu\lambda)} \, g_{\lambda\rho} \, C^{\rho} \, \phi^{\nu}+\frac{1}{3\ell} \, \xi^{(\mu\lambda)} \, C^{\nu} \, \phi_{\lambda} \nonumber \\
&& +\frac{2}{3\ell} \, \xi^{(\mu\lambda)} \, C^{\rho} \, {\phi^{\nu}}_{\lambda\rho}+\frac{1}{3\ell} \, {\phi^{\mu\nu}}_{\lambda} \, C^{\rho} \, g_{\rho\sigma} \, \xi^{(\lambda\sigma)}-\frac{1}{3\ell} \, g^{\mu\nu} \, C^{\lambda} \, \xi^{(\rho\sigma)} \, g_{\sigma\lambda} \, \phi_{\rho} \nonumber \\
&& -\frac{2}{3\ell} \, C^{\nu} \, {\phi^{\mu}}_{\lambda\rho} \, \xi^{(\lambda\rho)}+\frac{1}{6} \, C^{(\nu\lambda)} \, \xi^{(\rho\sigma)} \, \hat{\nabla}^{\mu} \, \phi_{\lambda\rho\sigma}+\frac{1}{6} \, C^{(\nu\lambda)} \, g_{\lambda\rho} \, \xi^{(\rho\sigma)} \, \hat{\nabla}^{\mu} \, \phi_{\sigma}\nonumber \\ &&
-\frac{1}{3} \, C^{(\nu\lambda)} \, \xi^{(\rho\sigma)} \, \hat{\nabla}_{\lambda} \, {\phi^{\mu}}_{\rho\sigma}+\frac{1}{6} \, C^{(\nu\lambda)} \, \xi^{(\mu\rho)} \, \hat{\nabla} \, \phi_{\rho}+\frac{1}{6} \, C^{(\nu\lambda)} \, \phi_{\rho} \hat{\nabla} \, \xi^{(\mu\rho)} \nonumber \\ &&
-\frac{1}{3} \, C^{(\nu\lambda)} \, {\phi^{\mu}}_{\rho\sigma} \, \hat{\nabla} \, \xi^{(\rho\sigma)}+\frac{1}{6} \, C^{(\nu\lambda)} \, g_{\lambda\rho} \, \xi^{(\rho\sigma)} \, \hat{\nabla}_{\sigma} \, \phi^{\mu} \nonumber \\ &&
+\frac{1}{18} \, g^{\mu\nu} \, C^{(\lambda\rho)} \, \xi^{(\sigma\kappa)} \, \hat{\nabla}_{\rho} \, \phi_{\lambda\sigma\kappa}-\frac{1}{9} \, g^{\mu\nu} 
C^{(\lambda\rho)} \, g_{\rho\sigma} \, \xi^{(\sigma\kappa)} \, \hat{\nabla}_{\lambda} \, \phi_{\kappa}+\frac{1}{6} \, C^{(\nu\lambda)} \, \xi^{(\mu\rho)} \, \hat{\nabla}_{\rho} \, \phi_{\lambda} \nonumber \\ &&
-\frac{1}{3} \, C^{(\mu\nu)} \, \xi^{(\lambda\rho)} \, \hat{\nabla}_{\rho} \phi_{\lambda}-\frac{1}{6} \, C^{(\lambda\rho)} \, \phi_{\lambda} \, \hat{\nabla}_{\rho} \, \xi^{(\mu\nu)}
+\frac{1}{6} \, C^{(\lambda\rho)} \, \phi^{\nu} \, \hat{\nabla}_{\rho} \, \xi^{(\mu\sigma)} \, g_{\sigma\lambda} \nonumber \\ &&
+\frac{1}{3} \, C^{(\lambda\rho)} \, {\phi^{\nu}}_{\lambda\sigma} \, \hat{\nabla}_{\rho} \, \xi^{(\mu\sigma)}+\frac{1}{6} \, C^{(\lambda\rho)} \, g^{\mu\nu} \, \phi_{\sigma} \, \hat{\nabla}_{\rho} \, \xi^{(\kappa\sigma)} \, g_{\lambda\kappa}\nonumber \\ &&
-\frac{1}{6} \, C^{(\lambda\rho)} \, g^{\mu\nu} \, \phi_{\sigma} \, \hat{\nabla} _{\rho} \, \xi^{(\kappa\sigma)} \, g_{\lambda\kappa}-\frac{1}{3} \, g^{\mu\nu} \, C^{(\lambda\rho)} \, \phi_{\lambda\sigma\kappa} \, \hat{\nabla}_{\rho} \, \xi^{(\sigma\kappa)} \nonumber \\ &&
-\frac{1}{6} \, C^{(\nu\lambda)} \, \xi^{(\rho\sigma)} \, \hat{\nabla}_{\sigma} \, {\phi^{\mu}}_{\lambda\rho}-\frac{1}{6} \, C^{(\nu\lambda)} \, g_{\lambda\rho} \, \xi^{\rho\sigma)} \, \hat{\nabla}^{\kappa} \, {\phi^{\mu}}_{\sigma\kappa}-\frac{1}{6} \, C^{(\nu\lambda)} \, \xi^{(\mu\rho)} \, \hat{\nabla)}^{\kappa} \, \phi_{\lambda\rho\kappa} \nonumber \\ &&
+\frac{1}{6} \, C^{(\mu\nu)} \, \xi^{(\lambda\rho)} \, \hat{\nabla}^{\sigma} \, \phi_{\lambda\rho\sigma}-\frac{1}{9} \, g^{\mu\nu} \, C^{(\lambda\rho)} \, g_{\lambda\sigma} \, \xi^{(\sigma\kappa)} \, \hat{\nabla}_{\kappa} \, \phi_{\rho}-\frac{1}{12} \, C^{(\nu\lambda)} \, g_{\lambda\rho} \, \hat{\nabla}_{\kappa} \, \phi^{\kappa} \nonumber \\ &&
+\frac{1}{18} \, g^{\mu\nu} \, C^{(\lambda\rho)} \, \xi^{(\sigma\kappa)} \, \hat{\nabla}_{\kappa} \, \phi_{\lambda\rho\sigma}+\frac{1}{9} \, g^{\mu\nu} \, C^{(\lambda\rho)} \, g_{\lambda\sigma} \, \xi^{(\sigma\kappa)} \, \hat{\nabla}_{\tau} \, {\phi_{\rho\kappa}}^{\tau}\nonumber \\ &&+\frac{1}{36} \, C^{(\lambda\rho)} \, g^{\mu\nu} \, \xi^{(\sigma\kappa)} \, g_{\lambda\sigma} \, g_{\rho\kappa} \, \hat{\nabla}_{\tau} \, \phi^{\tau} \qquad +(\mu \leftrightarrow \nu \ \text{interchanged}).
\end{eqnarray}
For $\delta \, C^0= [\delta \, C^0]_0 +[\delta \, C^0]_1   +{\cal O}(\phi^2) $, we have
\begin{eqnarray}
\, [\delta \, C^0]_0 &= &  -\frac{1}{6\ell} \, \xi^{(\mu\nu)} \, C^{(\lambda\rho)} \, g_{\mu\lambda} \, g_{\nu\rho}, \nonumber \\
\, [\delta \, C^0]_1 & = & - \frac{2}{3\ell} \, \xi^{(\mu\nu)} \,  C^{\lambda} \, \phi_{\mu\nu\lambda}.
\end{eqnarray}
 
\section{$[\Delta \, \Gamma^{( \tau\eta)}_{\mu\nu}]_1 $}
\hspace{5mm}
${\cal O}(\phi^1)$ corrections to $\Delta \Gamma^{(\tau\eta)}_{\mu\nu}$ in (\ref{delGamma0})  are presented in this appendix. 
Here the indices of $T_{\mu\nu,\lambda}$ are raised by $g^{\mu\nu}$. 
\begin{eqnarray}
[\Delta \, \Gamma^{( \tau\eta)}_{\mu\nu}]_1 
&=& \frac{1}{8} \,  J^{(\tau\eta)(\lambda\rho)}    \Bigl[\frac{1}{3} g_{\nu\rho} \phi_{\sigma} \, T_{\mu\lambda}{}^{,\sigma} + \frac{1}{3} \ g_{\nu\lambda} \phi_{\sigma} \, T_{\mu\rho}{}^{,\sigma} -  \phi_{\nu\rho\sigma} \
T_{\mu}{}^{\sigma}{}_{,\lambda} -  \phi_{\nu\lambda\sigma} T_{\mu}{}^{\sigma}{}_{,\rho}   \nonumber \\
&& + 2 g_{\lambda\rho} \phi_{\nu\sigma\kappa} \
T_{\mu}{}^{\sigma,\kappa} -  g_{\nu\rho} \phi_{\lambda\sigma\kappa} T_{\mu}{}^{\sigma, \kappa} -  g_{\nu\lambda} \phi_{\rho\sigma\kappa} \
T_{\mu}{}^{\sigma,\kappa} + \frac{1}{3} g_{\mu\rho} \phi_{\sigma} \, T_{\nu\lambda}{}^{,\sigma} \nonumber \\ &&+ \
\frac{1}{3} g_{\mu\lambda} \phi_{\sigma} \, T_{\nu\rho}{}^{,\sigma} -  \phi_{\mu\rho\sigma} \
T_{\nu}{}^{\sigma}{}_{,\lambda} -  \phi_{\mu\lambda\sigma} T_{\nu}{}^{\sigma}{}_{,\rho} + 2 g_{\lambda\rho} \phi_{\mu\sigma\kappa} \
T_{\nu}{}^{\sigma,\kappa} -  g_{\mu\rho} \phi_{\lambda\sigma\kappa} T_{\nu}{}^{\sigma,\kappa} \nonumber \\
&&-  g_{\mu\lambda} \phi_{\rho\sigma\kappa} \
T_{\nu}{}^{\sigma,\kappa} -  \phi_{\nu\rho\sigma} T_{\lambda}{}^{\sigma}{}_{,\mu} -  \phi_{\mu\rho\sigma} \
T_{\lambda}{}^{\sigma}{}_{,\nu} + 2 \phi_{\mu\nu\sigma} T_{\lambda}{}^{\sigma}{}_{,\rho} -  g_{\nu\rho} \phi_{\mu\sigma\kappa} \
T_{\lambda}{}^{\sigma,\kappa}   \nonumber \\
&&           -  g_{\mu\rho} \phi_{\nu\sigma\kappa} T_{\lambda}{}^{\sigma,\kappa} + 2 g_{\mu\nu} \phi_{\rho\sigma\kappa} \
T_{\lambda}{}^{\sigma,\kappa} -  \phi_{\nu\lambda\sigma} T_{\rho}{}^{\sigma}{}_{,\mu} -  \phi_{\mu\lambda\sigma} \
T_{\rho}{}^{\sigma}{}_{,\nu} + 2 \phi_{\mu\nu\sigma} T_{\rho}{}^{\sigma}{}_{,\lambda} \nonumber \\
&&-  g_{\nu\lambda} \phi_{\mu\sigma\kappa} \
T_{\rho}{}^{\sigma,\kappa} -  g_{\mu\lambda} \phi_{\nu\sigma\kappa} T_{\rho}{}^{\sigma,\kappa} + 2 g_{\mu\nu} \phi_{\lambda\sigma\kappa} \
T_{\rho}{}^{\sigma,\kappa} - 4 g_{\lambda\rho} \phi_{\mu\nu\kappa} T^{\sigma\kappa}{}_{,\sigma}     \nonumber \\
&&                   + 2 g_{\nu\rho} \phi_{\mu\lambda\kappa} \
T^{\sigma\kappa}{}_{,\sigma}     + 2 g_{\nu\lambda} \phi_{\mu\rho\kappa} T^{\sigma\kappa}{}_{,\sigma} + 2 g_{\mu\rho} \phi_{\nu\lambda\kappa} \
T^{\sigma\kappa}{}_{,\sigma} + 2 g_{\mu\lambda} \phi_{\nu\rho\kappa} T^{\sigma,\kappa}{}_{\sigma}   \nonumber \\
&&                    - 4 g_{\mu\nu} \phi_{\lambda\rho\kappa} \
T^{\sigma\kappa}{}_{,\sigma}      -  g_{\mu\rho} g_{\nu\lambda} \phi_{\kappa}{}^{\tau}{}_{\tau} T^{\sigma\kappa}{}_{,\sigma} -  \
g_{\mu\lambda} g_{\nu\rho} \phi_{\kappa}{}^{\tau}{}_{\tau} T^{\sigma\kappa}{}_{,\sigma} + 2 g_{\mu\nu} g_{\lambda\rho} \
\phi_{\kappa}{}^{\tau}{}_{\tau} T^{\sigma\kappa}{}_{,\sigma}\Bigr] \nonumber \\ &&
\end{eqnarray}


\end{document}